\documentclass[a4paper,11pt]{article}	
\pdfoutput=1 
\usepackage{jheppub} 

\usepackage[table]{xcolor}                  
\usepackage{colortbl}
\usepackage{booktabs}                       
\usepackage{multirow,bigdelim}              
\usepackage{longtable}                      
\usepackage{arydshln}                       

\usepackage{units}                          
\usepackage{amsmath,amssymb,mathtools}      
\usepackage{slashed}                        
\usepackage{amsthm}
\usepackage{bbm,dsfont}                     

\usepackage{bm}                             
\usepackage{upgreek}                        
\usepackage[vcentermath]{youngtab}          
\usepackage{ytableau}                       
\usepackage{empheq}                         
\usepackage{suffix}	                        
\usepackage{lscape}	                        

\usepackage{graphicx}                       
\usepackage{subfig}	                        
\usepackage{tikz}                           
\usepackage{float}
\usepackage[export]{adjustbox}              

\usepackage{url}                            
\usepackage{makeidx}                       
\usepackage[numbers,sort&compress]{natbib}  
\usepackage[colorlinks=true,urlcolor=blue,anchorcolor=blue,citecolor=blue,filecolor=blue,linkcolor=blue,menucolor=blue,pagecolor=blue,linktocpage=true,pdfproducer=medialab,pdfa=true]{hyperref}	                    

\usepackage{ifthen}	                        

\allowdisplaybreaks[2]          

\graphicspath{{figs/}}

\DeclareMathOperator{\Tr}{Tr}

\DeclareMathOperator{\Li}{Li}

\DeclareMathOperator{\diag}{diag}

\DeclareMathOperator{\real}{Re}
\DeclareMathOperator{\imag}{Im}

\def\mc{\mathcal}
\def\md{\mathbf}
\def\mf{\mathfrak}
\def\ms{\mathsf}

\newcommand{\bs}[1]{\boldsymbol{#1}}

\def\IC{\mathbb{C}}

\def\IQ{\mathbb{Q}}
\def\IR{\mathbb{R}}
\def\IZ{\mathbb{Z}}

\def\cO{\mathcal{O}}

\def\tq{\tilde{q}}
\def\tx{\tilde{x}}

\def\vev#1{\left\langle #1 \right\rangle}

\newcommand{\pd}{\partial}

\newcommand{\re}{{\rm e}}
\newcommand{\ri}{{\mathsf{i}}}
\newcommand{\rd}{{\rm d}}

\newcommand{\nn}{\nonumber \\}

\newcommand{\bb}{\ms{b}}
\newcommand{\s}{\sigma}
\newcommand{\knot}[1]{\md{#1}}
\DeclareMathOperator{\DD}{\text{D}}

\makeindex

\title{\boldmath Resurgence in complex Chern-Simons theory at generic levels}
\author[a]{Zhihao Duan}
\author[b]{Jie Gu}
\affiliation[a]{Korea Institute for Advanced Study\\
  85 Hoegiro, Dongdaemun-gu, Seoul 02455, Korea}

\affiliation[b]{School of Phyiscs and Shing-Tung Yau Center\\
  Southeast University, Nanjing 210096, China}%

\preprint{KIAS-P22064}
\emailAdd{xduanz@kias.re.kr}
\emailAdd{eij.ug.phys@gmail.com}

\abstract{In this note we study the resurgent structure of
  $sl(2,\IC)$ Chern-Simons state integral model on knot complements
  $S^3\backslash\knot{4}_1,S^3\backslash\knot{5}_2$ with generic
  discrete level $k\geq 1$ and with small boundary holonomy
  deformation. The coefficients of the saddle point expansions are in
  the trace field of the knot extended by the holonomy parameter.
  Despite increasing complication of the asymptotic series as the
  level $k$ increases, the resurgent structure of the asymptotic
  series is universal: both the distribution of Borel plane
  singularities and the associated Stokes constants are independent of
  the level $k$.}

\keywords{resurgence, complex Chern-Simonss theory, quantum knot invariants}

\begin{document}
\maketitle
\flushbottom

\section{Introduction}
\label{sc:intr}

Complex Chern-Simons theory remains an active field of research that
links up many fields in both physics and mathematics.  From the
physics side, $sl(2,\IC)$ Chern-Simons theory was first considered in
\cite{Witten:1988hc} as a means to describe three dimensional quantum
gravity on Lorentzian spacetime with positive cosmological constant.
Later a beautiful connection was revealed between $sl(N,\IC)$
Chern-Simons theory and 3d $N=2$ superconformal field theories by the
name of the 3d-3d correspondence
\cite{Dimofte:2010tz,Terashima:2011qi,Dimofte:2011ju,Dimofte:2013iv}
(see \cite{Dimofte:2014ija} for a review), which was used to define a
large class of 3d SCFTs.  Conversely, the BPS sector of the dual 3d
SCFT was used in \cite{Gukov:2016gkn,Gukov:2017kmk} to define
homological invariants $\widehat{Z}$ for the three manifold.

From the mathematical side, complex Chern-Simons theory plays an
important role in quantum topology.  It was found by Witten long time
ago \cite{Witten:1988hf} that Jones polynomials of knots, which are
Laurent polynomials in $q$, are vacuum expectation values of Wilson
loops along the knot in $SU(2)$ Chern-Simons theory with
$q = \exp(2\pi\ri/(k+2))$, where $k$ is the discrete level.  It is
then natural to consider more general complex values of $q$, which
translates to the complexification of the gauge group
\cite{Witten:2010cx}.
The complex Chern-Simons theory then provides a new topological
invariant for 3d manifolds.

For instance, when $M$ is a knot complement, the partition function of
the $sl(2,\IC)$ Chern-Simons theory on $M$ is generally believed to
reduce to a finite-dimensional integral called the state integral, as
proposed in
\cite{Andersen:2011bt,Andersen:2014aoa,Andersen:2016ugz,Dimofte:2011py,Dimofte:2014zga}
based on \cite{Hikami:2006cv,Dimofte:2009yn}, which is constructed
based on an ideal triangulation of the manifold.
The saddle points of the state integral in the weak coupling limit are
non-Abelian $sl(2,\IC)$ flat connections on the three manifold
\cite{Witten:1988hf}, and the associated asymptotic series
$\varphi^{(\s_j)}$ ($j\geq 1$) can be obtained by Gaussian expansion
\cite{Dimofte:2009yn}.  The state integral and thus the asymptotic
series are believed to be topological invariants as they do not change
with a different triangulation scheme.

On the other hand, the state integral model has the disadvantage that
it is incomplete in the sense that it misses the information of
Abelian flat connections \cite{Chung:2014qpa}.  Instead, the
asymptotic series $\varphi^{(\s_0)}(\tau)$ associated to the Abelian
flat connection $\sigma_0$ can be determined from the colored Jones
polynomials of the knot expanded in terms of $h = \log q$
\cite{Bar-Natan,Rozansky:1998urb,Kricker,GK:noncommutative}.

It is natural to study the resurgent structure of these asymptotic
series.  On the one hand, complex Chern-Simon theory is a special
quantum field theory with no renormalons, and therefore these
asymptotic series of saddle points must transform to each other by
Stokes automorphism, forming a so-called resurgent structure.  On the
other hand, the resurgence theory tells us that the Stokes constants
that control all the Stokes automorphisms define necessarily new
invariants which are non-perturbative in nature.  The resurgence
problem in Chern-Simons theory was first considered in
\cite{Garoufalidis:2007wa}, where it was found (and later emphasized
again in \cite{Garoufalidis:2021osl}) that the Borel transform of
$\varphi^{(\s_0)}(\tau)$ has poles at
$\text{Vol}(M)+\ri CS(M) +4\pi^2\IZ\ri$, which can be interpreted as a
resurgent formulation of the Volume Conjecture
\cite{Kashaev:1996kc,Murakami:2001cjp}.  Later the resurgent problems
of the series $\varphi^{(\s_j)}(\tau)$ ($j\neq 0$) for non-Abelian
saddle points were considered in
\cite{Gukov2016,Gang:2017hbs,Garoufalidis:2021lcp}.  See also
\cite{Garoufalidis:2020pax} on related results on Faddeev's quantum
dilogarithm which is a crucial ingredient of state integrals.

Recently the resurgent problem for the $sl(2,\IC)$ Chern-Simons theory
at level $k=1$ on the complement of the two simplest hyperbolic knots
$\knot{4}_1$ and $\knot{5}_2$ were completely solved in
\cite{Garoufalidis:2021osl,Garoufalidis:2020xec,Garoufalidis:2020nut}.
It was demonstrated that $\varphi^{(\s_0)}(\tau)$ of the Abelian flat
connection is related to $\varphi^{(\s_j)}(\tau)$ ($j\neq 0$) of
non-Abelian flat connections by Stokes automoprhisms but not the other
way around, while $\varphi^{(\s_j)}(\tau)$ ($j\neq 0$) transform to
themselves.  The Stokes constants of all these Stokes automorphisms
were computed explicitly.  Surprisingly the Stokes constants relating
$\varphi^{(\s_j)}(\tau)$ ($j\neq 0$) were found to coincide with the
BPS invariants in the dual 3d $N=2$ superconformal field theory on
$S^1$ \cite{Garoufalidis:2020xec,Garoufalidis:2020nut}, providing
another interesting entry of dictionary in the 3d-3d correspondence.
Furthermore, Stokes automorphisms were used to construct a new state
integral for the knot $\knot{4}_1$, which now includes the
contribution of the Abelian flat connection as well
\cite{Garoufalidis:2021osl}.  Finally, the statement that the series
$\varphi^{(\s_0)}(\tau)$ Borel resums to the $\widehat{Z}$ invariant
\cite{Gukov:2019mnk} was corrected to include additional contributions
from Borel resummation of $\varphi^{(\s_j)}(\tau)$
\cite{Garoufalidis:2021osl}.

In this work we continue this line of research, and generalise the
results of \cite{Garoufalidis:2020xec,Garoufalidis:2020nut} to
$sl(2,\IC)$ Chern-Simons theories with levels $k\geq 1$ with small
boundary holonomy deformation $x$.
We first demonstrate that the asymptotic series $\varphi^{(k,\s_j)}(x;\tau)$
($j\geq 1$) associated to non-Abelian flat connections are such that
their coefficients always live in the trace field of the knot extended
by the holonomy parameter $x$.
We then show that these asymptotic series enjoy universal resurgent
structures:
Both the distribution of Borel plane singularities and the associated
Stokes constants are independent of the level $k$, despite that the
asymptotic series themselves depend highly non-trivally on the level.

The remainder of the paper is structured as follows.  We summarise our
results, both on the properties of the asymptotic series and their
resurgent structure in section~\ref{sc:resurg}.  We then give details
in two example sections~\ref{sc:41},\ref{sc:52}.  Finally we conclude
and list open questions\ in section~\ref{sc:con}.

\section{Resurgent structure of complex Chern-Simons theory at generic
  level}
\label{sc:resurg}



\subsection{Perturbative complex Chern-Simons theory at generic level}

The $sl(2,\IC)$ Chern-Simons theory on a three manifold $M$ has action
\begin{equation}
  I_{k,s} = t S_{\text{CS}}(A) + \tilde{t} S_{\text{CS}}(\overline{A})
\end{equation}
where $S_{\text{CS}}(A)$ is the three dimensional Chern-Simons action
\begin{equation}
  S_{\text{CS}}(A) = \frac{1}{4\pi}\int_M \Tr (A\wedge\rd A + \frac{2}{3}A\wedge
  A\wedge A),
\end{equation}
$A$ is a connection of the $sl(2,\IC)$ bundle over $M$, and
$\overline{A}$ its complex conjugation.
The couplings $t,\tilde{t}$ can be split as
\begin{equation}
  t = \frac{1}{2}(k+\ri s),\quad \tilde{t} =\frac{1}{2}(k-\ri s).
\end{equation}
In a consistent quantum theory, the level $k$ must be an integer so
that the partition function of the Chern-Simons theory is invariant
under large gauge transformations.  The other level $s$ is constrained
to be either real or imaginary in order for the CS theory to be
unitary, which, however, we do not impose in this work.

Suppose the 3-manifold $M$ is a knot complement $M = S^3\backslash K$
with torus boundary.  The $sl(2,\IC)$ connection on the boundary torus
is parametrised by the holonomies $x,y$ along the meridian and the
longitude.  The condition that the $sl(2,\IC)$ connection can be
extended to the interior of $M$ constrains that $x,y$ are not
independent but are related by a polynomial equation
\begin{equation}
  A(x,y) = 0
\end{equation}
called the $A$-polynomial.
Furthermore, in a quantum theory, the holonomy $x$ should be
parametrised as
\begin{equation}
  x = \exp\left(\frac{2\pi \bb\mu}{k} - \frac{2\pi\ri n}{k}\right),
\end{equation}
where $\mu\in\IC$, $n\in\IZ_k$, and the parameter $\bb$ is related to
the levels $k,s$ by
\begin{equation}
  \bb^2 = \frac{k-\ri s}{k + \ri s}.
\end{equation}

By the 3d-3d correspondence, the $sl(2,\IC)$ Chern-Simons theory at
discrete level $k\geq 1$ on $M$ is dual to a 3d $N=2$ SCFT denoted by
$T_2[M]$ put on an orbifold of the squashed 3-sphere $S^3_\bb/\IZ_k$
\cite{Dimofte:2014zga},
\begin{equation}
  S^3_\bb/\IZ_k = \left\{(z,w)\in \IC^2 \,|\, \bb^2|z|^2 +
    \bb^{-2}|w|^2 = 1\right\}/ (z,w)\sim (\re^{2\pi\ri/k}z,\re^{-2\pi\ri/k}w).
\end{equation}
Here $\bb$ has the geometric meaning of the squashing parameter.  When
$M$ is a knot complement, the $T_2[M]$ theory has a $u(1)$ flavor
symmetry.  The parameters $\mu,n$ are then respectively the flavor
mass, and the holonomy of the flavor vector field.  The latter takes
value in $\pi_1(S^3_\bb/ \IZ_k) = \IZ_k$.

When $M = S^3\backslash K$ is a hyperbolic manifold, it can be
triangulated, namely decomposed to $N$ ideal tetrahedra glued along
their faces.  In \cite{Dimofte:2011ju}, the $T_2[M]$ theory was
constructed based on the triangulation data from the $T_2[\Delta]$
associated with an ideal tetrahedron $\Delta$, which is well
understood.  Based on this construction, the partition function of
$T_2[M]$ on $S^3_\bb/\IZ_k$ was computed \cite{Dimofte:2014zga}, and
it was used to give the state integral model for the $sl(2,\IC)$
Chern-Simons theory at level $k$.

It is important to note that, as first pointed out in
\cite{Chung:2014qpa}, the construction in \cite{Dimofte:2011ju} misses
an entire sector related to the Abelian flat connection, and thus the
partition function computed in \cite{Dimofte:2014zga} is also not
complete.  Rather, the state integral model should be regarded as the
reduced partition function of the $sl(2,\IC)$ Chern-Simons theory that
does not include the contribution from the Abelian flat connection.

Let us discuss in a bit detail the state integral model.  The
triangulation of $M$ in terms of $N$ tetrahedra is described by the
small Neumann-Zagier data\footnote{The complete Neumann-Zagier data
  introduced in \cite{Dimofte:2015kkp,Dimofte:2012qj} also includes a
  choice of the solution to the Neumann-Zagier equation.} that can be
encoded in a tuple $\gamma = (\pmb{A},\pmb{B},\nu)$ consisting
of 
two matrices $\pmb{A},\pmb{B}\in GL(N,\IZ)$ and a vector $\nu
\in\IZ^N$. They encode the coefficients of Thurston's gluing equations
for the triangulation including $N-1$ independent equations imposing
condition on internal edges, and one equation describing holonomy on
external edges.  We order rows of $\pmb{A},\pmb{B}$ and elements of
$\nu$ so that the first rows of $\pmb{A},\pmb{B}$ and the first element
of $\nu$ corresponds to the external edge.

It was shown in \cite{Neumann:1985vht} that $(\pmb{A}\;\pmb{B})$ forms
the top half of a symplectic matrix, so that $\pmb{A}\pmb{B}^T$ is
symmetric and $(\pmb{A}\;\pmb{B})$ is of full rank.  On the other hand,
giving a triangulation of $M$, the choice of Neumann-Zagier data is
not unique.  Following \cite{Dimofte:2015kkp,Dimofte:2012qj} we always
choose a set of Neumann-Zagier data where $\pmb{B}$ is invertible over
integers, and then $\pmb{B}^{-1}\pmb{A}$ must be symmetric.

Let us introduce in addition
$\bs{\sigma} = (\s_1,\ldots,\s_N) \in \IC^N$,
$\pmb{m} = (m_1,\ldots,m_N)\in(\IZ_k)^N$ as well as
$\bs{\mu} = (\mu,0,\ldots,0) \in \IC\times 0^{N-1}$,
$\pmb{n} = (n,0,\ldots,0) \in \IZ_k\times 0^{N-1}$.
By following the procedure in \cite{Dimofte:2014zga} and extending
slightly the formulas in \cite{Dimofte:2015kkp}, the state integral
model can be written as 
\begin{align}
  &\mc{Z}_\gamma^{(k)}(\mu,n;\bb)  =
    \frac{
    1
    }{k^N\sqrt{\det \pmb{B}}}
    \sum_{m\in (\IZ_k)^N}\int
    \rd^N \bs\sigma
    \textcolor{blue}{\re^{\frac{2\pi\ri}{k}(-\bs\sigma \pmb{B}^{-1}\bs\mu+\pmb{m} \pmb{B}^{-1}\pmb{n})}}
    \re^{\frac{\pi\ri}{k}(-\bs\sigma \pmb{B}^{-1}\pmb{A}\bs\sigma +
    \pmb{m} \pmb{B}^{-1}\pmb{A} \pmb{m} + 2 c_\bb\bs{\sigma}
    \pmb{B}^{-1}\bs{\nu})}\nn 
  &\phantom{===============}      
    (-1)^{\pmb{m} \pmb{B}^{-1}\pmb{A} \pmb{m}}
    \prod_{i=1}^N \mc{Z}_\bb^{(k)}[\Delta](\sigma_i,m_i).
      \label{eq:ZNgmMu0}
\end{align}
The part in blue is due to non-vanishing boundary holonomy and thus
new compared to the results in \cite{Dimofte:2015kkp}. Here
$\mc{Z}_\bb^{(k)}[\Delta](\sigma_i,m_i)$ is the partition function of
the Chern-Simons theory on a tetrahedron, which can be expressed in
terms of Faddeev's quantum dilogarithm $\Phi_\bb(x)$
\begin{equation}
  \mc{Z}_\bb^{(k)}[\Delta](\mu,n) =
  \prod_{(r,s)\in \Gamma(k;n)}\Phi_\bb(c_\bb
  -\frac{1}{k}(\mu+\ri \bb r + \ri\bb^{-1}s))
\end{equation}
where
\begin{equation}
  \Gamma(k;n) = \{(r,s)\in\IZ^2\;|\; 0\leq r,s<k, r-s \equiv n
  (\text{mod}\;k)\}
\end{equation}
and $c_\bb = \frac{\ri}{2}(\bb+\bb^{-1})$.  This defines a meromorphic
function of $\mu\in\IC$ for each $n\in\IZ_k$, and it is defined for
all values of $\bb$ with $\bb^2$ in the cut plane
$\IC' = \IC\backslash \IR_{\leq 0}$.
When $\imag \bb >0$ or $\imag\bb <0$, it has the factorisation form
\begin{equation}
  \mc{Z}_{\bb}^{(k)}[\Delta](\mu,n) = (qx^{-1};q)_\infty (\tq^{-1}\tx^{-1};\tq^{-1})_\infty
\end{equation}
where we use the notation
\begin{equation}
  \begin{gathered}
    q = \exp \frac{2\pi\ri}{k}(\bb^2+1),\quad \tq = \exp
    -\frac{2\pi\ri}{k}(\bb^{-2}+1),\\
    x = \exp \left(\frac{2\pi\bb\mu}{k} - \frac{2\pi\ri n}{k}\right),\quad
    \tx = \exp \left(\frac{2\pi\bb^{-1}\mu}{k} + \frac{2\pi\ri n}{k}\right),
  \end{gathered}
\end{equation}
and $(a,q)_\infty$ is the $q$-Pochhammer symbol defined to be
$\prod_{j=1}^\infty(1-a q^j)$ if $|q|<1$ or
$1/(q^{-1}a;q^{-1})_\infty$ if $|q|>1$.

One can give explicitly the contour of integral to make the $N$
dimensional integral convergent \cite{Dimofte:2014zga}.  But it is a
bit complicated and we specify the contours in individual examples in
Sections~\ref{sc:41}, \ref{sc:52}.

We are interested in the double scaling limit
\begin{equation}
  \bb\rightarrow 0,\quad\mu \rightarrow\infty, \quad
  \bb\mu\;\text{fixed},
\end{equation}
where the last condition guarantees that the holonomy parameter $x$
remains finite.
Let us introduce
\begin{equation}
  \zeta = \re^{2\pi\ri/k},\quad \tau = \bb^2
\end{equation}
By scaling the variables
\begin{equation}
  Z_i = 2\pi\bb\sigma_i, \quad u_i = 2\pi\bb\mu_i =
  \begin{cases}
    2\pi\bb\mu,\quad &i=1\\
    0,\quad &i\geq 2
  \end{cases}
\end{equation}
and using the asymptotic expansion \eqref{eq:ZDasymp} of the
tetrahedron partition function, the integral \eqref{eq:ZNgmMu0} can be
written as
\begin{align}
  \mc{Z}_\gamma^{(N)}(\mu,n;\tau) =
  &
    \frac{1
    }
    {k^N\sqrt{\det \pmb{B}}(4\pi^2\tau)^{N/2}}\nn
  &\sum_{\pmb{m}\in(\IZ_k)^N}(-1)^{\pmb{m} \pmb{B}^{-1}\pmb{A}\pmb{m}}
    \zeta^{\frac{1}{2}\pmb{m}\pmb{B}^{-1}\pmb{A}\pmb{m}+\pmb{m}\pmb{B}^{-1}\pmb{n}}
    \int\rd^N \pmb{Z} \exp \sum_{\ell=0}^\infty (2\pi\ri\tau)^{\ell-1}
    U^{(\pmb{m})}_\ell(\pmb{u},\pmb{Z})
\end{align}
where the coefficient functions are
\begin{subequations}
  \begin{align}
    &U_0(\pmb{u},\pmb{Z}) = \textcolor{blue}{\frac{1}{k}\pmb{Z}\pmb{B}^{-1}\pmb{u} +}
      \frac{1}{2k}\pmb{Z} \pmb{B}^{-1}\pmb{A}\pmb{Z}
      -\frac{\pi\ri}{k}\pmb{Z}\pmb{B}^{-1}\bs\nu
      +\frac{1}{k}\sum_{i=1}^N\Li_2(\re^{-Z_i})\label{eq:U0},\\
    &U^{(\pmb{m})}_{\ell\geq 1}(\pmb{u},\pmb{Z}) =
      -\frac{1}{2k}\pmb{Z}\pmb{B}^{-1}\bs\nu\delta_{\ell,1} + 
      \sum_{i=1}^N\sum_{s=1}^k
      \frac{B_\ell(s/k)}{\ell!}
      \Li_{2-\ell}(\re^{\frac{2\pi\ri}{k}(s+m_i)}\re^{-\frac{Z_i}{k}}).\label{eq:Ul}
  \end{align}
\end{subequations}
We drop the superscript $(\pmb{m})$ for $U_0$ since it does not depend
on the index $\pmb{m}$. We have also simplified the expression of
$U_0$ using the identity
\begin{equation}
  \sum_{s=1}^k \Li_2(\re^{2\pi \ri s/k}x) = \frac{1}{k}\Li_2(x^k),
\end{equation}
which can be easily proved by the series expansion of $\Li_2(x)$.  The
saddle point equations, i.e. the critical point equations for $U_0$
are
\begin{align}
  u_j + \sum_{i=1}^N A_{ji}Z_i + \sum_{i=1}^N B_{ji}
  \log(1-\re^{-Z_i}) = \pi\ri \nu_j,\quad j=1,\ldots,N,
  \label{eq:cr-eq-u}
\end{align}
or equivalently with
\begin{equation}
  z_i = \re^{Z_i},\quad w_i = \re^{u_i}
\end{equation}
we find
\begin{align}
  w_j \prod_{i=1}^N z_i^{A_{ji}}(1-z_i^{-1})^{B_{ji}} =
  (-1)^{\nu_j},\quad
  j=1,\ldots,N.
  \label{eq:cr-eq-w}
\end{align}
Note that at the end of the day, we must set $u_j=0$ or equivalently
$w_j = 1$ for $j=2,\ldots,N$.
Since
\begin{equation}
  x = \exp\frac{2\pi\ri}{k}\left(-\ri\bb\mu-n\right) =
  \exp \left(\frac{u}{k} - \frac{2\pi\ri n}{k}\right)
  = \zeta^{-n} w_1^{1/k},
\end{equation}
the critical equations can also be written as
\begin{subequations}
  \begin{align}
    &x^k\prod_{i=1}^N z_i^{A_{ji}}(1-z_i^{-1})^{B_{ji}} =
      (-1)^{\nu_j},\quad
      j=1,
      \label{eq:cr-eq-x1}\\
    &\prod_{i=1}^N z_i^{A_{ji}}(1-z_i^{-1})^{B_{ji}} =
      (-1)^{\nu_j},\quad
      j=2,\ldots,N.
      \label{eq:cr-eq-x2}
  \end{align}
\end{subequations}
They can be regarded as a deformation of the Neumann-Zagier
equations. Each solution $\pmb{z}^* = \exp \pmb{Z}^*$ to the critical
equations corresponds to a non-Abelian $sl(2,\IC)$ flat connection
$\sigma_*$ over $M$.

We can choose one of the critical points (saddle points)
$\pmb{Z}^* = (Z_i^*)$ and expand around the critical point by
$Z_j = Z_j^* + (2\pi\ri\tau)^{1/2} \delta Z_j$.  By an expansion first
in terms of $\tau$ and then performing Gaussian integral order by
order, we obtain a trans-series
$\Phi_{\gamma}^{(k,\s_*)}(\mu,n;\tau)$.  This was worked out in
\cite{Dimofte:2015kkp} when $\mu=0$ and $n=0$.  In the generic case
the calculation is similar, and we present the results here.

Let us introduce $z = \re^{Z}$, as well as
\begin{equation}
  z' = (1-z)^{-1},\quad z'' = 1-z^{-1}
\end{equation}
so that
\begin{equation}
  z z' z'' = -1.
\end{equation}
From \eqref{eq:Ul} it is necessary to choose a $k$-th
root of $Z^*$. We introduce $\theta_i$ such that
\begin{equation}
  (\theta_i)^k = z_i^*.
\end{equation}
%
%
Let us also define the Average and Vev of a function $g$
\begin{align}
  \mathop{\text{Av}}(g(\pmb{m})) =
  \frac{\sum_{\pmb{m}\in(\IZ_k)^N} a_{\pmb{m}}(x,\theta) g(\pmb{m})}
  {\sum_{\pmb{m}\in(\IZ_k)^N} a_{\pmb{m}}(x,\theta)}
\end{align}
with
\begin{align}
  a_{\pmb{m}}(x,\theta) =
  \textcolor{blue}{x^{-\pmb{m}\pmb{B}^{-1}}}
  (-1)^{\pmb{m}\pmb{B}^{-1}\pmb{A}\pmb{m}}
  \zeta^{\frac{1}{2}(\pmb{m}\pmb{B}^{-1}\pmb{A}\pmb{m}+\pmb{m}\pmb{B}^{-1}\bs\nu)}
  \theta^{-\pmb{B}^{-1}\pmb{A}\pmb{m}}
  \prod_{i=1}^N(\zeta\theta_i^{-1};\zeta)_{m_i}^{-1}.
  \label{eq:am}
\end{align}
and
\begin{align}
  \vev{g(\delta \pmb{Z})} = \frac{\int \rd^N \delta \pmb{Z}\,
  \re^{-\frac{1}{2}\delta \pmb{Z}^T \pmb{H} \delta \pmb{Z}}g(\delta \pmb{Z})}
  {\int\rd^N \delta \pmb{Z}\,
  \re^{-\frac{1}{2}\delta \pmb{Z}^T \pmb{H} \delta \pmb{Z}}}
\end{align}
with
\begin{align}
  \pmb{H} = \frac{1}{k}\left(-\pmb{B}^{-1}\pmb{A} + \Delta_{{z^*}'}\right),
\end{align}
and
\begin{equation}
  \Delta_z = \diag(z_1,\ldots,z_N).
\end{equation}

The perturbative expansion of the state integral \eqref{eq:ZNgmMu0}
near the critical point $\pmb{Z}^*$ then reads
\begin{align}
  \Phi_\gamma^{(k,\sigma_*)}(u,n;\tau) = 
  \re^{U^{(k)}_{0,0}(\sigma_*)/(2\pi\ri\tau)}
  \omega_{\gamma}^{(k,\sigma_*)}\varphi_{\gamma}^{(k,\sigma_*)}(u,n;\tau).
  \label{eq:Zpert}
\end{align}
The 1-loop contribution is
\begin{align}
  \omega_{\gamma}^{(k,\sigma_*)} = \frac{1}
  {(\ri k)^{N/2}\sqrt{\det(A\Delta_{{z^*}''}+B\Delta_{z^*}^{-1})
  {\pmb{z}^*}^{\pmb{B}^{-1}\pmb{\nu}/k}
  }}
  \prod_{i=1}^N D_k^*(\theta_i^{-1})^{1/k} \sum_{\pmb{m}\in(\IZ_k)^N}
  a_{\pmb{m}}(x,\theta).
  \label{eq:omega}
\end{align}
Here recall the \emph{cyclic quantum dilogarithms} are defined by
\begin{equation}
  D_k(z) = \prod_{s=1}^{k-1}(1-\zeta^s z)^s,\quad
  D_{k}^*(z) = \prod_{s=1}^{k-1}(1-\zeta^{-s}z)^s.
\end{equation}
Higher loop contributions are given in
\begin{align}
  \varphi_{\gamma}^{(k,\sigma_*)}(u,n;\tau) =
  \mathop{\text{Av}}\left(
  \vev{
  g(\delta \pmb{Z},\pmb{m};x,\theta)
  }
  \right) = 1 + \tau \,\IC[[\tau]]
  \label{eq:varphi}
\end{align}
and
\begin{align}
  g(\delta \pmb{Z},\pmb{m};x,\theta) =
  \exp\left(
  \sum_{\ell\geq 0}(2\pi\ri\tau)^{\ell+d/2-1}
  {\sum_{d\geq  0}}'U_{k,(i^d)}^{(\ell,\pmb{m})}\delta Z_i^d\right),
\end{align}
where ${\sum_{d\geq 0}}'$ means $d\geq 3$ for $\ell=0$, $d\geq 1$ for
$\ell =1$, and $d\geq 0$ for $\ell\geq 2$.  Besides, in each sum a
summation over $i=1,\ldots,N$ is implicit.  The coefficients are
\begin{subequations}
  \begin{align}
    U^{(k)}_{0,0} =
    &\frac{1}{k}\pmb{Z}^*\pmb{B}^{-1}\pmb{u}
      -\frac{\pi\ri}{k}\pmb{Z}^*\pmb{B}^{-1}\bs\nu
      +\frac{1}{2k}\pmb{Z}^*\pmb{B}^{-1}\pmb{A}\pmb{Z}^*+
      \frac{1}{k}\sum_{j=1}^N\Li_2(\re^{-Z_j^*}),
      \label{eq:U00}
    \\
    U^{(k)}_{0,(i^d)} =
    &\frac{(-1)^d}{k d!}\Li_{2-d}(\re^{-Z_i^*}),\quad d\geq 3,\\
    U^{(k,\pmb{m})}_{\ell\geq 1,(i^d)} =
    &\delta_{\ell,1}\delta_{d,0}
      \left(-\frac{1}{2k}\pmb{Z}^*\pmb{B}^{-1}\bs\nu\right)
      +\delta_{\ell,1}\delta_{d,1}
      \left(-\frac{1}{2k}(\pmb{B}^{-1}\bs\nu)_i\right)\nn+
    &\frac{(-1)^d}{k^d d!\ell!}\sum_{j=1}^N\sum_{s=1}^k
      B_\ell(s/k)
      \Li_{2-\ell-d}(\re^{\frac{2\pi\ri}{k}(s+m_j)}\re^{-\frac{Z_j^*}{k}})
      (\delta_{ij})^d.
  \end{align}
\end{subequations}

The trans-series $\Phi_\gamma^{(k,\sigma_*)}(u,n;\tau)$ has some very
important properties.
First of all, unlike the other coefficients, $U_{0,0}(\sigma_*)$ in
\eqref{eq:U00} depends on the level $k$ only as an overall factor
$1/k$,
\begin{equation}
  U^{(k)}_{0,0}(\sigma_*) = \frac{1}{k} U_{0,0}(\sigma_*).
  \label{eq:U00k}
\end{equation}
And $U_{0,0}(\sigma_*)$ only depends on the Neumann-Zagier data
$\gamma$, the choice of solution $\sigma_*$, and the holonomy
parameter $u = \log(x^k)$ (but not on the other holonomy parameter
$n$).  When $|u|\ll 1$, $U_{0,0}(\sigma_*)$ is the deformed
complexified hyperbolic volume of the knot complement
$S^3\backslash K$ \cite{Gukov:2003na}.

Next, even though the expression of the trans-series
$\Phi_\gamma^{(k,\s_*)}(u,n;\tau)$ depends explicitly on
$\theta_i^*$, a choice of the $k$-th root of $z_i^*$, one can show
that a different choice of $\theta_i^*$ merely amounts to an overall
constant factor.
The exponential $\exp U_{0,0}/(2\pi\ri \tau)$ manifestly
depends on $z_i^*$ instead of $\theta_i^*$.
Besides, one can prove that \emph{both
  $(\omega_{\gamma}^{(k,\sigma_*)})^{2k}$ and
  $\varphi_{\gamma}^{(k,\sigma_*)}(\tau)$ are invariant under the
  transformation $\theta_i\rightarrow \zeta\theta_i$}.  This has
already been proved when the holonomy $x$ is turned off in
\cite{Dimofte:2015kkp}, and the coefficients of the power series
$\varphi_{\gamma}^{(k,\sigma_*)}(\tau)$ are said to live in the trace
field of the knot.
When the holonomy $x$ is turned on, a similar proof following closely
\cite{Dimofte:2015kkp} can be written down with all the necessary
ingredients provided in \eqref{eq:omega},\eqref{eq:varphi}, and we do
not repeat the proof here.

In addition, the dependence on the holonomy of the trans-series
$\Phi_\gamma^{(k,\s_*)}(u,n;\tau)$ is quite elegant.  \emph{Both
  $(\omega_{\gamma}^{(k,\sigma_*)})^{2k}$ and
  $\varphi_{\gamma}^{(k,\sigma_*)}(\tau)$ depend on the combination
  $x = \exp(u/k - 2\pi\ri n/k)$ but not on $u$ and $n$ individually}.
This is because the NZ solution $z^*$ depends on $u$ and $n$
implicitly through the modified NZ equations \eqref{eq:cr-eq-x1},
\eqref{eq:cr-eq-x2}, which only depend on $x^k$.  On top of that,
$(\omega_{\gamma}^{(k,\s_*)})^{2k}$ and
$\varphi_{\gamma}^{(k,\s_*)}(\tau)$ depend on $u,n$ through the NZ
solution $z^*$ as well as $a_{\pmb{m}}(x,\theta)$, which is also a
function of $x$.
Combining the discussion of these two paragraphs, we thus claim that
the coefficients of the power series
$\varphi_\gamma^{(k,\s_*)}(u,n;\tau)$ are understood to live in
$F_{k,x}$, the trace field of the knot extended by holonomy $x$,
defined by
\begin{equation}
  F_{k,x} = F_k(x),\quad F_k = F(\zeta), \quad F = \IQ(z_1,\ldots,z_N).
\end{equation}

Finally, although a universal expression \eqref{eq:omega} of the
one-loop contribution $\omega_\gamma^{(k,\s_*)}$ at any level is
possible, the power series $\varphi_\gamma^{(k,\s_*)}$ depend on the
level in a very complicated way, as we will see in example
sections~\ref{sc:41.asymp},\ref{sc:52.asymp}.

We also comment that the trans-series \eqref{eq:Zpert} can be computed
from radial asymptotics of $q$-series that come from the evaluation of
the state integral model for $q$ at complex roots of unity, similar to
the discussion in \cite{Garoufalidis:2018qds}.
Examples of the $q$-series, also known as holomorphic blocks, are
shown in Sections~\ref{sc:41.blocks} and \ref{sc:52.blocks}.
They can be regarded as $x$-feformation of the Nahm sums.

Even though the form of trans-series \eqref{eq:Zpert} is elegant and
it allows us to infer the above universal properties, it is
computationally inefficient as it involves an $N$-dimensional
integral.  Often the state integral \eqref{eq:ZNgmMu0} can be
simplified.
For example in the case of complements of knots $\knot{4}_1$ and
$\knot{5}_2$, in Thurston's scheme of triangulation the number $N$ of
ideal tetrahedra is respectively $2$ and $3$, and their state integral
models involve a two- and a three-dimensional integral respectively.
However using integral identities of the tetrahedron partition
function $\mc{Z}_\bb^{(k)}[\Delta](\sigma,m)$, both of them can be
reduced to one-dimensional integrals
\cite{Dimofte:2014zga,Andersen:2016ugz}, and their perturbative
expansion in terms of Gaussian integral is much simpler.  This will be
demonstrated in Sections~\ref{sc:41}, \ref{sc:52}.
Nevertheless the universal properties of the asymptotic series
discussed in this section still hold.

\subsection{Resurgent structure of complex Chern-Simons theory}

The power series $\varphi_{\gamma}^{(k,\s_*)}(u,n;\tau)$ are
asymptotic.  To make sense of them, we need to apply the resurgence
theory \cite{Ecalle,Mitschi:2016fxp} (see reviews
\cite{Marino:2012zq,Dorigoni:2014hea,Aniceto:2018bis} by physicists).
Generically in physics perturbation series are asymptotic.  The
coefficients grow factorially due to the proliferation of Feynman
diagrams\footnote{Renormalons can also contribute to factorial growth
  of coefficients in a general QFT, but they are absent in complex
  Chern-Simons theory.},
\begin{equation}
  f(\tau) = \sum_{\ell=0}^\infty a_n\tau^n,\quad a_n\sim \cO(A^nn!).
\end{equation}
The series is divergent and it cannot be summed to an exact finite
value in the traditional sense.  Nevertheless, we can convert it into
an analytic function of $\tau$ by the means of Borel resummation.

We first construct the Borel transform of the
original series
\begin{equation}
  \mc{B}[f](t) = \sum_{\ell=0}^\infty \frac{a_n}{n!} t^n,
\end{equation}
which is now a convergent series with a finite radius of convergence
$1/A$.  The boundary of the disk of convergence is punctuated by
singular points.  If the singular points are sparse, the convergent
series can be analytically continued to the entire complex plane,
known as the Borel plane.  The Borel transform is said to be
\emph{resurgent} if its singular points are isolated and it can be
analytically continued to infinity in all directions bypassing its
singular points.  In this case, we can perform the Laplace transform
and define the Borel resummation
\begin{equation}
  s[f](\tau) = \int_0^\infty \mc{B}[f](\tau t) \re^{-t}\rd t.
\end{equation}
This definition involves integration along a ray $\rho_\theta$ of
angle $\theta =\arg\tau$.  If the Borel transform has no singular
points along the ray, the integral is well-defined and it gives us a
function of $\tau$.  If, however, the Borel transform does have a
singular point $w$ on the ray $\rho_\theta$, the integration is
obstructed, and we are missing non-perturbative corrections of the
order $\re^{-w/\tau}$ coming from another saddle point, and the action
of the new saddle point is $V_w = V_0 + w$, where $V_0$ is the action
at the original saddle point associated to the series $f$ we start
with.
We can define a pair of lateral Borel resummations
\begin{equation}
  s_{\pm}[f](\tau) = \int_0^{\re^{\pm \ri 0}\infty}
  \mc{B}[f](\tau t)\re^{-t} \rd t,
\end{equation}
and their difference, called the Stokes discontinuity
\begin{equation}
  \text{disc}_\theta[f](\tau) = s_{+}[f](\tau) - s_{-}[f](\tau),
\end{equation}
is of the order $\re^{-w/\tau}$.

The Borel transform $\mc{B}[f]$ is in addition called a \emph{simple
  resurgent function} if all its singular points are simple poles or
branch points.
In this paper, we assume there are only logarithmic branch points.
Then at the vicinity of a singular point at $w$, the resurgent
function $\mc{B}[f](t)$ has the form
\begin{equation}
  \mc{B}[f](t+w) = 
  -\frac{\mc{S}_w}{2\pi\ri} \mc{B}[g](t)\log(t) + \text{reg}(t).
\end{equation}
where both $\mc{B}[g](t)$ and $\text{reg}(t)$ are regular at $t=0$,
and the constant $\mc{S}_w$ is called the Stokes constant.  Here we
make it manifest that the regular function $\mc{B}[g](t)$ can be
regarded as the Borel transform of another asymptotic series
$g(\tau)$, which in fact is the perturbation series at the new saddle
point $w$.  Suppose no other singular points of $\mc{B}[f](t)$ share
the same argument $\theta = \arg w$, the Stokes discontinuity of the
original Borel resummation across the Stokes ray $\rho_\theta$ is
\begin{equation}
  \text{disc}_\theta[f](\tau) = \mc{S}_w\re^{-w/\tau} s_-[g](\tau).
  \label{eq:discStokes}
\end{equation}
To put it in a more democratic form, we can introduce elementary
trans-series
\begin{equation}
  F(\tau) = \re^{-V_0/\tau}f(\tau),\quad
  G(\tau) = \re^{-V_w/\tau}g(\tau).
\end{equation}
Then the formula of Stokes discontinuity reads
\begin{equation}
  \text{disc}_\theta[F](\tau) = s_+[F](\tau) - s_-[F](\tau)
  = \mc{S}_w s_-[G](\tau).
\end{equation}
Abstractly, it can also be represented as a linear transformation of
trans-series, whose coefficient is the Stokes constant,
\begin{equation}
  \mf{S}_\theta F = F+\mc{S}_w G.
\end{equation}
This is called the Stokes automorphism.

\begin{figure}
  \centering
  \includegraphics[height=3.5cm]{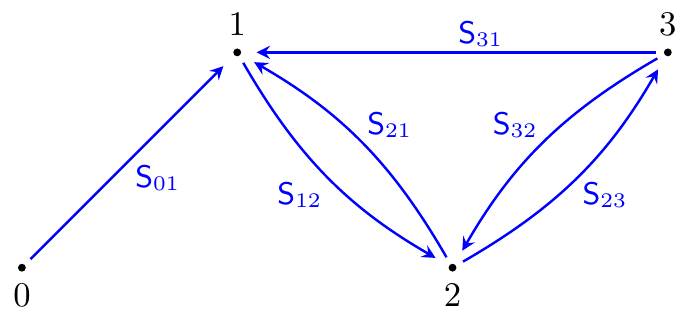}
  \caption{Demonstration of resurgent structure.}
  \label{fg:reg-net}
\end{figure}

In summary, starting from a single perturbation series $\varphi_0$ in
a theory, one can find new saddle points and their actions by looking
for singular points of the Borel transform $\mc{B}[\varphi_0]$, and
furthermore explore the asymptotic series associated to the new saddle
points by computing the Stokes discontinuity (or expansion of Borel
transform at the singular point) of $\varphi_0$.  The same procedure
can be repeated on the new asymptotic series to uncover additional
saddle points. In general, all the saddle points of the theory are
interconnected in this way, which is called the resurgent structure of
the theory, cf.~Fig.\ref{fg:reg-net}. These relationships are
completely controlled by Stokes constants, which define for us new
invariants of the theory.
We note that Stokes constants are not necessarily symmetric,
i.e.~$\mc{S}_{ij}\neq \mc{S}_{ji}$.

In practise we usually do not have the entire asymptotic series
$f(\tau)$ but only the truncated power series up to certain
(relatively high) order.  The Borel transform of the truncated power
series is only a polynomial which has no poles.
In order to mimic the singularity structure of the Borel transform of
the full series, we
make use of the Pad\'{e} approximant to approximate
$\mc{B}[f](t)$,
\begin{equation}
  \mathcal{P}_N[f](t) = \frac{p_0 + p_1 t +\cdots+ p_{N} t^{N}}
  {1 + q_1 t + \cdots + q_{N} t^{N}}\,.
\end{equation}
$p_i$ and $q_i$ are determined by requiring that the power series
expansion of $\mc{P}_N[f](t)$ in $t$ coincides with $\mc{B}[f](t)$ up
to order $2N$.  Thus the poles of $\mc{P}_N[f](t)$ can be used to
infer the singularities of the Borel transform.  Then we numerically
Laplace transform the Pad\'{e} approximant to obtain the Borel
resummation. The precision of the Pad\'{e} approximant can be improved
by utilizing, say, the conformal mapping method. One can find more
numerical techniques in \cite{Caliceti_2007}.

Suppose that the theory in question has finitely many saddle points
$\sigma_i$, with the associated trans-series
\begin{equation}
  F_i(\tau) = \re^{-V_i/\tau} f_i(\tau),\quad i =1,\ldots, r.
\end{equation}
We represent all the trans-series at once by an $r$-dimensional vector
$F(\tau)$
\begin{equation}
  F(\tau) =
  \begin{pmatrix}
    F_1(\tau)\\
    \vdots\\
    F_r(\tau)
  \end{pmatrix}.
\end{equation}
If saddle point $j$ appears as a singular point at $V_j - V_i$ of the
Borel transform of $F_i(\tau)$ and the Stokes constant is
$\mc{S}_{ij}$, and it is the only singular point on the Stokes ray
$\rho_{\theta_{ij}}$ where $\theta_{ij} = \arg(V_j-V_i)$, then the
Borel resummation above and below $\rho_{\theta_{ij}}$ are related by
the Stokes automorphism
\begin{equation}
  s_+[F](\tau) = \mf{S}_{\theta_{ij}} s_-[F](\tau),
  \label{eq:Sthij}
\end{equation}
where $\mf{S}_{\theta_{ij}}$ is an $r\times r$ matrix called the Stokes
matrix, and it has the form
\begin{equation}
  \mf{S}_{\theta_{ij}} = I + \mc{S}_{ij} E_{ij}
\end{equation}
with $I$ the identity matrix, and $E_{i,j}$ the elementary matrix with
$(i,j)$-entry $1$ and all other entries zero.

If there are multiple singular points on the ray $\rho_{\theta}$, we
assume the \emph{locality condition}: the Stokes matrices of any two
Borel plane singular points commute if they are on the same Stokes
ray.  In this case, \eqref{eq:Sthij} should be generalised to
\begin{equation}
  s_+[F](\tau) = \mf{S}_\theta s_-[F](\tau),\quad
  \mf{S}_\theta = \prod_{\arg\iota=\theta} \mf{S}_\iota.
\end{equation}
The order of the product of the local Stokes matrices is irrelevant
due to the locality condition.  Furthermore, given two rays
$\rho_{\theta^+}$ and $\rho_{\theta^-}$ whose arguments satisfy
$0<\theta^+-\theta^- \leq \pi$, we define the \emph{global Stokes
  automorphism}
\begin{equation}
  s_{\theta^+}[F](\tau) = \mf{S}_{\theta^-\rightarrow \theta^+} s_{\theta^-}[F](\tau)
\end{equation}
where both sides should be analytically continued to the same value of
$\tau$.  The global Stokes automorphism
$\mf{S}_{\theta^-\rightarrow \theta^+}$ has the property of unique
factorisation
\begin{equation}
  \mf{S}_{\theta^-\rightarrow \theta^+} =
  \prod_{\theta^-<\theta<\theta^+}^{\leftarrow}
  \mf{S}_\theta.
  \label{eq:Sfac}
\end{equation}
The superscript $\leftarrow$ indicates that the product is ordered so
that $\theta$ increases from right to left.

We can apply this theory to the trans-series
$\Phi_\gamma^{(k,\s_i)}(u,n;\tau)$ with $(i=1,\ldots,r)$ in
\eqref{eq:Zpert} coming from the state integral of $sl(2,\IC)$ Chern-Simons theory.
We also define the vector of trans-series
\begin{equation}
  \Phi_\gamma^{(k)}(u,n;\tau) =
  \begin{pmatrix}
    \Phi_\gamma^{(k,\s_1)}\\
    \vdots\\
    \Phi_\gamma^{(k,\s_r)}
  \end{pmatrix}(u,n;\tau).
  \label{eq:Zkgamma}
\end{equation}
We will be concerned with the regime where $|u|\ll 1$, which is
equivalent to $|x| \sim 1$.  We note that in this regime both holonomy
parameters $(u,n)$ can be read off uniquely from $x$: $u$ is given by
\begin{equation}
  u = \log(x^k),
\end{equation}
after which $n\in\IZ_k$ is chosen such that
\begin{equation}
  \zeta^n x = \re^u.
\end{equation}

The Borel transform $\mc{B}[\Phi_\gamma^{(k,\sigma_i)}](u,n;t)$ has
singularities at
\begin{equation}
  \Lambda_{i}^{(k),0} = \{\iota_{i,j}/k \;|\; j=1,\ldots,r,\; j\neq i\}
\end{equation}
where
\begin{equation}
  \iota_{i,j} = \frac{U_{0,0}(\s_i)-U_{0,0}(\s_j)}{2\pi\ri}.
\end{equation}
They correspond to the flat connections $j\neq i$ as predicted by the
resurgence theory.  Furthermore, there are additional singularities on
top of and slightly away from these singular points and on the
imaginary axis as well; the full set of singular points is
\begin{equation}
  \Lambda_{i}^{(k)} = \left\{
    \frac{\iota_{i,j}+2\pi\ri \ell + \log(x^k) m}{k}    \;\Big|\;
    j=1,\ldots,r,\;
    \ell\in\
    \begin{cases}
      \IZ\quad j\neq i\\
      \IZ_{\neq 0}\quad j=i
    \end{cases}\hspace{-2ex},\; m\in I(i,j,\ell)
  \right\}. \label{eq:Lmdu}
\end{equation}
Here $I(i,j,\ell)$ are finite and continuous sets of integers centered
on $0$.
These singularities form infinite towers in the imaginary direction
(see Figs.~\ref{fg:sing.k1.41}, \ref{fg:sing.k1.52} for
illustrations), and the Stokes rays passing through them are said to
form ``peacock patterns'' \cite{Garoufalidis:2020xec}.  This type of
distribution of Borel plane singularities is in fact quite
univeral. Similar patterns have already been seen in large $N$
expansion of Chern-Simons matrix integral
\cite{Pasquetti2010,Aniceto:2014hoa}, in closedly related topological
string free energies
\cite{Couso-Santamaria:2016vwq,Gu:2021ize,Alim:2021mhp,Grassi:2022zuk},
in earlier studies of complex Chern-Simons theory
\cite{Garoufalidis:2007wa}, and in other related works
\cite{Garoufalidis:2020pax}.
In the context of complex Chern-Simons theory, this reflects the fact
that due to ambiguity of the Chern-Simons action\footnote{For instance
  by a large gauge transformation the Chern-Simons action changes by
  $2\pi$.  This will change the complexified hyperbolic volume
  $U_{0,0}/(2\pi\ri)$ in \eqref{eq:U00k}, whose imaginary part is the
  Chern-Simons action.}, each trans-series
$\Phi_\gamma^{(k,\s_i)}(u,n;\tau)$ should be upgraded to a family of
trans-series with the same power series but with shifted instanton
action.  Recall the definition
\begin{equation}
  \tq = \exp \left(-\frac{2\pi\ri}{k\tau}- \frac{2\pi\ri}{k}\right),\quad
  \tx = \exp \left(\frac{\log (x^k)}{k\tau} + \frac{2\pi\ri n}{k}\right).
\end{equation}
It is convenient to parametrise the trans-series in each family as
\begin{equation}
  \Phi_\gamma^{(k,\s_i)_{\ell,m}}(u,n;\tau) = \tx^m\tq^\ell
  \Phi_\gamma^{(k,\s_i)}(u,n;\tau),\quad \ell,m\in\IZ.
\end{equation}
On the other hand, the Stokes constant that relates two trans-series
labelled by $(\s_i)_{\ell,m}$ and $(\s_j)_{\ell',m'}$ only depends on
$\s_i,\s_j$ and the differences $\ell-\ell'$, $m-m'$, and we denote it
by $\mc{S}_{i,j;\ell-\ell',m-m'}$. As a consequence, we only need to
study the resurgent properties of the vector of trans-series
$\Phi_\gamma^{(k)}(u,n;\tau)$ defined in \eqref{eq:Zkgamma}.  The Stokes
automorphism associated to the Borel plane singularity
$\iota_{i,j;\ell,m}/k = (\iota_{i,j}+2\pi\ri\ell+\log(x^k)m)/k$ is
given by
\begin{equation}
  s_+[\Phi_\gamma^{(k)}](u,n;\tau) = \mf{S}_{\iota_{i,j;\ell,m}/k}(\tx,\tq)
  s_-[\Phi_\gamma^{(k)}](u,n;\tau)
\end{equation}
where the Stokes matrix is
\begin{equation}
  \mf{S}_{\iota_{i,j;\ell,m}/k}(\tx,\tq) = I +
  \mc{S}_{i,j,\ell,m}\tx^m\tq^\ell E_{i,j}.
\end{equation}
Due to the factorisation property $\eqref{eq:Sfac}$, in order to
compute all the Stokes constants, it suffices to compute a suitable
choice of finite number of global Stokes automorphisms, and extract
the Stoks constants from their factorised form.  For instance we can
choose to compute
\begin{equation}
  \mf{S}^{(k)}_{+}(\tx,\tq) =
  \mf{S}^{(k)}_{0_{-}\rightarrow \pi+0_{-}}
  (\tx,\tq),\quad
  \mf{S}^{(k)}_{-}(\tx,\tq) =
  \mf{S}^{(k)}_{\pi+0_{-}\rightarrow 2\pi+0_{-}}
  (\tx,\tq).
\end{equation}
The entries of the global Stokes automorphisms are no longer
constants, but elements in $\IZ[\tx^{\pm 1}][[\tq]]$ (upper half
plane) or $\IZ[\tx^{\pm 1}][[1/\tq]]$(lower half plane).  See
\cite{Garoufalidis:2021osl,Garoufalidis:2020xec} for concrete
examples.

In the case of the two simplest hyperbolic knots
$\knot{4}_1, \knot{5}_2$, at level $k=1$ and for $|u|\ll 1$, the
global Stokes automorphisms $\mf{S}^{(1)}_{\pm}(\tx,\tq)$ have been
conjectured in \cite{Garoufalidis:2020xec,Garoufalidis:2020nut}.  In
this paper, we generalise the studies of these two knots to generic
level $k\geq 1$, and the main result of this paper is that the Stokes
automorphisms are \emph{independent} of the level $k$
\begin{equation}
  \mf{S}^{(k)}_{\pm}(\tx,\tq) = \mf{S}^{(1)}_{\pm}(\tx,\tq).
  \label{eq:Sk}
\end{equation}
This is rather surprising because the power series
$\varphi_{\gamma}^{(k,\s_*)}(u,n;\tau)$ are very different at
different levels.  We will demonstrate the complexity of power series
at higher levels and provide numerical evidence of \eqref{eq:Sk} in
Sections~\ref{sc:41}, \ref{sc:52}.  But before ending this section, we
give some philosophical argument for \eqref{eq:Sk}.  The following
subsection is more speculative.

\subsection{Stokes constants and BPS invariants}

An interesting discovery concerning the Stokes constants extracted
from the Stokes automorphisms $\mf{S}_{\pm}^{(k=1)}(\tx,\tq)$ was made
in \cite{Garoufalidis:2020xec,Garoufalidis:2020nut}: they were
identified with the BPS invariants of the 3d SCFT $T_2[M]$ related to
the $sl(2,\IC)$ Chern-Simons theory on $M= S^3\backslash K$ by the
3d-3d correspondence \cite{Dimofte:2011ju}.  In particular, the
entries of $\mf{S}_{+}^{(k=1)}(\tx,\tq)$ (respectively
$\mf{S}_{-}^{(k=1)}(\tx,\tq)$), which are power series in
$\IZ[\tx^{\pm 1}][[\tq]]$ (respectively
$\IZ[\tx^{\pm 1}][[\tq^{-1}]]$), were identified with linear
combinations of the rotated superconformal indices
$\text{Ind}_K^{\text{rot}}(m,\zeta;q)$
\cite{Dimofte:2011py,Beem:2012mb} with different magnetic fluxes $m$.
The identification is most clean in the $(1,1)$ entry associated to
the geometry flat connection $\s_1$, where one found
\begin{equation}
  \mf{S}_{+}^{(k=1)}(x,q)_{(1,1)} = \text{Ind}_K^{\text{rot}}(0,x;q).
  \label{eq:SInd}
\end{equation}
The original motivation of this work was to generalise this
relationship.

The identification \eqref{eq:SInd} has not been proved. Nevertheless,
it should certainly be understood in the framework of 3d-3d
correspondence, which in particular claims that the supersymmetric
vacua of the dual 3d SCFT are given by $sl(2,\IC)$ flat connections on
$M$ \cite{Dimofte:2010tz}, which are precisely the saddle points of
the complex Chern-Simons theory.
At a generic level $k\geq 1$, on the one hand, the Hilbert space of the
quantised Chern-Simons theory is merely \cite{Dimofte:2014zga}
\begin{equation}
  \mc{H}^{(k)} = \mc{H}^{(k=1)}\otimes_\IC \IC^k.
\end{equation}
On the other hand, there are no more BPS states on the side of the
SCFT \footnote{There are missing BPS states concerning the Abelian
  flat connection $\s_0$. But they are related to the Stokes
  automorphisms of the asymptotic series associated to $\s_0$, which
  is invisible in the state integral model.  See some discussion in
  Section~\ref{sc:con}.}.  It then seems natural that the Stokes
automorphism at generic level $k\geq 1$ should still be identified
with the 3d BPS invariants, and therefore be independent of the level
$k$ of the Chern-Simons theory.

\section{Figure eight}
\label{sc:41}

\subsection{Asymptotic series}
\label{sc:41.asymp}

The state integral of figure eight knot at level $k$ is already given
as a one-dimensional integral in \cite{Andersen:2016ugz}
\begin{align}
  \chi^{(k)}_{\knot{4}_1}(\mu,n;\tau) =
  &\frac{\eta_k}{k}\sum_{m\in \IZ_k}\int_{\IR+\ri\imag(c_\bb)-\ri|\mu|-\ri 0}\rd \sigma
    \mc{Z}^{(k)}_\bb[\Delta](\sigma-\mu,m-n)
    \mc{Z}^{(k)}_\bb[\Delta](\sigma,m)\nn
  &\phantom{==}\times(-1)^m
    \re^{\frac{\pi\ri}{k}((\sigma-c_\bb)^2-2(\mu-\sigma+c_\bb)^2-m^2+2(n-m)^2)},
      \label{eq:chi-41}
\end{align}
where $\eta_k = \exp \frac{\pi\ri}{6}(k+2 c_\bb^2/k)$.
This can be derived from the generic formula \eqref{eq:ZNgmMu0}.
We can choose the Neumann-Zagier data
\begin{equation}
  \pmb{A} =
  \begin{pmatrix}
    1&0\\1&1
  \end{pmatrix},\quad
  \pmb{B} =
  \begin{pmatrix}
    0&-1\\-1&-1
  \end{pmatrix},\quad
  \bs\nu =
  \begin{pmatrix}
    0\\0
  \end{pmatrix},
\end{equation}
with which \eqref{eq:ZNgmMu0} yields
\begin{align}
  &\mc{Z}_\gamma^{(k)}(\mu,n;\bb)\nn =
  &\frac{1}{\ri k^2}\sum_{m\in (\IZ_k)^2}\int \rd^2\sigma
    \re^{\frac{2\pi\ri}{k}(-\sigma_1\mu_1+\sigma_2\mu_1+m_1n_1-m_2n_1)}
    \re^{\frac{\pi\ri}{k}(2\sigma_1\sigma_2-2m_1m_2)}
    \mc{Z}_\bb^{(k)}[\Delta](\sigma_1,m_1)
    \mc{Z}_\bb^{(k)}[\Delta](\sigma_2,m_2) \nn=
  &\frac{\re^{\frac{\pi\ri}{12}(k+\frac{8c_\bb^2}{k})}}{\ri k}
    \sum_{m\in\IZ_k}
    \int\rd \sigma
    \mc{Z}_\bb^{(k)}[\Delta](\sigma,m)
    \mc{Z}_\bb^{(k)}[\Delta](\sigma+\mu,m+n)\nn
  &\phantom{=========}
    (-1)^{m+n}
    \exp\frac{\pi\ri}{k}\left(
    -(\sigma-c_\bb)^2-4\mu\sigma-\mu^2+2c_\bb\mu+m^2+4mn+n^2
    \right).
\end{align}
In the last step, we have integrated over $\sigma_2$ and summed
over $m_2$ using the Fourier transformation \eqref{eq:FourierFQD}.
We then relabeled $(\sigma_1,m_1)$ to $(\sigma,m)$ as well as
$(\mu_1,n_1)$ to $(\mu,n)$.  After changing the signs of $\mu,n$ we
find the same expression as \eqref{eq:chi-41} up to an overall
irrelevant factor
\begin{equation}
  \chi^{(k)}_{\knot{4}_1}(\mu,n;\tau) = \ri
  \re^{\frac{\pi\ri}{12}(k-\frac{4c_{\bb}^2}{k})}(-1)^n
  \re^{\frac{\pi\ri}{k}(-\mu^2-2c_\bb\mu+n^2)}
  \mc{Z}_{\gamma}^{(k)}(-\mu,-n;\hbar).
\end{equation}

As a one-dimensional integral, it is much easier to perform saddle
point analysis on \eqref{eq:chi-41}.  We first introduce an
alternative representation of the tetrahedron partition
function\footnote{This version of tetrahedron partition function was
  introduced in \cite{Andersen:2016ugz}.}
\begin{equation}
  \DD_\bb^{(k)}(u,m) = \mc{Z}_\bb^{(k)}[\Delta](c_\bb-\sqrt{k}u,m),
  \label{eq:Dk}
\end{equation}
in terms of which, the state integral \eqref{eq:chi-41} reads
\begin{align}
  \chi^{(k)}_{\knot{4}_1}(\mu,n;\tau) =
  &\frac{\eta_k}{k}\sum_{m\in \IZ_k}
    (-1)^m \re^{\frac{\pi\ri}{k}(2n^2-4nm+m^2)}
    \nn
  &\phantom{=}\times
    \int_{\IR+\ri 0}\rd \sigma
    \DD^{(k)}_\bb(\frac{1}{\sqrt{k}}(\sigma+\mu),m-n)
    \DD^{(k)}_\bb(\frac{1}{\sqrt{k}}\sigma,m)
    \re^{\frac{\pi\ri}{k}(-2\mu^2-4\mu\sigma-\sigma^2)}.
\end{align}
We scale the variables
\begin{equation}
  Z = 2\pi\bb \sigma,\quad u =2\pi\bb \mu
\end{equation}
and then perform asymptotic expansion of the tetrahedron partition
functions in the small $\bb$ limit using \eqref{eq:ZDasymp},
\begin{equation}
  \chi_{\knot{4}_1}^{(k)}(\mu,n;\bb) \sim
  \frac{\eta_k}{2\pi\bb k}\sum_{m\in\IZ_k}
  (-1)^m \re^{\frac{\pi\ri}{k}(2n^2-4nm+m^2)}
  \int \rd Z \exp\sum_{\ell\geq 0}(2\pi\ri\bb^2)^{\ell-1} U_\ell(u,Z)
\end{equation}
where the potential functions are
\begin{subequations}
  \begin{align}
    U_0(u,Z) =
    &\frac{1}{k}\left(
    u^2+2uZ +\frac{Z^2}{2} + 
    \Li_2(-\re^{Z}) + \Li_2(-\re^{u+Z})\right),\\
    U_{\ell\geq 1}(u,Z) =
    &\frac{1}{\ell!}\sum_{j\in\IZ_k}
      B_\ell(1-1/(2k)-j/k)
      \left(
      \Li_{2-\ell}(\zeta^{m-j-1/2} \re^{Z/k})
      +\Li_{2-\ell}(\zeta^{m-j-1/2}\re^{Z/k}x)
      \right).
  \end{align}
\end{subequations}
The critical equation of the leading order potential function reads
\begin{equation}
  0 = \frac{\pd}{\pd Z}U_0(u,Z) = \frac{1}{k}\left(2u+Z -\log(1+\re^{Z})
    -\log(1+\re^{u+Z})\right),
\end{equation}
which can be simplified to the algebraic equation
\begin{equation}
  -X^2 Y = (1-XY)(1-Y),
  \label{eq:alcr.41}
\end{equation}
with
\begin{equation}
  X = \re^u = x^k,\quad Y = -\re^{Z}.
\end{equation}
We also introduce the variable $\theta$ such that
\begin{equation}
  Y = -\theta^k.
\end{equation}

The critical point equation \eqref{eq:alcr.41} has two solutions in
terms of $Y$, corresponding to the geometric and conjugate flat
connections $\s_1, \s_2$ on the knot complement.
Numerically, they are the ones that in the limit $X\rightarrow 1$
reduce respectively to the solutions $\re^{\pi\ri/3}$ and
$\re^{-\pi\ri/3}$ of \eqref{eq:alcr.41}.
Near each critical point $\s_j$, the integrand is approximated by the
exponential
\begin{equation}
  \exp\left(\frac{V^{(k)}(\s_j)}{2\pi\ri\tau}\right),
  \label{eq:exp}
\end{equation}
where
\begin{align}
  V^{(k)}(\s_j) = U_0(\log(X),\log(-Y_j))+\frac{\pi^2}{6k}.
\end{align}
The extra factor $\pi^2/(6k)$ comes from the factor $\eta_k$.
Note the interesting relation $V^{(k)}(X,Y) = V^{(1)}(X,Y)/k$.

In addition, by expanding near the critical point and performing the
Gaussian integration order by order, for each critical point $\s_j$ we
can find an asymptotic power series $\varphi^{(k,\s_j)}(x;\tau)$,
which, together with the exponential factor \eqref{eq:exp}, forms the
trans-series
\begin{equation}
  \Phi_{\knot{4}_1}^{(k,\sigma_j)}(x;\tau) =
  \re^{V^{(k)}(\s_j)/(2\pi\ri\tau)}
  \varphi_{\knot{4}_1}^{(k,\s_j)}(x;\tau),\quad
  j=1,2.
\end{equation}
The one-loop contribution
$\varphi_{\knot{4}_1}^{(k,\s_j)}(x;0) =
\omega^{(k,\s_j)}_{\knot{4}_1}(x)$ has a universal
expression
\begin{align}
  \omega^{(k,\s_j)}_{\knot{4}_1}(x)=
  &\frac{\re^{\frac{\pi\ri}{4}+\frac{\pi\ri  k}{6}+\frac{2\pi\ri}{k}(n^2-1/12)}}
    {\sqrt{k(x^k\theta^{2k}-1)}}\theta^{1/2}x
    D_k^{*}(\zeta^{-1/2}\theta)^{1/k}
    D_k^{*}(\zeta^{-1/2}\theta x)^{1/k}\nn
  &\times\sum_{m\in\IZ_k}
    (-1)^m\zeta^{m^2/2}\theta^m x^{2m}
    (\zeta^{1/2}\theta;\zeta)_m^{-1}(\zeta^{1/2}\theta x;\zeta)_m^{-1}.
\end{align}
The power series
$\varphi_{\knot{4}_1}^{(k,\s_j)}(x;\tau)/\omega^{(k,\s_j)}_{\knot{4}_1}(x)$
on the other hand, depends in a very complicated way on the level $k$.

The case of $k=1$ is particularly simple. The dependence on
$n\in\IZ_1$ is trivial, and the state integral reduces to
\cite{Garoufalidis:2020xec}
\begin{align}
  \chi_{\knot{4}_1}(\mu;\tau) =
  \chi_{\knot{4}_1}^{(1)}(\mu;\tau) =
  \Phi_\bb(0)^{-2}\int_{\IR+\ri 0} \rd v \Phi_\bb(\mu+v)\Phi_\bb(v)
  \re^{\pi\ri(v^2-2(\mu+v)^2)}.
\end{align}
where we have used $\eta_1 = \Phi_\bb(0)^{-2}$.  The power series
$\varphi_{\knot{4}_1}^{(\s_i)}(x;\tau) =
\varphi_{\knot{4}_1}^{(k=1,\s_i)}(x;\tau)$ are given by
\cite{Garoufalidis:2020xec}
\begin{align}
  (\omega_{\knot{4}_1}^{(\s_1)})^{-1}
  \varphi_{\knot{4}_1}^{(\s_1)}(x;\frac{\tau}{2\pi\ri}) =
  &1-
    \frac{1}{24\gamma(x)^3}\left(
    x^{-3}-x^{-2}-2x^{-1}+15-2x-x^2+x^3\right)\tau\nn  +
  &\frac{1}{1152\gamma^6(x)}\left(
    x^{-6}-2x^{-5}-3x^{-4}+610x^{-3}-606x^{-2}-1210x^{-1}\right.\nn+
  &\left.3117 -1210x-606x^2+610x^3-3x^4-2x^5+x^6
    \right)\tau^2+\cO(\tau^3).
    \label{eq:vphi1.k1}
\end{align}
with
\begin{equation}
  \gamma(x) = \sqrt{x^{-2}-2x^{-1}-1-2x+x^2}.
\end{equation}
while $\varphi_2^{(\s_2)}(x;\tau) = \ri \varphi_1^{(\s_2)}(x;-\tau)$.
At level $k=2$, the power series are
\begin{align}
  &(\omega_{\knot{4}_1}^{(k=2,\s_j)})^{-1}\varphi_{\knot{4}_1}^{(k=2,\s_j)}(x;\tau)\nn
    =
  &1 -((1 + 6 x + 5 x^2 - 30 x^3 + 10 x^4 + 30 x^5 - 3 x^6 + 30 x^7 + 
    10 x^8 - 30 x^9 + 5 x^{10} + 6 x^{11} + x^{12}) \nn
  &(-1 - x^2 + x^4 + 2 x^2 Y_j))/(48 (-1 - x + x^2)^2 (1 - x + x^2)^2 (-1 + x + 
    x^2)^2 (1 + x + x^2)^2)\tau + \cO(\tau^2).
    \label{eq:phi41k2}
\end{align}
%
Additional terms as well as first few terms of the power series at
level $k=3$ can be found in Appendix~\ref{sc:phi.41}.

Several conclusions can be drawn from these data.  It is clear that
the coefficients of the power series depend only on the solution $Y_j$
to the NZ equation, but not on a $k$-th root of $Y_j$.  Furthermore,
they depend on the deformation parameters $u,n$ only through the
holonomy parameter $x$.  Finally, it is evident that the power series
become increasingly more complicated at higher levels $k$.

\subsection{Holomorphic blocks}
\label{sc:41.blocks}

One happy fact of state integral model of knot $\knot{4}_1$ at level 1
is that when $\imag{\bb}\neq 0$, it can be evaluated by summing over
residues and it factorises to products of $q$- and $\tq$-series
\cite{Garoufalidis:2013rca}, i.e., the holomorphic blocks
\cite{Beem:2012mb}.  In addition, variants known as the descendants of
the state integral can be written, which also enjoys factorisation.
This fact turns out to be very useful in the resurgence analysis, as
the Stokes matrices can be written in terms of them
\cite{Garoufalidis:2020xec,Garoufalidis:2020nut}.  We find here at
generic level $k$, we can also write down descendants of the state
integral and they also factorise to holomorphic blocks.  In fact, the
holomorphic blocks are independent of $k$ (only implicitly through
$x,\tx,q,\tq$, cf.~\cite{Dimofte:2014zga}), while the factorisation
formulas are not.

We introduce the following descendant state integral for knot
$\knot{4}_1$
\begin{align}
  \chi^{(k)}_{\knot{4}_1;r,s}(\mu,n;\tau) = \frac{\eta_k}{k}
  &\sum_{m\in\IZ_k}\int_{\mc{C}}
    \rd \sigma \mc{Z}_\bb^{(k)}[\Delta](\sigma-\mu,m-n)
    \mc{Z}_{\bb}^{(k)}[\Delta](\sigma,m)\nn
  &\times(-1)^m
    \re^{\frac{\pi\ri}{k}((\sigma-c_\bb)^2-2(\mu-\sigma+c_\bb)^2-m^2+2(n-m)^2)}\nn
  &\textcolor{blue}
    {\times\re^{\frac{2\pi\ri}{k}(-\ri(r\bb+s\bb^{-1})(c_\bb-\sigma)+(r-s)m)}}.
\end{align}
The motivation for the particular form of the descendant state
integral is that both the tetrahedron partition function
$\mc{Z}_\bb^{(k)}[\Delta](\sigma,m)$ and the extra factor colored in blue
\begin{equation}
  f(\sigma,m) = \exp\left(\frac{2\pi}{k}(r\bb+s\bb^{-1})(c_\bb-\sigma)
    +\frac{2\pi\ri}{k}(r-s)m\right)
\end{equation}
in the integrand enjoys nice quasi-periodicity property under the
shift
\begin{equation}
  (\sigma,m) \rightarrow (\sigma-\ri\bb^{\pm 1},m \pm 1),\quad
  (\sigma,m) \rightarrow (\sigma+\ri\bb^{\pm 1},m \mp 1).
\end{equation}
The former is given in \eqref{eq:Db-per}, and the latter is simply
\begin{subequations}
\begin{align}
  &f(\sigma\mp\ri\bb,m \pm 1) = f(\sigma,m) q^{\pm r},\\
  &f(\sigma\pm\ri\bb^{-1},m \pm 1) = f(\sigma,m)\tq^{\pm s}.
\end{align}
\end{subequations}
In addition, the form of the descendant state integral may be related
to insertion of a light knot from the Chern-Simons theory point of
view \cite{Agarwal:2022cdm}, or to turning on FI parameters from the
3d SCFT point of view \cite{Yoshida:2014ssa}.

Note that compared to the state integral
$\chi_{\knot{4}_1}^{(k)}(\mu,n;\tau)$ defined in \eqref{eq:chi-41},
here the contour of integral $\mc{C}$ has to be modified such that it
asymptotes at both $\pm \infty$ to the horizontal line
$\imag{\sigma} = \sigma_0$ with
$\sigma_0 <\imag(c_\bb)-|\imag(\mu)|-|\real(r\bb+s\bb^{-1})|$ but is
deformed near the origin so that all the poles of the integrand are
below the contour.  The integrand only has poles coming from the two
tetrahedron partition functions
$\mc{Z}^{(k)}_\bb[\Delta](\sigma-\mu,m-n)$ and
$\mc{Z}^{(k)}_\bb[\Delta](\sigma,m)$, and they are located
respectively at
\begin{align}
  &v = \mu
    -\ri\bb\alpha-\ri\bb^{-1},\quad
    \alpha-\beta \equiv m-n \;(\text{mod}\; k),\\
  &\sigma =
    -\ri\bb \,\alpha -\ri\bb^{-1} \beta,\quad
    \alpha-\beta \equiv m \;(\text{mod}\; k).
\end{align}
On the other hand, since the integrand decays exponentially fast
towards infinity below the contour $\mc{C}$, we can evaluate the
descendant state integrand by completing the contour from below and
adding up residues from these poles.
Then we find in this way that
\begin{align}
  \chi^{(k)}_{\knot{4}_1;r,s}(\mu,n;\tau)=
  &\re^{-\frac{\pi\ri k}{4}-\frac{\pi\ri}{2k}-\frac{\pi\ri}{4k}(\tau+\tau^{-1})}
    A_r(x,q)A_{-s}(\tx,\tq^{-1}) \nn
  &+\re^{\frac{\pi\ri          k}{4}+\frac{\pi\ri}{2k}+\frac{\pi\ri}{4k}(\tau+\tau^{-1})}
    B_r(x,q)B_{-s}(\tx,\tq^{-1}).
\end{align}
Here $A_r(x,q), B_r(x,q)$ are descendant holomorphic blocks defined by
\begin{align}
  &A_r(x,q) = \theta(-q^{1/2}x;q)^{-2}x^{2r} J(q^rx^2,x;q),\\
  &B_r(x,q) = \theta(-q^{-1/2}x;q)x^r J(q^r x,x^{-1};q),
\end{align}
where $J(x,y;q^\varepsilon):= J^{\varepsilon}(x,y;q)$ for $|q|<1$ and
$\varepsilon = \pm$ is the $q$-Hahn Bessel function
\begin{subequations}
  \begin{align}
    &J^+(x,y;q) = (qy;q)_\infty \sum_{n=0}^\infty
      (-1)^n\frac{q^{\frac{1}{2}n(n+1)}x^n}{(q;q)_n(qy;q)_n}\\
    &J^-(x,y;q) = \frac{1}{(y;q)_\infty} \sum_{n=0}^\infty
      (-1)^n\frac{q^{\frac{1}{2}n(n+1)}x^ny^{-n}}{(q;q)_n(qy^{-1};q)_n}.  
  \end{align}
\end{subequations}
Note that the definition of descendant holomorphic blocks as function
of $x,q$ are independent of $k$.  They enjoy the same properties as
discussed in \cite{Garoufalidis:2020xec}.  If we define the descendant Wronskian matrix
\begin{equation}
  W_r(x;q) =
  \begin{pmatrix}
    A_r(x;q) & B_{r}(x;q)\\
    A_{r+1}(x;q) & B_{r+1}(x;q)
    \label{eq:W.41}
  \end{pmatrix},
\end{equation}
it enjoys the inversion property that
\begin{equation}
  W_m(x;q)
  \begin{pmatrix}
    1&0\\0&-1
  \end{pmatrix}
  W_{\ell}(x;q^{-1})^T \in GL(2,\IZ[q^{\pm 1}x^{\pm 1}]).
\end{equation}
In particular
\begin{equation}
  W_{-1}(x;q)
  \begin{pmatrix}
    1&0\\0&-1
  \end{pmatrix}
  W_{-1}(x;q^{-1})^T =
  \begin{pmatrix}
    x^{-2}+x^{-1}-1 & 1 \\ 1 & 0
  \end{pmatrix}.
  \label{eq:Wcont.41}
\end{equation}

\subsection{Resurgent structure}
\label{sc:41.lvlk}

\begin{figure}
  \centering
  \includegraphics[height=6cm]{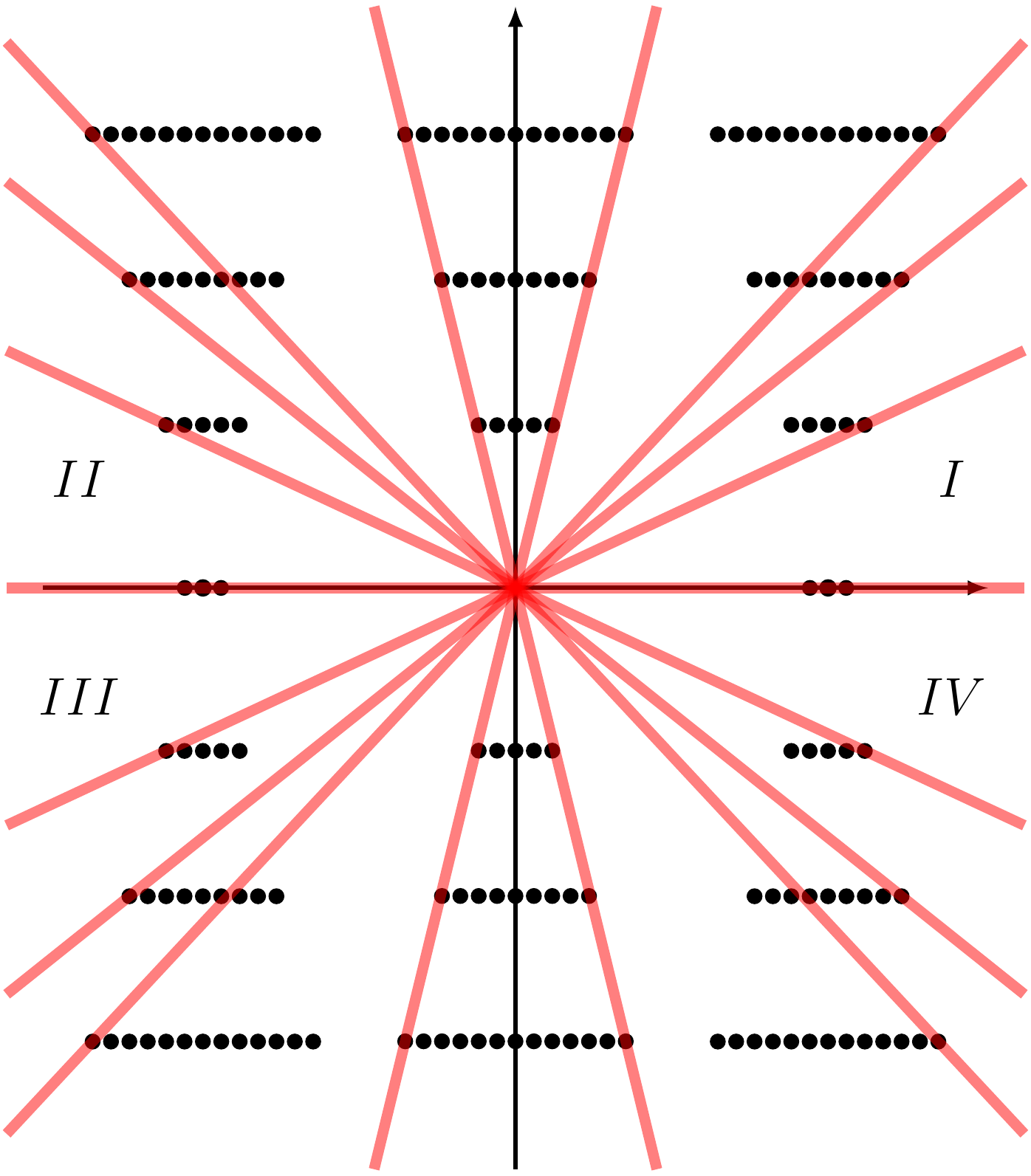}
  \caption{Singular points of the Borel transform of the vector of
    trans-series of the $sl(2,\IC)$ Chern-Simons theory at level $k=1$
    on the complement of the knot $\knot{4}_1$.}
  \label{fg:sing.k1.41}
\end{figure}

We collect the trans-series associated to the geometric and conjugate
flat connections in a vector
\begin{equation}
  \Phi_{\knot{4}_1}^{(k)}(x;\tau) =
  \begin{pmatrix}
    \Phi_{\knot{4}_1}^{(k,\s_1)}\\
    \Phi_{\knot{4}_1}^{(k,\s_2)}
  \end{pmatrix}(x;\tau).
\end{equation}
The singularities of the Borel transform of each component
trans-series are located in $\Lambda_{i}^{(k)}$ given in
\eqref{eq:Lmdu}.  In the case of level $k=1$, we superimpose the Borel
singularities of the components of $\Phi_{\knot{4}_1}^{(1)}(x;\tau)$
and plot them in Fig.~\ref{fg:sing.k1.41}.
%
%
At a generic level $k\geq 1$, the
distrubtion of the Borel plane singularities is the same and their
actual locations are reduced by a factor of $1/k$ according to
\eqref{eq:Lmdu}.

As explained in Section~\ref{sc:resurg}, Stokes constants of
individual Borel plane singularities can be extracted from global
Stokes automorphisms.
The entire Borel plane is divided by Stokes rays of components of
$\Phi_{\knot{4}_1}^{(k)}(x;\tau)$ into infinitely many cones.  We
label the four cones above and below the positive and negative real
axes by $I,II,III,IV$, ordered in counter-clockwise direction, as
illustrated in Fig.~\ref{fg:sing.k1.41}.  We choose to calculate the
global Stokes automorphism
$\mf{S}_+^{(k)}(\tx;\tq) = \mf{S}^{(k)}_{IV\rightarrow II}(\tx;\tq)$
and
$\mf{S}_-^{(k)}(\tx;\tq) = \mf{S}^{(k)}_{II \rightarrow IV}(\tx;\tq)$,
both in the anti-clockwise direction, which encode all the Stokes
constants.

In order to derive the Stokes automorphisms, we first calculate the
Borel resummation of the trans-series.  Following
\cite{Garoufalidis:2020xec} and with numerical evidence presented
shortly after, We claim that they can be written as bilinear products
of holomorphic blocks in $q$ and $\tq$. Concretely,
\begin{subequations}
\begin{align}
  s_{I}(\Phi^{(k)}_{\knot{4}_1})(x;\tau) =
  &\begin{pmatrix}
    -\tx & 1 + \tx^{-1}\\
    0&1
  \end{pmatrix}W_{-1}(\tx;\tq^{-1})
       \Delta^{(k)}(\tau) B(x;q),\quad |\tq|<1,
       \label{eq:s1.41}\\
  s_{II}(\Phi^{(k)}_{\knot{4}_1})(x;\tau) =
  &\begin{pmatrix}
    0 & -\tx \\ 1 & -\tx - \tx^2
  \end{pmatrix}W_{-1}(\tx^{-1};\tq^{-1})
                    \begin{pmatrix}
                      1&0\\0&-1
                    \end{pmatrix}\Delta^{(k)}(\tau) B(x;q),\quad |\tq|<1,
                              \label{eq:s2.41}\\
  s_{III}(\Phi^{(k)}_{\knot{4}_1})(x;\tau) =
  &\begin{pmatrix}
    0&-\tx\\1&1
  \end{pmatrix} W_{-1}(\tx^{-1};\tq^{-1})
               \begin{pmatrix}
                 1&0\\0&-1
               \end{pmatrix}\Delta^{(k)}(\tau)B(x;q),\quad |\tq|>1,
  \label{eq:s3.41}\\
  s_{IV}(\Phi^{(k)}_{\knot{4}_1})(x;\tau) =
  &\begin{pmatrix}
    -\tx&-\tx\\0&1
  \end{pmatrix} W_{-1}(\tx;\tq^{-1})\Delta^{(k)}(\tau)B(x;q),\quad |\tq|<1.
                  \label{eq:s4.41}
\end{align}
\end{subequations}
Here $W_{-1}(x,q)$ is the Wronskian of holomorphic blocks defined in
\eqref{eq:W.41}, and
\begin{equation}
  B(x;q) =
  \begin{pmatrix}
    A_0(x;q)\\
    B_0(x;q)
  \end{pmatrix}
\end{equation}
$\Delta^{(k)}(\tau)$ is the diagonal matrix defined by
\begin{equation}
  \Delta^{(k)}(\tau) =
  \diag(\re^{-\frac{\pi\ri
      k}{4}-\frac{\pi\ri}{2k}-\frac{\pi\ri}{4k}(\tau+\tau^{-1})},
  \re^{\frac{\pi\ri k}{4}+\frac{\pi\ri}{2k}+\frac{\pi\ri}{4k}(\tau+\tau^{-1})}
  ).
\end{equation}
From \eqref{eq:s1.41} and \eqref{eq:s2.41}, we can derive the Stokes
automorphism $\mf{S}_{I\rightarrow II}^{(k)}(\tq)$ which encompasses all
the Stokes constants in the upper half plane.  This step involves only
multiplication of $\tq$-series.  On the other hand, from
\eqref{eq:s1.41} and \eqref{eq:s4.41} we can derive the Stokes
constants associated to the three Borel plane singularities on the
positive real axis, and this step involves analytic continuation of
the Wronskian $W_{-1}(\tx^{-1};\tq^{-1})$ from inside to outside the
unit circle with the help of \eqref{eq:Wcont.41}.  We can combine the
two results to write down the global Stokes automorphism
$\mf{S}_+^{(k)}(\tx;\tq)$ from the ray $\rho_{0_-}$ to
$\rho_{\pi+0_-}$.  Likewise, we can compute the global Stokes
automorphism $\mf{S}_-^{(k)}(\tx;\tq)$ from the ray $\rho_{\pi+0_-}$
to $\rho_{2\pi+0_-}$.  Together they encompass the Stokes constants of
all the Borel plane singularities.
\begin{align}
  \mf{S}_+^{(k)}(\tx;\tq) =
  &\begin{pmatrix}
    0&-1\\\tx^{-1}&-1-\tx
  \end{pmatrix}W_{-1}(\tx^{-1};\tq^{-1})W_{-1}(\tx;\tq)^T
                    \begin{pmatrix}
                      0&\tx\\-1&-1-\tx^{-1}
                    \end{pmatrix},\quad |\tq|<1\\
  \mf{S}_-^{(k)}(\tx;\tq) =
  &\begin{pmatrix}
    \tx&\tx\\0&-1
  \end{pmatrix}W_{-1}(\tx;\tq) W_{-1}(\tx^{-1};\tq^{-1})^T
                \begin{pmatrix}
                  \tx^{-1}&0\\\tx^{-1}&-1
                \end{pmatrix},\quad |\tq|<1.
\end{align}
In particular, we notice that the Stokes automorphisms and therefore
the Stokes constants do not depend on the level $k$.

At level $k=1$ numerical evidences for the position of Borel plane
singularities as well as
\eqref{eq:s1.41},\eqref{eq:s2.41},\eqref{eq:s3.41},\eqref{eq:s4.41}
were given in detail in \cite{Garoufalidis:2020xec}.

\begin{figure}[h!]
  \subfloat[]{\includegraphics[width=0.5\textwidth]
    {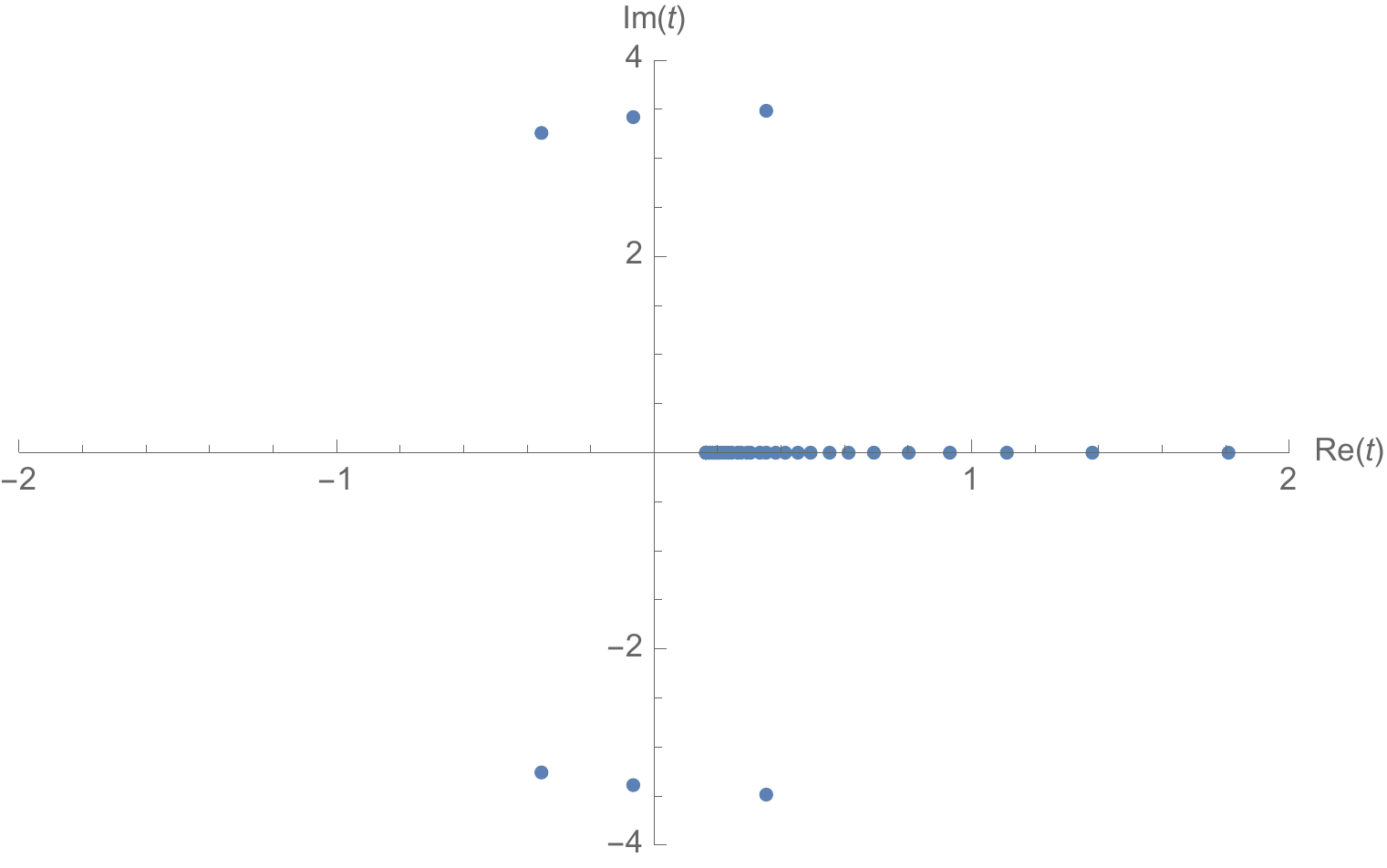}} \hspace{2ex}
  \subfloat[]{\includegraphics[width=0.5\textwidth]
    {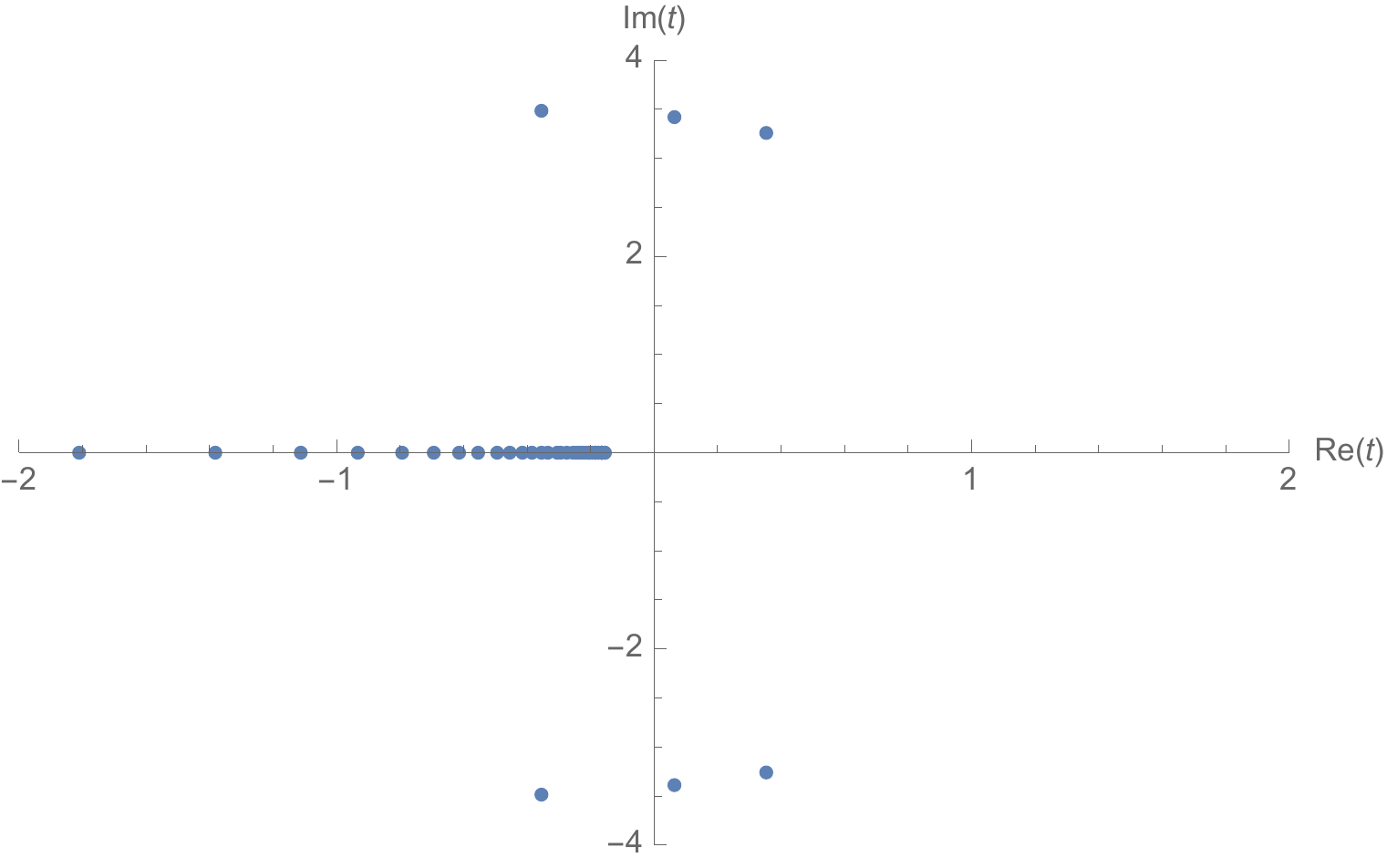}}
  \caption{The distribution of poles of the Pad\'{e} approximant for
    $(a): \varphi_{\knot{4}_1}^{(k=2,\sigma_1)}(x,\tau)$ and
    $(b): \varphi_{\knot{4}_1}^{(k=2,\sigma_2)}(x,\tau)$ with
    $x = 6/5$ and $N = 100$ terms.}
  \label{fig:41level2n0all}
\end{figure}

For higher levels, the computation is similar, although the expression
becomes more complicated.
At level $k=2$, we first 
choose $x = 6/5$, corresponding to $u=2\log(6/5)$ and $n=0$.  We
truncate the power series to $N = 100$ terms. The poles of the
Pad\'{e} approximant for $\Phi_{\knot{4}_1}^{(2)}(x,\tau)$ are shown
in Fig. \ref{fig:41level2n0all}. Notice that the accumulating poles on
the real axis hint at a branch cut on the corresponding exact Borel
plane.
    
Clearly, if we consider the vector $\Phi(x,\tau)$, the Borel plane is
divided into four regions as promised.  It remains to check
\eqref{eq:s1.41},\eqref{eq:s2.41},\eqref{eq:s3.41},\eqref{eq:s4.41}.
The comparison is tabulated in Table \ref{tab:41k2n0}.

\begin{table}
  \centering%
  \subfloat[Region $I$: $\tau = \frac{1}{8}\re^{\frac{\pi\ri}{4}}$]
  {\begin{tabular}{*{5}{>{$}c<{$}}}\toprule
     &|\frac{s_I(\Phi)(x;\tau)}{F_I(x;\tau)}- 1|
     &|\frac{s_I(\Phi)(x;\tau)}{s_I'(\Phi)(x;\tau)}- 1|
     &|\tq(\tau)|& \text{Min}(|\tx(x,\tau)|,|\tx(x,\tau)^{-1}|)\\\midrule
     \sigma_1& 8.6\times 10^{-16} & 3.5\times10^{-15}
     & \multirow{2}{*}{$1.9\times 10^{-8}$}
      & \multirow{2}{*}{$0.36$}\\
     \sigma_2& 3.7\times 10^{-23} & 2.4\times 10^{-22}&&\\\bottomrule
   \end{tabular}}\\
 \subfloat[Region $II$: $\tau = \frac{1}{8}\re^{\frac{3\pi\ri}{4}}$]
 {\begin{tabular}{*{5}{>{$}c<{$}}}\toprule
    &|\frac{s_{II}(\Phi)(x;\tau)}{F_{II}(x;\tau)}- 1|
    &|\frac{s_{II}(\Phi)(x;\tau)}{s_{II}'(\Phi)(x;\tau)}- 1|
    &|\tq(\tau)|& \text{Min}(|\tx(x,\tau)|,|\tx(x,\tau)^{-1}|)\\\midrule
    \sigma_1& 3.7\times 10^{-23} & 2.4\times 10^{-22}
    & \multirow{2}{*}{$1.9\times 10^{-8}$}
      & \multirow{2}{*}{$0.36$}\\
     \sigma_2& 8.6\times 10^{-16} & 3.5\times 10^{-15}&&\\\bottomrule
   \end{tabular}}\\
 \subfloat[Region $III$: $\tau = \frac{1}{8}\re^{-\frac{3\pi\ri}{4}}$]
 {\begin{tabular}{*{5}{>{$}c<{$}}}\toprule
    &|\frac{s_{III}(\Phi)(x;\tau)}{F_{III}(x;\tau)}- 1|
    &|\frac{s_{III}(\Phi)(x;\tau)}{s_{III}'(\Phi)(x;\tau)}- 1|
    & |\tq(\tau)^{-1}|& \text{Min}(|\tx(x,\tau)|,|\tx(x,\tau)^{-1}|)\\\midrule
    \sigma_1& 3.7\times 10^{-23} &2.4\times 10^{-22}
    & \multirow{2}{*}{$1.9\times 10^{-8}$}
      & \multirow{2}{*}{$0.36$}\\
     \sigma_2&8.6\times 10^{-16}  & 3.5\times 10^{-15}&&\\\bottomrule
  \end{tabular}}\\
 \subfloat[Region $IV$: $\tau = \frac{1}{8}\re^{-\frac{\pi\ri}{4}}$]
 {\begin{tabular}{*{5}{>{$}c<{$}}}\toprule
    &|\frac{s_{IV}(\Phi)(x;\tau)}{F_{IV}(x;\tau)}- 1|
    &|\frac{s_{IV}(\Phi)(x;\tau)}{s_{IV}'(\Phi)(x;\tau)}- 1|
    & |\tq(\tau)^{-1}|& \text{Min}(|\tx(x,\tau)|,|\tx(x,\tau)^{-1}|)\\\midrule
    \sigma_1& 8.6\times 10^{-16} & 3.5\times 10^{-15}
    & \multirow{2}{*}{$1.9\times 10^{-8}$}
      & \multirow{2}{*}{$0.36$}\\
     \sigma_2& 3.7\times 10^{-23} & 2.4\times 10^{-22}&&\\\bottomrule
   \end{tabular}}
 \caption{We perform the numerical Borel resummation for
   $\Phi_{\knot{4}_1}^{(k=2)}(x;\tau)$ at $x = 6/5$
   ($u=2\log 6/5, n=0$) with 100 terms, after choosing suitable $\tau$
   in four different regions. We compare them with equations
   \eqref{eq:s1.41},\eqref{eq:s2.41},\eqref{eq:s3.41},\eqref{eq:s4.41}
   whose right hand side is denoted as $F_R(x;\tau)$. Meanwhile, we
   estimate the contribution of higher order terms by resumming
   $\Phi_{\knot{4}_1}^{(k=2)}(x;\tau)$ with 95 terms. The values of
   $|\tilde{q}(\tilde{q}^{-1})|$ and $|\tilde{x}^{\pm 1}|$ are also
   provided for comparison.}
  \label{tab:41k2n0}
\end{table}

\begin{figure}[h!]
  \centering
  \subfloat[]
  {\includegraphics[width=0.5\textwidth]{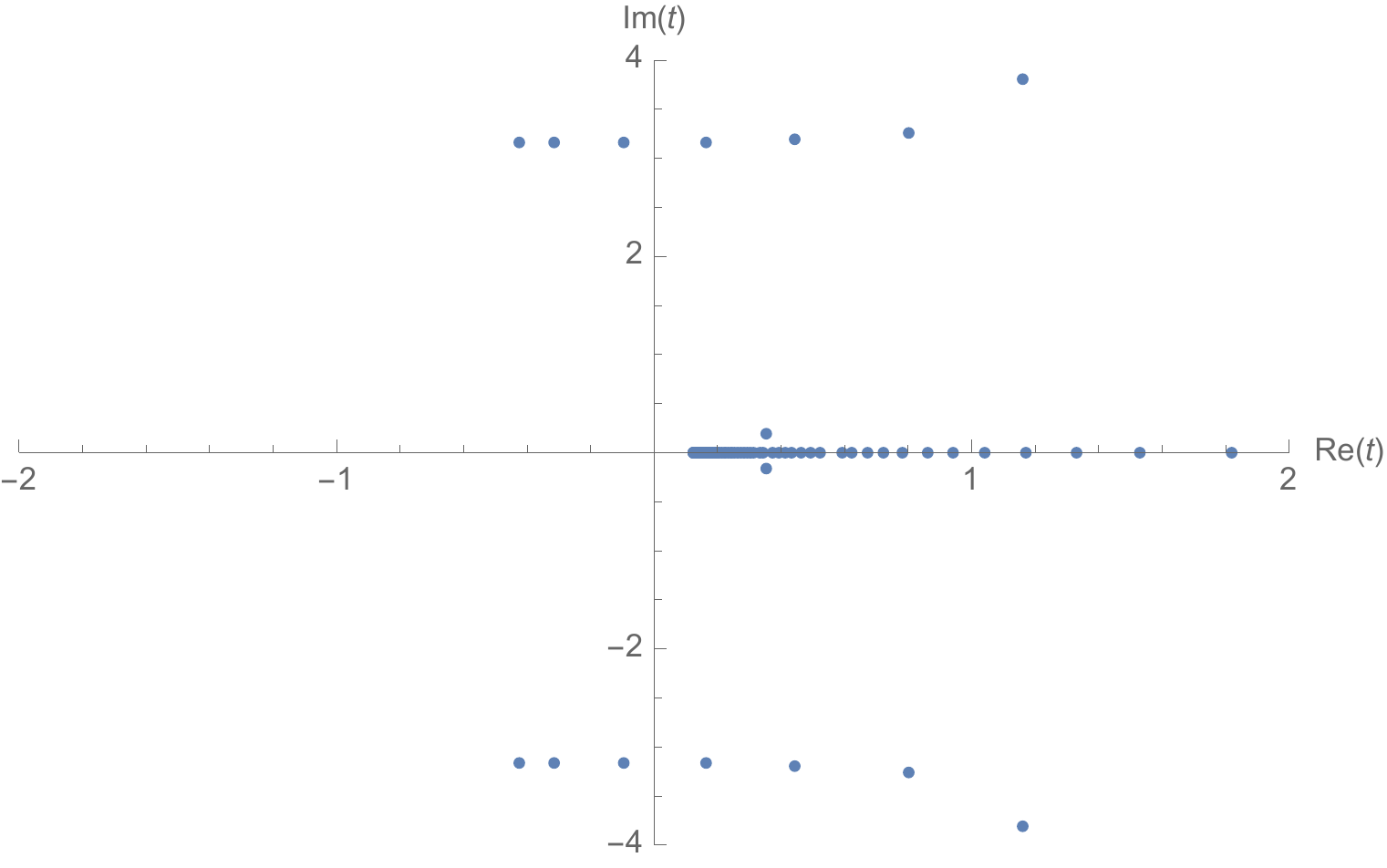}}
  \centering
  \subfloat[]
  {\includegraphics[width=0.5\textwidth]{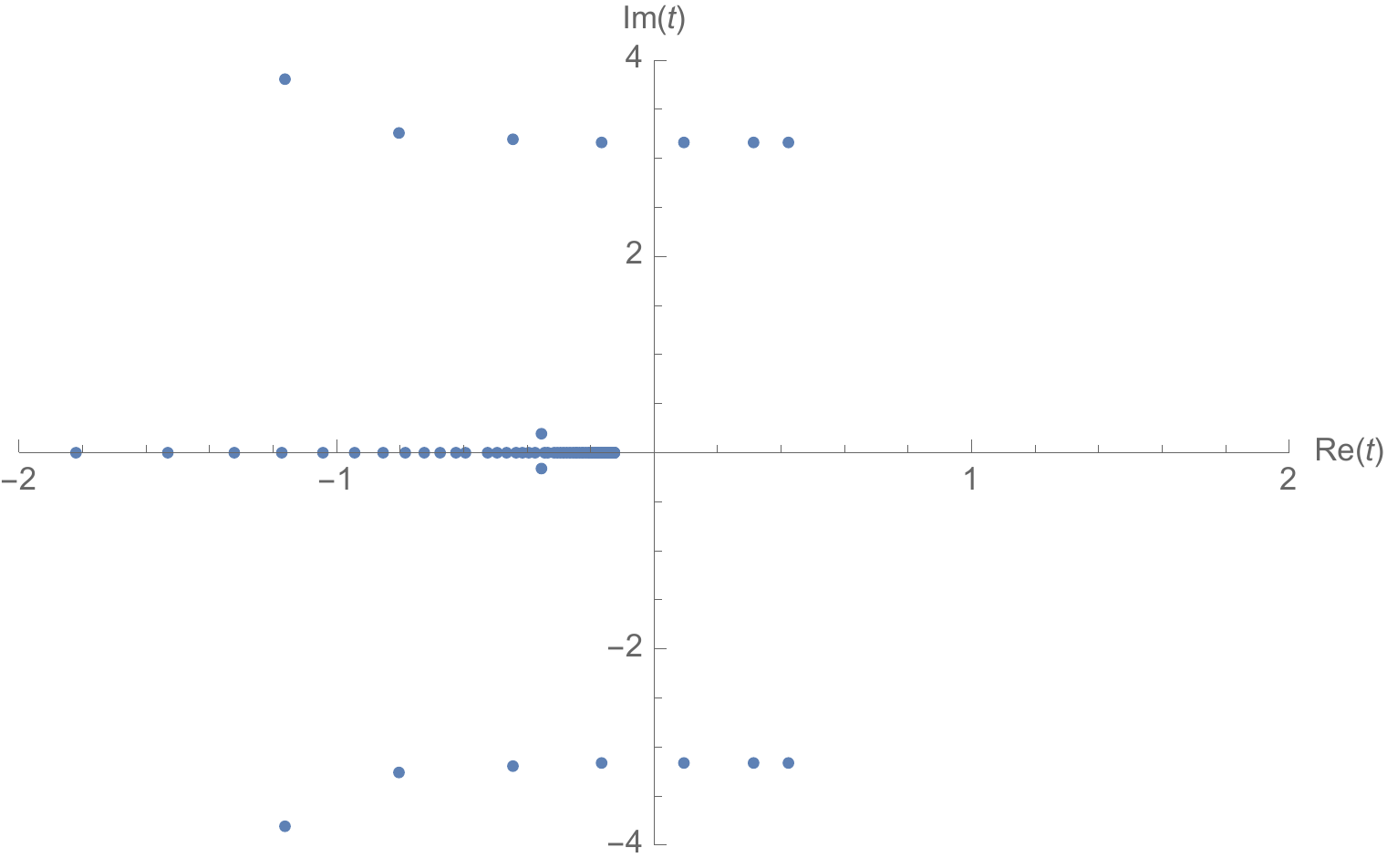}}
  \caption{The distribution of poles of the Pad\'{e} approximant for
    $(a): \varphi_{\knot{4}_1}^{(k=2,\sigma_1)}(x,\tau)$ and
    $(b): \varphi_{\knot{4}_1}^{(k=2,\sigma_2)}(x,\tau)$ with
    $x = -5/4$ and $N = 200$ terms.}
  \label{fig:41level2n1all}
\end{figure}

We consider next
$x = -5/4$, corresponding to $u = 2\log(5/4)$ and $n=1$, and we
truncate the power series up to $N = 200$ terms. The positions of
poles for the Pad\'{e} approximant $\Phi_{\knot{4}_1}^{(k=2)}(x,\tau)$
are shown in Fig. \ref{fig:41level2n1all}. Again one observes the
emergent branch cut on the real axis.
We then present the numerical evidence of
\eqref{eq:s1.41},\eqref{eq:s2.41},\eqref{eq:s3.41},\eqref{eq:s4.41}
for each region in Tab.~\ref{tab:41k2n1}.  As we can see, the relative
errors between two sides of \eqref{eq:s1.41}, \eqref{eq:s2.41},
\eqref{eq:s3.41}, \eqref{eq:s4.41} (the first column) are always
within the precision of the Borel resummation (the second column), and
are always far smaller than a potential $\tq$ (or $1/\tq$ in the lower
half plane) or a $\tx^{\pm 1}$ correction (the third and the fourth
columns).

\begin{table}
  \centering%
  \subfloat[Region $I$: $\tau = \frac{1}{10}\re^{\frac{\pi\ri}{5}}$]
  {\begin{tabular}{*{5}{>{$}c<{$}}}\toprule
     &|\frac{s_I(\Phi)(x;\tau)}{F_I(x;\tau)}- 1|
     &|\frac{s_I(\Phi)(x;\tau)}{s_I'(\Phi)(x;\tau)}- 1|
     &|\tq(\tau)|& \text{Min}(|\tx(x,\tau)|,|\tx(x,\tau)^{-1}|)\\\midrule
     \sigma_1& 8.2\times 10^{-22} & 8.6\times10^{-22}
     & \multirow{2}{*}{$9.6\times 10^{-9}$}
      & \multirow{2}{*}{$0.16$}\\
     \sigma_2& 3.9\times 10^{-36} & 5.1\times 10^{-36}&&\\\bottomrule
   \end{tabular}}\\
 \subfloat[Region $II$: $\tau = \frac{1}{10}\re^{\frac{4\pi\ri}{5}}$]
 {\begin{tabular}{*{5}{>{$}c<{$}}}\toprule
    &|\frac{s_{II}(\Phi)(x;\tau)}{F_{II}(x;\tau)}- 1|
    &|\frac{s_{II}(\Phi)(x;\tau)}{s_{II}'(\Phi)(x;\tau)}- 1|
    &|\tq(\tau)|& \text{Min}(|\tx(x,\tau)|,|\tx(x,\tau)^{-1}|)\\\midrule
    \sigma_1& 3.9\times 10^{-36} & 5.1\times 10^{-36}
    & \multirow{2}{*}{$9.6\times 10^{-9}$}
      & \multirow{2}{*}{$0.16$}\\
     \sigma_2& 8.2\times 10^{-22} & 8.6\times 10^{-22}&&\\\bottomrule
   \end{tabular}}\\
 \subfloat[Region $III$: $\tau = \frac{1}{10}\re^{-\frac{4\pi\ri}{5}}$]
 {\begin{tabular}{*{5}{>{$}c<{$}}}\toprule
    &|\frac{s_{III}(\Phi)(x;\tau)}{F_{III}(x;\tau)}- 1|
    &|\frac{s_{III}(\Phi)(x;\tau)}{s_{III}'(\Phi)(x;\tau)}- 1|
    & |\tq(\tau)^{-1}|& \text{Min}(|\tx(x,\tau)|,|\tx(x,\tau)^{-1}|)\\\midrule
    \sigma_1& 3.9\times 10^{-36} &5.1\times 10^{-36}
    & \multirow{2}{*}{$9.6\times 10^{-9}$}
      & \multirow{2}{*}{$0.16$}\\
     \sigma_2&8.2\times 10^{-22}  & 8.6\times 10^{-22}&&\\\bottomrule
  \end{tabular}}\\
 \subfloat[Region $IV$: $\tau = \frac{1}{10}\re^{-\frac{\pi\ri}{5}}$]
 {\begin{tabular}{*{5}{>{$}c<{$}}}\toprule
    &|\frac{s_{IV}(\Phi)(x;\tau)}{F_{IV}(x;\tau)}- 1|
    &|\frac{s_{IV}(\Phi)(x;\tau)}{s_{IV}'(\Phi)(x;\tau)}- 1|
    & |\tq(\tau)^{-1}|& \text{Min}(|\tx(x,\tau)|,|\tx(x,\tau)^{-1}|)\\\midrule
    \sigma_1& 8.2\times 10^{-22} & 8.6\times 10^{-22}
    & \multirow{2}{*}{$9.6\times 10^{-9}$}
      & \multirow{2}{*}{$0.16$}\\
     \sigma_2& 3.9\times 10^{-36} & 5.1\times 10^{-36}&&\\\bottomrule
   \end{tabular}}
 \caption{We perform numerical Borel resummation for
   $\Phi_{\knot{4}_1}^{(k=2)}(x;\tau)$ at $x = -5/4$
   ($u=2\log 5/4, n=1$) with 200 terms, after choosing suitable $\tau$
   in four different regions. We compare them with equations
   \eqref{eq:s1.41},\eqref{eq:s2.41},\eqref{eq:s3.41},\eqref{eq:s4.41}
   whose right hand side is denoted as $F_R(x;\tau)$. Meanwhile, we
   estimate the contribution of higher order terms by resumming
   $\Phi_{\knot{4}_1}^{(k=2)}(x;\tau)$ with 195 terms. The values of
   $|\tilde{q}(\tilde{q}^{-1})|$ and $|\tilde{x}^{\pm 1}|$ are also
   provided for comparison.}
  \label{tab:41k2n1}
\end{table}

We also perform numerical analysis at level $k = 3$.
We choose $x = \frac{6}{5} \re^{2\pi n \ri/3}$ ($n=0,1,2$)
corresponding to $u=3\log(6/5)$ and $n=0,1,2$ respectively.
The power series are truncated up to $N=220$ terms.
The poles of the Pad\'{e} approximant have a similar structure to the
cases $k=1,2$, so we omit them here.  The numerical evidence for the
Borel resummation of the trans-series
\eqref{eq:s1.41},\eqref{eq:s2.41},\eqref{eq:s3.41},\eqref{eq:s4.41} in
each region are given in Tabs. \ref{tab:41k3n0}, \ref{tab:41k3n1} and
\ref{tab:41k3n2} respectively.

\begin{table}
  \centering%
  \subfloat[Region $I$: $\tau = \frac{1}{9}\re^{\frac{\pi\ri}{5}}$]
  {\begin{tabular}{*{5}{>{$}c<{$}}}\toprule
     &|\frac{s_I(\Phi)(x;\tau)}{F_I(x;\tau)}- 1|
     &|\frac{s_I(\Phi)(x;\tau)}{s_I'(\Phi)(x;\tau)}- 1|
     &|\tq(\tau)|& \text{Min}(|\tx(x,\tau)|,|\tx(x,\tau)^{-1}|)\\\midrule
     \sigma_1& 3.1\times 10^{-15} & 1.6\times10^{-15}
     & \multirow{2}{*}{$1.5\times 10^{-5}$}
      & \multirow{2}{*}{$0.27$}\\
     \sigma_2& 4.6\times 10^{-27} & 5.2\times 10^{-26}&&\\\bottomrule
   \end{tabular}}\\
 \subfloat[Region $II$: $\tau = \frac{1}{9}\re^{\frac{4\pi\ri}{5}}$]
 {\begin{tabular}{*{5}{>{$}c<{$}}}\toprule
    &|\frac{s_{II}(\Phi)(x;\tau)}{F_{II}(x;\tau)}- 1|
    &|\frac{s_{II}(\Phi)(x;\tau)}{s_{II}'(\Phi)(x;\tau)}- 1|
    &|\tq(\tau)|& \text{Min}(|\tx(x,\tau)|,|\tx(x,\tau)^{-1}|)\\\midrule
    \sigma_1& 5.8\times 10^{-27} & 1.7\times 10^{-27}
    & \multirow{2}{*}{$1.5\times 10^{-5}$}
      & \multirow{2}{*}{$0.27$}\\
     \sigma_2& 1.1\times 10^{-15} & 8.3\times 10^{-16}&&\\\bottomrule
   \end{tabular}}\\
 \subfloat[Region $III$: $\tau = \frac{1}{9}\re^{-\frac{4\pi\ri}{5}}$]
 {\begin{tabular}{*{5}{>{$}c<{$}}}\toprule
    &|\frac{s_{III}(\Phi)(x;\tau)}{F_{III}(x;\tau)}- 1|
    &|\frac{s_{III}(\Phi)(x;\tau)}{s_{III}'(\Phi)(x;\tau)}- 1|
    & |\tq(\tau)^{-1}|& \text{Min}(|\tx(x,\tau)|,|\tx(x,\tau)^{-1}|)\\\midrule
    \sigma_1& 5.8\times 10^{-27} &1.7\times 10^{-27}
    & \multirow{2}{*}{$1.5\times 10^{-5}$}
      & \multirow{2}{*}{$0.27$}\\
     \sigma_2&1.1\times 10^{-15}  & 8.3\times 10^{-16}&&\\\bottomrule
  \end{tabular}}\\
 \subfloat[Region $IV$: $\tau = \frac{1}{9}\re^{-\frac{\pi\ri}{5}}$]
 {\begin{tabular}{*{5}{>{$}c<{$}}}\toprule
    &|\frac{s_{IV}(\Phi)(x;\tau)}{F_{IV}(x;\tau)}- 1|
    &|\frac{s_{IV}(\Phi)(x;\tau)}{s_{IV}'(\Phi)(x;\tau)}- 1|
    & |\tq(\tau)^{-1}|& \text{Min}(|\tx(x,\tau)|,|\tx(x,\tau)^{-1}|)\\\midrule
    \sigma_1& 3.1\times 10^{-15} & 1.6\times 10^{-15}
    & \multirow{2}{*}{$1.5\times 10^{-5}$}
      & \multirow{2}{*}{$0.27$}\\
     \sigma_2& 4.6\times 10^{-27} & 5.2\times 10^{-26}&&\\\bottomrule
   \end{tabular}}
 \caption{We perform numerical Borel resummation for $\Phi_{\knot{4}_1}^{(k=3)}(x,\tau)$ at $x = 6/5$ with 220 terms, after choosing suitable $\tau$ in four different regions. We compare them with equations \eqref{eq:s1.41},\eqref{eq:s2.41},\eqref{eq:s3.41},\eqref{eq:s4.41} whose right hand side is denoted as $F_R(x,\tau)$. Meanwhile, we estimate the contribution of higher order terms by resumming $\Phi_{\knot{4}_1}^{(k=3)}(x,\tau)$ with 215 terms. The values of $|\tilde{q}(\tilde{q}^{-1})|$ and $|\tilde{x}^{\pm 1}|$ are also provided for comparison.}
  \label{tab:41k3n0}
\end{table}

\begin{table}
  \centering%
  \subfloat[Region $I$: $\tau = \frac{1}{9}\re^{\frac{\pi\ri}{5}}$]
  {\begin{tabular}{*{5}{>{$}c<{$}}}\toprule
     &|\frac{s_I(\Phi)(x;\tau)}{F_I(x;\tau)}- 1|
     &|\frac{s_I(\Phi)(x;\tau)}{s_I'(\Phi)(x;\tau)}- 1|
     &|\tq(\tau)|& \text{Min}(|\tx(x,\tau)|,|\tx(x,\tau)^{-1}|)\\\midrule
     \sigma_1& 4.5\times 10^{-16} & 4.9\times10^{-16}
     & \multirow{2}{*}{$1.5\times 10^{-5}$}
      & \multirow{2}{*}{$0.27$}\\
     \sigma_2& 6.5\times 10^{-27} & 3.8\times 10^{-27}&&\\\bottomrule
   \end{tabular}}\\
 \subfloat[Region $II$: $\tau = \frac{1}{8}\re^{\frac{4\pi\ri}{5}}$]
 {\begin{tabular}{*{5}{>{$}c<{$}}}\toprule
    &|\frac{s_{II}(\Phi)(x;\tau)}{F_{II}(x;\tau)}- 1|
    &|\frac{s_{II}(\Phi)(x;\tau)}{s_{II}'(\Phi)(x;\tau)}- 1|
    &|\tq(\tau)|& \text{Min}(|\tx(x,\tau)|,|\tx(x,\tau)^{-1}|)\\\midrule
    \sigma_1& 3.7\times 10^{-26} & 1.1\times 10^{-25}
    & \multirow{2}{*}{$5.3\times 10^{-5}$}
      & \multirow{2}{*}{$0.31$}\\
     \sigma_2& 6.1\times 10^{-15} & 5.7\times 10^{-15}&&\\\bottomrule
   \end{tabular}}\\
 \subfloat[Region $III$: $\tau = \frac{1}{8}\re^{-\frac{4\pi\ri}{5}}$]
 {\begin{tabular}{*{5}{>{$}c<{$}}}\toprule
    &|\frac{s_{III}(\Phi)(x;\tau)}{F_{III}(x;\tau)}- 1|
    &|\frac{s_{III}(\Phi)(x;\tau)}{s_{III}'(\Phi)(x;\tau)}- 1|
    & |\tq(\tau)^{-1}|& \text{Min}(|\tx(x,\tau)|,|\tx(x,\tau)^{-1}|)\\\midrule
    \sigma_1& 3.2\times 10^{-26} &1.3\times 10^{-25}
    & \multirow{2}{*}{$5.3\times 10^{-5}$}
      & \multirow{2}{*}{$0.31$}\\
     \sigma_2&3.0\times 10^{-15}  & 9.8\times 10^{-15}&&\\\bottomrule
  \end{tabular}}\\
 \subfloat[Region $IV$: $\tau = \frac{1}{9}\re^{-\frac{\pi\ri}{5}}$]
 {\begin{tabular}{*{5}{>{$}c<{$}}}\toprule
    &|\frac{s_{IV}(\Phi)(x;\tau)}{F_{IV}(x;\tau)}- 1|
    &|\frac{s_{IV}(\Phi)(x;\tau)}{s_{IV}'(\Phi)(x;\tau)}- 1|
    & |\tq(\tau)^{-1}|& \text{Min}(|\tx(x,\tau)|,|\tx(x,\tau)^{-1}|)\\\midrule
    \sigma_1& 4.5\times 10^{-16} & 4.9\times 10^{-16}
    & \multirow{2}{*}{$1.5\times 10^{-5}$}
      & \multirow{2}{*}{$0.27$}\\
     \sigma_2& 6.5\times 10^{-27} & 3.8\times 10^{-27}&&\\\bottomrule
   \end{tabular}}
 \caption{We perform numerical Borel resummation for $\Phi_{\knot{4}_1}^{(k=3)}(x,\tau)$ at $x = \frac{6}{5}\re^{4\pi\ri/3}$ with 220 terms, after choosing suitable $\tau$ in four different regions. We compare them with equations \eqref{eq:s1.41},\eqref{eq:s2.41},\eqref{eq:s3.41},\eqref{eq:s4.41} whose right hand side is denoted as $F_R(x,\tau)$. Meanwhile, we estimate the contribution of higher order terms by resumming $\Phi_{\knot{4}_1}^{(k=3)}(x,\tau)$ with 215 terms. The values of $|\tilde{q}(\tilde{q}^{-1})|$ and $|\tilde{x}^{\pm 1}|$ are also provided for comparison.}
  \label{tab:41k3n1}
\end{table}

\begin{table}
  \centering%
  \subfloat[Region $I$: $\tau = \frac{1}{9}\re^{\frac{\pi\ri}{5}}$]
  {\begin{tabular}{*{5}{>{$}c<{$}}}\toprule
     &|\frac{s_I(\Phi)(x;\tau)}{F_I(x;\tau)}- 1|
     &|\frac{s_I(\Phi)(x;\tau)}{s_I'(\Phi)(x;\tau)}- 1|
     &|\tq(\tau)|& \text{Min}(|\tx(x,\tau)|,|\tx(x,\tau)^{-1}|)\\\midrule
     \sigma_1& 2.4\times 10^{-16} & 7.5\times10^{-16}
     & \multirow{2}{*}{$1.5\times 10^{-5}$}
      & \multirow{2}{*}{$0.27$}\\
     \sigma_2& 4.4\times 10^{-27} & 7.9\times 10^{-27}&&\\\bottomrule
   \end{tabular}}\\
 \subfloat[Region $II$: $\tau = \frac{1}{8}\re^{\frac{4\pi\ri}{5}}$]
 {\begin{tabular}{*{5}{>{$}c<{$}}}\toprule
    &|\frac{s_{II}(\Phi)(x;\tau)}{F_{II}(x;\tau)}- 1|
    &|\frac{s_{II}(\Phi)(x;\tau)}{s_{II}'(\Phi)(x;\tau)}- 1|
    &|\tq(\tau)|& \text{Min}(|\tx(x,\tau)|,|\tx(x,\tau)^{-1}|)\\\midrule
    \sigma_1& 3.2\times 10^{-26} & 1.3\times 10^{-25}
    & \multirow{2}{*}{$5.3\times 10^{-5}$}
      & \multirow{2}{*}{$0.31$}\\
     \sigma_2& 3.0\times 10^{-15} & 9.8\times 10^{-15}&&\\\bottomrule
   \end{tabular}}\\
 \subfloat[Region $III$: $\tau = \frac{1}{8}\re^{-\frac{4\pi\ri}{5}}$]
 {\begin{tabular}{*{5}{>{$}c<{$}}}\toprule
    &|\frac{s_{III}(\Phi)(x;\tau)}{F_{III}(x;\tau)}- 1|
    &|\frac{s_{III}(\Phi)(x;\tau)}{s_{III}'(\Phi)(x;\tau)}- 1|
    & |\tq(\tau)^{-1}|& \text{Min}(|\tx(x,\tau)|,|\tx(x,\tau)^{-1}|)\\\midrule
    \sigma_1& 3.7\times 10^{-26} &1.1\times 10^{-25}
    & \multirow{2}{*}{$5.3\times 10^{-5}$}
      & \multirow{2}{*}{$0.31$}\\
     \sigma_2&6.1\times 10^{-15}  & 5.7\times 10^{-15}&&\\\bottomrule
  \end{tabular}}\\
 \subfloat[Region $IV$: $\tau = \frac{1}{9}\re^{-\frac{\pi\ri}{5}}$]
 {\begin{tabular}{*{5}{>{$}c<{$}}}\toprule
    &|\frac{s_{IV}(\Phi)(x;\tau)}{F_{IV}(x;\tau)}- 1|
    &|\frac{s_{IV}(\Phi)(x;\tau)}{s_{IV}'(\Phi)(x;\tau)}- 1|
    & |\tq(\tau)^{-1}|& \text{Min}(|\tx(x,\tau)|,|\tx(x,\tau)^{-1}|)\\\midrule
    \sigma_1& 4.5\times 10^{-16} & 7.5\times 10^{-16}
    & \multirow{2}{*}{$1.5\times 10^{-5}$}
      & \multirow{2}{*}{$0.27$}\\
     \sigma_2& 4.4\times 10^{-27} & 7.9\times 10^{-27}&&\\\bottomrule
   \end{tabular}}
 \caption{We perform numerical Borel resummation for
   $\Phi_{\knot{4}_1}^{(k=3)}(x,\tau)$ at
   $x = \frac{6}{5}\re^{2\pi\ri/3}$ with 220 terms, after choosing
   suitable $\tau$ in four different regions. We compare them with
   equations
   \eqref{eq:s1.41},\eqref{eq:s2.41},\eqref{eq:s3.41},\eqref{eq:s4.41}
   whose right hand side is denoted as $F_R(x,\tau)$. Meanwhile, we
   estimate the contribution of higher order terms by resumming
   $\Phi_{\knot{4}_1}^{(k=3)}(x,\tau)$ with 215 terms. The values of
   $|\tilde{q}(\tilde{q}^{-1})|$ and $|\tilde{x}^{\pm 1}|$ are also
   provided for comparison.}
  \label{tab:41k3n2}
\end{table}

\section{Three twists}
\label{sc:52}

\subsection{Asymptotic series}
\label{sc:52.asymp}

The state integral of knot $\knot{5}_2$ at level $k$ is given as a
one-dimensional integral in \cite{Andersen:2016ugz}
\begin{align}
  \chi^{(k)}_{\knot{5}_2}(\mu,n;\bb) =
  &\frac{(\eta_k)^3}{k}\sum_{m\in\IZ_k}
    \int_{\IR+\ri\imag(c_\bb)-\ri 0}\rd \sigma
    \mc{Z}^{(k)}_\bb[\Delta](\sigma,m)                                          \mc{Z}^{(k)}_\bb[\Delta](\sigma+\mu,m+n)
    \mc{Z}^{(k)}_\bb[\Delta](\sigma-\mu,m-n)\nn
  &\phantom{===}\times(-1)^{n}
    \re^{\frac{\pi\ri}{k}(-(\sigma+\mu-c_\bb)^2-(\sigma-\mu-c_\bb)^2-\mu^2+(m+n)^2+(m-n)^2+n^2)}.      \label{eq:chi-52}
\end{align}
This can also be derived from the generic formula \eqref{eq:ZNgmMu0}.
We choose the Neumann-Zagier data
\begin{equation}
  \pmb{A} = 
  \begin{pmatrix}
    0 & 1 & 0\\
    -1&0&-1\\
    0&2&0
  \end{pmatrix},\quad \pmb{B} =
  \begin{pmatrix}
    -1&0&0\\
    0&1&0\\
    -1&0&-1
  \end{pmatrix},\quad \nu =
  \begin{pmatrix}
    0\\0\\0
  \end{pmatrix},
\end{equation}
with which \eqref{eq:ZNgmMu0} yields
\begin{align}
  &\mc{Z}_\gamma^{(k)}(\mu,n;\bb)\nn =
  &\frac{1}{k^3}\sum_{m\in(\IZ_k)^3}\int\rd^3\s\,
    \re^{\frac{2\pi\ri}{k}(\s_1(\s_2+\mu_1)-m_1(m_2+n_1))}
    \mc{Z}_\bb^{(k)}[\Delta](\s_1,m_1)\nn
  &\phantom{=========}
    \re^{\frac{2\pi\ri}{k}(\s_3(\s_2-\mu_1)-m_3(m_2-n_1))}
    \mc{Z}_\bb^{(k)}[\Delta](\s_3,m_3)\nn
  &\phantom{=========}
    \mc{Z}_\bb^{(k)}[\Delta](\s_2,m_2)
    \nn=
  &\frac{\re^{\frac{\pi\ri}{6}(k+\frac{8c_\bb^2}{k})}}{k}
    \sum_{m\in\IZ_k}\int\rd \s
    \mc{Z}_\bb^{(k)}[\Delta](\s,m)
    \mc{Z}_\bb^{(k)}[\Delta](\s+\mu,m+n)
    \mc{Z}_\bb^{(k)}[\Delta](\s-\mu,m-n)
    \nn
  &\phantom{========}
    \re^{-\frac{2\pi\ri}{k}((\s-c_\bb)^2+\mu^2-m^2-n^2)}.
\end{align}
where we have used again the formula of Fourier transformation of the
tetrahedron partition function \eqref{eq:FourierFQD}.
This is the same as \eqref{eq:chi-52} up to an overall factor
\begin{equation}
  \chi^{(k)}_{\knot{5}_2}(\mu,n;\bb) =
  \re^{\frac{\pi\ri}{3}(k-\frac{c_\bb^2}{k})} (-1)^n
  \re^{\frac{\pi\ri}{k}(-\mu^2+n^2)}\mc{Z}_\gamma^{(k)}(\mu,n;\bb).
\end{equation}

We now also perform the saddle point expansion of the one-dimensional
integral \eqref{eq:chi-52}.  We introduce the alternative
representation of the tetrahedron partition function \eqref{eq:Dk}, in
terms of which, the state integral \eqref{eq:chi-52} reads
\begin{align}
  \chi^{(k)}_{\knot{5}_2}(\mu,n;\bb) =
  &\frac{(\eta_k)^3}{k}(-1)^n\re^{\frac{3\pi\ri}{k}(n^2-\mu^2)}
    \sum_{m\in\IZ_k}\re^{\frac{2\pi\ri m^2}{k}}\nn
  &\phantom{=}\times\int_{\IR+\ri 0}\rd \sigma
    \DD^{(k)}_\bb(\frac{1}{\sqrt{k}}\sigma,m)
    \DD^{(k)}_\bb(\frac{1}{\sqrt{k}}(\sigma-\mu),m+n)
    \DD^{(k)}_\bb(\frac{1}{\sqrt{k}}(\sigma+\mu),m-n) \re^{-\frac{2\pi\ri\sigma^2}{k}}.
\end{align}
We scale the variables
\begin{equation}
  Z = 2\pi\bb \sigma,\quad u =2\pi\bb \mu
\end{equation}
and then perform asymptotic expansion of the tetrahedron partition
functions in the small $\bb$ limit using \eqref{eq:ZDasymp}.
\begin{equation}
  \chi_{\knot{5}_2}^{(k)}(\mu,n;\bb) \sim
  \frac{(\eta_k)^3}{2\pi\bb k}(-1)^n \re^{\frac{3\pi\ri}{k}(n^2-\mu^2)}
  \sum_{m\in\IZ_k}
  \re^{\frac{2\pi\ri}{k}m^2}
  \int \rd Z \exp\sum_{\ell\geq 0}(2\pi\ri\bb^2)^{\ell-1} U_\ell(u,Z)
\end{equation}
where the potential functions are
\begin{subequations}
  \begin{align}
    U_0(u,Z) =
    &\frac{1}{k}\left(
      Z^2  +
      \Li_2(-\re^{Z}) + \Li_2(-\re^{Z-u}) + \Li_2(-\re^{Z+u})\right),\\
    U_{\ell\geq 1}(u,Z) =
    &\frac{1}{\ell!}\sum_{j\in\IZ_k}B_\ell(1-1/(2k)-j/k)
      \Big(
      \Li_{2-\ell}(\zeta^{m-j-1/2} \re^{\frac{Z}{k}})\nn
    &\phantom{===}+\Li_{2-\ell}(\zeta^{m+n-j-1/2}\re^{\frac{Z-u}{k}})
      +\Li_{2-\ell}(\zeta^{m-n-j-1/2}\re^{\frac{Z+u}{k}})
      \Big).
  \end{align}
\end{subequations}
The critical equation of the leading order potential function reads
\begin{equation}
  0 = \frac{\pd}{\pd Z}U_0(u,Z)
  =\frac{1}{k}\left( 2Z - \log(1+\re^{Z})
   - \log(1+\re^{-u+Z}) - \log(1+\re^{u+Z})\right),
\end{equation}
which can be simplified to the algebraic equation
\begin{equation}
  Y^2 = (1-Y)(1-XY)(1-X^{-1}Y),
  \label{eq:alcr.52}
\end{equation}
with
\begin{equation}
  X = \re^u = x^k,\quad Y = - \re^{Z}.
\end{equation}
We also introduce the variable $\theta$ such that
\begin{equation}
  Y  = -\theta^k.
\end{equation}

The critical point equation \eqref{eq:alcr.52} has three solutions in
terms of $Y$, corresponding to the geometric, conjugate, and the real flat
connections $\s_1, \s_2, \s_3$ on the knot complement $S^3\backslash
\knot{5}_2$.
Numerically they are the ones which in the limit $X\rightarrow 1$
reduce respectively to the solutions $0.78492+1.30714\ldots\ri$,
$0.78492-1.30714\ldots\ri$ and $0.43016\ldots$ to \eqref{eq:alcr.52}.
Near each critical point $\s_j$, the integrand is approximated by the
exponential
\begin{equation}
  \exp\left(\frac{V^{(k)}(\s_j)}{2\pi\ri\tau}\right),
  \label{eq:exp}
\end{equation}
where
\begin{align}
  V^{(k)}(\s_j) = U_0(\log(X),-\log(Y_j))+ \frac{3u^2}{2k} + \frac{\pi^2}{2k}.
\end{align}
The last two terms come from the prefactors outside the summation over
$m\in\IZ_k$.
We again notice the curious fact that $V^{(k)}(\sigma_j) = V^{(1)}(\s_j)/k$.

By expanding near the critical point and performing the Gaussian
integration order by order, for each critical point $\s_j$ we can find
an asymptotic power series $\varphi^{(k,\s_j)}(x;\tau)$, which,
together with the exponential factor \eqref{eq:exp}, forms the
trans-series
\begin{equation}
  \Phi_{\knot{5}_2}^{(k,\sigma_j)}(x;\tau) =
  \re^{V^{(k)}(\s_j)/(2\pi\ri\tau)}
  \varphi_{\knot{5}_2}^{(k,\s_j)}(x;\tau),\quad
  j=1,2,3.
\end{equation}
The one-loop contribution  $\varphi_{\knot{5}_2}^{(k,\s_j)}(x;0) =
\omega_{\knot{5}_2}^{(k,\s_j)}(x)$ has a universal
expression
\begin{align}
  \omega_{\knot{5}_2}^{(k,\s_j)}(x) =
  &\frac{(-1)^{n-1}\re^{\frac{\pi\ri}{4}+\frac{\pi\ri k}{2}+\frac{\pi\ri}{2k}(6n^2-1)}}
    {\sqrt{-k(3+2s(x^k)\theta^k+(s(x^k)-1)\theta^{2k})}}
    \theta
    D_k^*(\zeta^{-1/2}\theta)^{1/k}
    D_k^*(\zeta^{-1/2}\theta x)^{1/k}
    D_k^*(\zeta^{-1/2}\theta/x)^{1/k}
    \nn
  &\times\sum_{m\in\IZ_k}\zeta^{m^2}\theta^{2m}(\zeta^{1/2}\theta;\zeta)_m^{-1}
    (\zeta^{1/2}\theta x;\zeta)_m^{-1}
    (\zeta^{1/2}\theta/x;\zeta)_m^{-1},
\end{align}
where
\begin{equation}
  s(x) = 1+x+1/x.
\end{equation}
The power series
$\varphi_{\knot{5}_2}^{(k,\s_j)}(x;\tau)/\omega_{\knot{5}_2}^{(k,\s_j)}(x)$
on the other hand depends in a complicated way on the level $k$.

The case of $k=1$ is particularly simple.  The dependence on
$n\in\IZ_1$ drops out, and the state integral reduces to
\cite{Garoufalidis:2020xec}
\begin{equation}
  \chi_{\knot{5}_2}(\mu;\tau) = \chi_{\knot{5}_2}^{(1)}(\mu,1;\tau)
  =\Phi_{\bb}(0)^{-6}\int_{\IR+\ri 0}\rd v \Phi_\bb(v)\Phi_\bb(v+\mu)\Phi_\bb(v-\mu)
  \re^{-\pi\ri(2v^2+3\mu^2)}
\end{equation}
where we have used $\eta_1 = \Phi_\bb(0)^{-2}$.  The power series
$\varphi_1^{(\s_i)}(x;\tau)$ are given by \cite{Garoufalidis:2020xec}
\begin{align}
  &(\omega_{\knot{5}_2}^{(\s_j)})^{-1}\varphi_{\knot{5}_2}^{(\s_1)}
    (x;\frac{\tau}{2\pi\ri})
    \nn=
  &1 + \frac{Y_j^4}
    {12\gamma_j^3}
    \Big(-12+18s-4s^2-5s^3+s^4+(27-16s-s^2+6s^3-s^4)Y_j+s(-19+2s)Y_j^2\Big)\tau
    \nn
  &+\frac{Y_j^{10}}{288\gamma_j^6}\Big(
    24201-34032s+7438s^2+8872s^3-7337s^4+5128s^5-3479s^6+1135s^7-93s^8\nn
  &-13s^9+s^{10}
    +(4680+2562s-1132s^2-8688s^3+6400s^4-4338s^5+3353s^6-1145s^7\nn
  &+94s^8+13s^9-s^{10})Y_j
    +(-5832+6972s+4921s^2-6026s^3+3034s^4-2454s^5+1030s^6\nn
  &-105s^7-12s^8+s^9)Y_j^2 \Big)\tau^2+\cO(\tau^3),
\end{align}
where
\begin{equation}
  s=s(x) = 1+x+x^{-1},
\end{equation}
and
\begin{equation}
  \gamma_j(x) = 3-2s(x)Y_j(x) + (s(x)-1)Y_j(x)^2
\end{equation}
At level $k=2$, the first few terms of the power series are given in
Appendix~\ref{sc:phi.52}.  It is clear that the coefficients of the
power series depend only on the solution $Y_j$ to the NZ equation, but
not on a $k$-th root of $Y_j$.  Furthermore, they depend on the
deformation parameters $u,n$ only through the holonomy parameter $x$.
Finally, the power series are increasingly more complicated at higher
levels $k$.

\subsection{Holomorphic blocks}
\label{sc:52.blocks}

As in the example of knot $\knot{4}_1$, here we introduce descendants
of the state integral and demonstrate their property of factorisation
into holomorphic blocks.

The descendant state integral for the knot $\knot{5}_2$ takes the form,
\begin{align}
  \chi^{(k)}_{\knot{5}_2;r,s}(\mu,n;\bb) =
  &\frac{(\eta_k)^3}{k}\sum_{m\in\IZ_k}\int_{\mc{C}}
    \rd \sigma
    \mc{Z}^{(k)}_\bb[\Delta](\sigma,m)
    \mc{Z}^{(k)}_\bb[\Delta](\sigma+\mu,m+n)
    \mc{Z}^{(k)}_\bb[\Delta](\sigma-\mu,m-n)\nn
  &\phantom{===}\times(-1)^{n}
    \re^{\frac{\pi\ri}{k}(-(\sigma+\mu-c_\bb)^2-(\sigma-\mu-c_\bb)^2-\mu^2+(m+n)^2+(m-n)^2+n^2)}\nn
  &\phantom{===}{\textcolor{blue}{\times\re^{\frac{2\pi\ri}{k}(-\ri(r\bb+s\bb^{-1})(c_\bb-\s)+(r-s)m)}}},
\end{align}
where the contour of integral $\mc{C}$ is such that it asymptotes at
both $\pm\infty$ to the horizontal line $\imag \s = \s_0$ with
$\sigma_0 <\imag(c_\bb)-|\real(r\bb+s\bb^{-1})|$ but is deformed near
the origin so that all the poles of the integrand are below the
contour.

The integrand only has poles coming from the tetrahedron partition
functions and they are located at
\begin{align}
  &v =
    -\ri\bb\alpha-\ri\bb^{-1}\beta,\quad
    \alpha-\beta \equiv m \;(\text{mod}\; k), \\
  &v = \mu-\ri\bb\alpha-\ri\bb^{-1}\beta,\quad
    \alpha-\beta \equiv m-n \;(\text{mod}\; k),\\
  &v = -\mu -\ri\bb\alpha - \ri\bb^{-1}\beta,\quad
    \alpha-\beta \equiv m+n \;(\text{mod}\; k).
\end{align}
We can evaluate the descendant state integral by completing the
contour from below and summing up residues from all these poles.  We
find in this way that
\begin{align}
  \chi^{(k)}_{\knot{5}_2;r,s}(\mu,n;\bb) = 
    &(-1)^n q^{r/2}\tq^{-s/2}
      \Big(
      \re^{-\frac{\pi\ri}{12}+\frac{5\pi\ri}{6k}+\frac{5\pi\ri}{12k}(\tau+\tau^{-1})}
      A_r(x;q)A_{s}(\tx,\tq^{-1})\nn
    &\phantom{========}
      +\re^{-\frac{5}{12}\pi\ri k+\frac{\pi\ri}{6k}+\frac{\pi\ri}{12k}(\tau+\tau^{-1})}
      B_r(x;q)B_{s}(\tx,\tq^{-1})\nn
    &\phantom{========}
      +\re^{-\frac{5}{12}\pi\ri k+\frac{\pi\ri}{6k} +\frac{\pi\ri}{12k}(\tau+\tau^{-1})}
      C_r(x;q)C_{s}(\tx,\tq^{-1})
      \Big).      
\end{align}
Here $A_r(x,q), B_r(x,q), C_r(x,q)$ are descendant holomorphic blocks
defined by
\begin{subequations}
  \begin{align}
    &A_r(x;q) = \mc{H}(x,x^{-1},q^r;q),\\
    &B_r(x;q) = \theta(-q^{1/2}x;q)^{-2} x^r
      \mc{H}(x,x^2,q^rx^2;q),\\
    &C_r(x;q) = \theta(-q^{-1/2}x;q)^{-2}x^{-r}
      \mc{H}(x^{-1},x^{-2},q^rx^{-2};q),
  \end{align}
\end{subequations}
where $H(x,y,z;q^\varepsilon)$ for $|q|<1$ and $\epsilon = \pm 1$ and
\begin{subequations}
  \begin{align}
    H^+(x,y,z;q) =
    &(qx;q)_\infty (qy;q)_\infty
    \sum_{n=0}^\infty \frac{q^{n(n+1)}z^n}{(q;q)_n(qx;q)_n(qy;q)_n}\\
    H^-(x,y,z;q) =
    &\frac{1}{(x;q)_\infty(y;q)_\infty}
      \sum_{n=0}^\infty (-1)^n
      \frac{q^{\frac{1}{2}n(n+1)}x^{-n}y^{-n}z^n}
      {(q;q)_n(qx^{-1};q)_n(qy^{-1};q)_n}.
  \end{align}
\end{subequations}
Note again that the definition of descendant holomorphic blocks as
functions of $x,q$ are independent of $k$.
They thus enjoy the same properties as discussed in
\cite{Garoufalidis:2020xec}.
If we define the descendant Wronskian matrix
\begin{align}
  W_r(x;q) =
  \begin{pmatrix}
    A_r(x;q) & B_r(x;q) & C_r(x;q)\\
    A_{r+1}(x;q) & B_{r+1}(x;q) & C_{r+1}(x;q)\\
    A_{r+2}(x;q) & B_{r+2}(x;q) & C_{r+2}(x;q)
  \end{pmatrix},
\end{align}
it enjoys the identity
\begin{equation}
  W_{m}(x^{-1};q) = W_m(x;q)
  \begin{pmatrix}
    1&0&0\\
    0&0&1\\
    0&1&0
  \end{pmatrix},
\end{equation}
as well as the inversion property
\begin{equation}
  W_m(x;q)W_{\ell}(x;q^{-1})^T \in PSL(3,\IZ[x^{\pm 1},q^{\pm 1}]).
\end{equation}
In particular
\begin{equation}
  W_{-1}(x;q) W_{-1}(x;q^{-1})^T =
  \begin{pmatrix}
    1 &0&0\\
    0&0&1\\
    0&1&x+x^{-1}
  \end{pmatrix}.
  \label{eq:Wcont.52}
\end{equation}

\subsection{Resurgent structure}
\label{sc:52.lvlk}

\begin{figure}
  \centering
  \includegraphics[height=6cm]{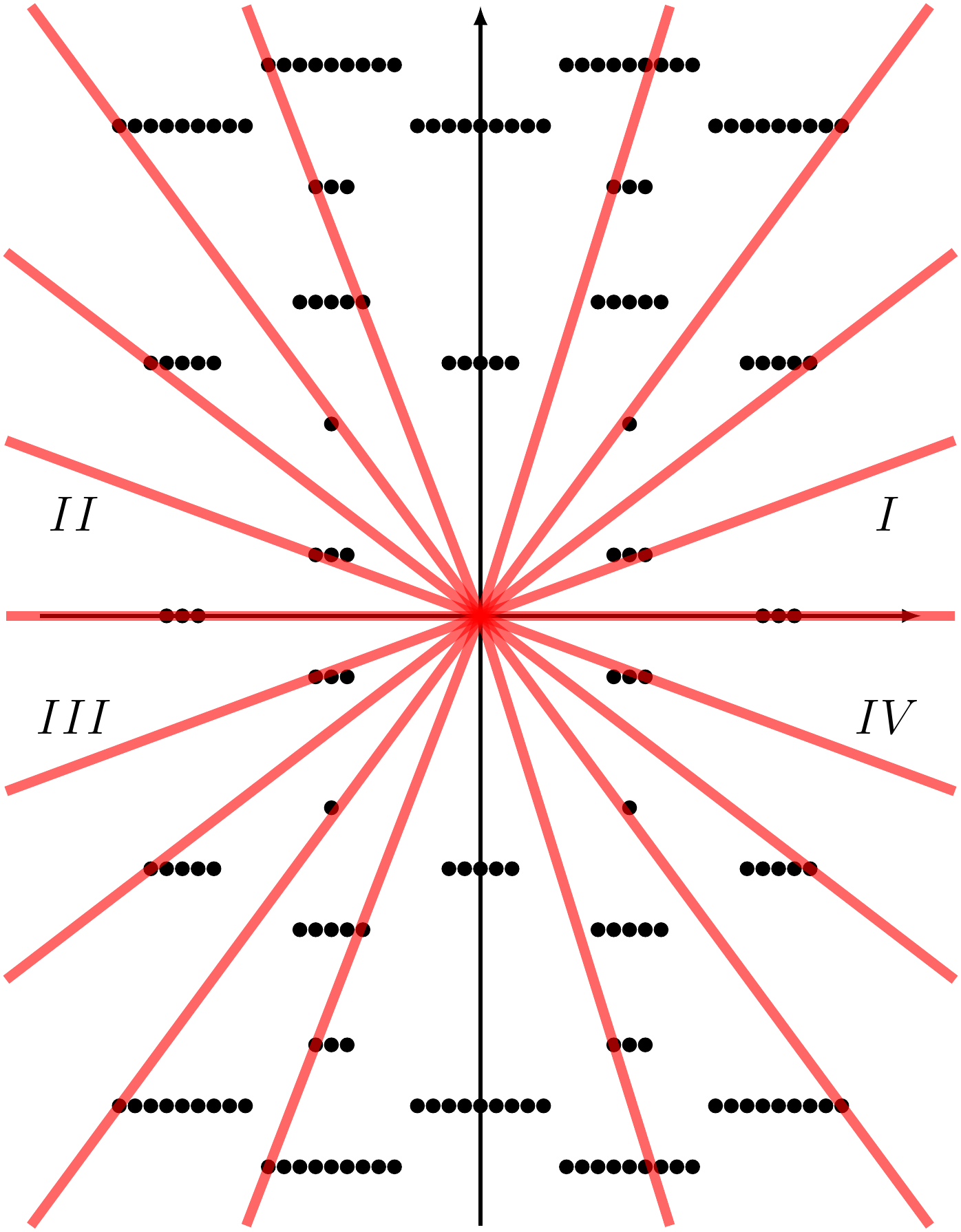}
  \caption{Singular points of the Borel transform of the vector of
    trans-series of the $sl(2,\IC)$ Chern-Simons theory at level $k=1$
    on the complement of the knot $\knot{5}_2$.}
  \label{fg:sing.k1.52}
\end{figure}

We collect the trans-series associated to the geometric and conjugate
flat connections in a vector
\begin{equation}
  \Phi_{\knot{5}_2}^{(k)}(x;\tau) =
  \begin{pmatrix}
    \Phi_{\knot{5}_2}^{(k,\s_1)}\\
    \Phi_{\knot{5}_2}^{(k,\s_2)}\\
    \Phi_{\knot{5}_3}^{(k,\s_2)}
  \end{pmatrix}(x;\tau).
\end{equation}
The singularities of the Borel transform of each component
trans-series are located in $\Lambda^{(k)}_{j}$ given in
\eqref{eq:Lmdu}.  In the case of level $k=1$, we superimpose the Borel
singularities of the components of $\Phi_{\knot{5}_2}^{(1)}(x;\tau)$
and plot them in Fig.~\ref{fg:sing.k1.52}.  At a generic level
$k\geq 1$, the distrubtion of the Borel plane singularities is the
same and their actual locations are reduced by a factor of $1/k$
according to \eqref{eq:Lmdu}.

As explained in Section~\ref{sc:resurg}, Stokes constants of
individual Borel plane singularities can be extracted from global
Stokes automorphisms.
The entire Borel plane is divided by Stokes rays of components of
$\Phi_{\knot{5}_2}^{(k)}(x;\tau)$ into infinitely many cones.  We
label the four cones above and below the positive and negative real
axes by $I,II,III,IV$, ordered in counter-clockwise direction, as
illustrated in Fig.~\ref{fg:sing.k1.52}.  We choose to calculate the
global Stokes automorphism
$\mf{S}_+^{(k)}(\tx;\tq) = \mf{S}^{(k)}_{IV\rightarrow II}(\tx;\tq)$
and
$\mf{S}_-^{(k)}(\tx;\tq) = \mf{S}^{(k)}_{II \rightarrow IV}(\tx;\tq)$,
both in the anti-clockwise direction, which encode all the Stokes
constants.

As in the previous section, in order to derive the Stokes
automorphisms, we first calculate the Borel resummation of the
trans-series.  Following \cite{Garoufalidis:2020xec} and with
numerical evidence presented shortly after, We claim that the Borel
resummation of trans-series can be written as bilinear products of
holomorphic blocks in $q$ and $\tq$. Concretely,
\begin{subequations}
\begin{align}
  s_{I}(\Phi)(x;\tau) =
  &\begin{pmatrix}
    0&1&1\\
    0&1&0\\
    -1&0&0
  \end{pmatrix}W_{-1}(\tx;\tq^{-1})
       \Delta^{(k)}(\tau) B(x;q), \quad |\tq| <1,
       \label{eq:s1.52}\\
  s_{II}(\Phi)(x;\tau) =
  &\begin{pmatrix}
    0&1&0\\
    0&1&1\\
    -1&0&0
  \end{pmatrix}W_{-1}(\tx^{-1};\tq^{-1})
          \Delta^{(k)}(\tau) B(x;q), \quad |\tq| <1,
          \label{eq:s2.52}\\
  s_{III}(\Phi)(x;\tau) =
  &\begin{pmatrix}
    0&1&0\\
    0&-s(\tx)&1\\
    -1&0&0
  \end{pmatrix} W_{-1}(\tx^{-1};\tq^{-1})
          \Delta^{(k)}(\tau)B(x;q), \quad |\tq| >1,
  \label{eq:s3.52}\\
  s_{IV}(\Phi)(x;\tau) =
  &\begin{pmatrix}
    0&-s(\tx)&1\\
    0&1&0\\
    -1&0&0
  \end{pmatrix} W_{-1}(\tx;\tq^{-1})\Delta^{(k)}(\tau)B(x;q),  \quad |\tq| >1.
                  \label{eq:s4.52}
\end{align}
\end{subequations}
Here $W_{-1}(x,q)$ is the Wronskian of holomorphic blocks defined in
\eqref{eq:W.41}, and
\begin{equation}
  B(x;q) =
  \begin{pmatrix}
    A_0(x;q)\\
    B_0(x;q)\\
    C_0(x;q)
  \end{pmatrix}
\end{equation}
$\Delta^{(k)}(\tau)$ is the diagonal matrix defined by
\begin{equation}
  \Delta^{(k)}(\tau) =
  \diag(\re^{-\frac{\pi\ri}{12}+\frac{5\pi\ri}{6k}+\frac{5\pi\ri}{12k}(\tau+\tau^{-1})},
  \re^{-\frac{5}{12}\pi\ri k+\frac{\pi\ri}{6k}+\frac{\pi\ri}{12k}(\tau+\tau^{-1})},
  \re^{-\frac{5}{12}\pi\ri k+\frac{\pi\ri}{6k}+\frac{\pi\ri}{12k}(\tau+\tau^{-1})}
  ).
\end{equation}
As in the example of knot $\knot{4}_1$, we use \eqref{eq:s1.52} and
\eqref{eq:s2.52} to derive Stokes automorphism in the upper half
plane, and use \eqref{eq:s1.52} and \eqref{eq:s4.52} to calculate
Stokes constants for Borel plane singularities in the positive axis,
which involves analytic continuation of the holomorphic blocks with
the help of \eqref{eq:Wcont.52}.  Combining these results, we can
write down the global Stokes automorphism $\mf{S}_+^{(k)}$ from sector
$IV$ to sector $II$ in anti-clockwise diretion.  Similar calculations
can be done to write down $\mf{S}_-^{(k)}$ from sector $II$ to sector
$IV$ in anti-clockwise direction.
\begin{subequations}
\begin{align}
  \mf{S}_+^{(k)}(\tx;\tq) =
  &\begin{pmatrix}
    0&1&0\\
    0&1&1\\
    -1&0&0
  \end{pmatrix}W_{-1}(\tx^{-1};\tq^{-1})W_{-1}(\tx;\tq)^T
                    \begin{pmatrix}
                      0&0&-1\\
                      1&1&0\\
                      0&1&0
                    \end{pmatrix},\quad |\tq|<1\\
  \mf{S}_-^{(k)}(\tx;\tq) = &\begin{pmatrix}
    0&-s(\tx)&1\\
    0&1&0\\
    -1&0&0
  \end{pmatrix}W_{-1}(\tx;\tq) W_{-1}(\tx^{-1};\tq^{-1})^T
                \begin{pmatrix}
                  0&0&-1\\
                  -s(\tx)&1&0\\
                  1&0&0
                \end{pmatrix},\quad |\tq|<1.
\end{align}
\end{subequations}
we notice that the Stokes automorphisms and therefore
the Stokes constants do not depend on the level $k$.

At level $k=1$, the numerical evidences for the Borel plane
singularities as well the Borel resummation formulas
\eqref{eq:s1.52},\eqref{eq:s2.52},\eqref{eq:s3.52},\eqref{eq:s4.52}
were provided in detail in \cite{Garoufalidis:2020xec}.

\begin{figure}[h!]
 $\begin{array}{rl}
  \subfloat[]{\includegraphics[width=0.5\textwidth]
    {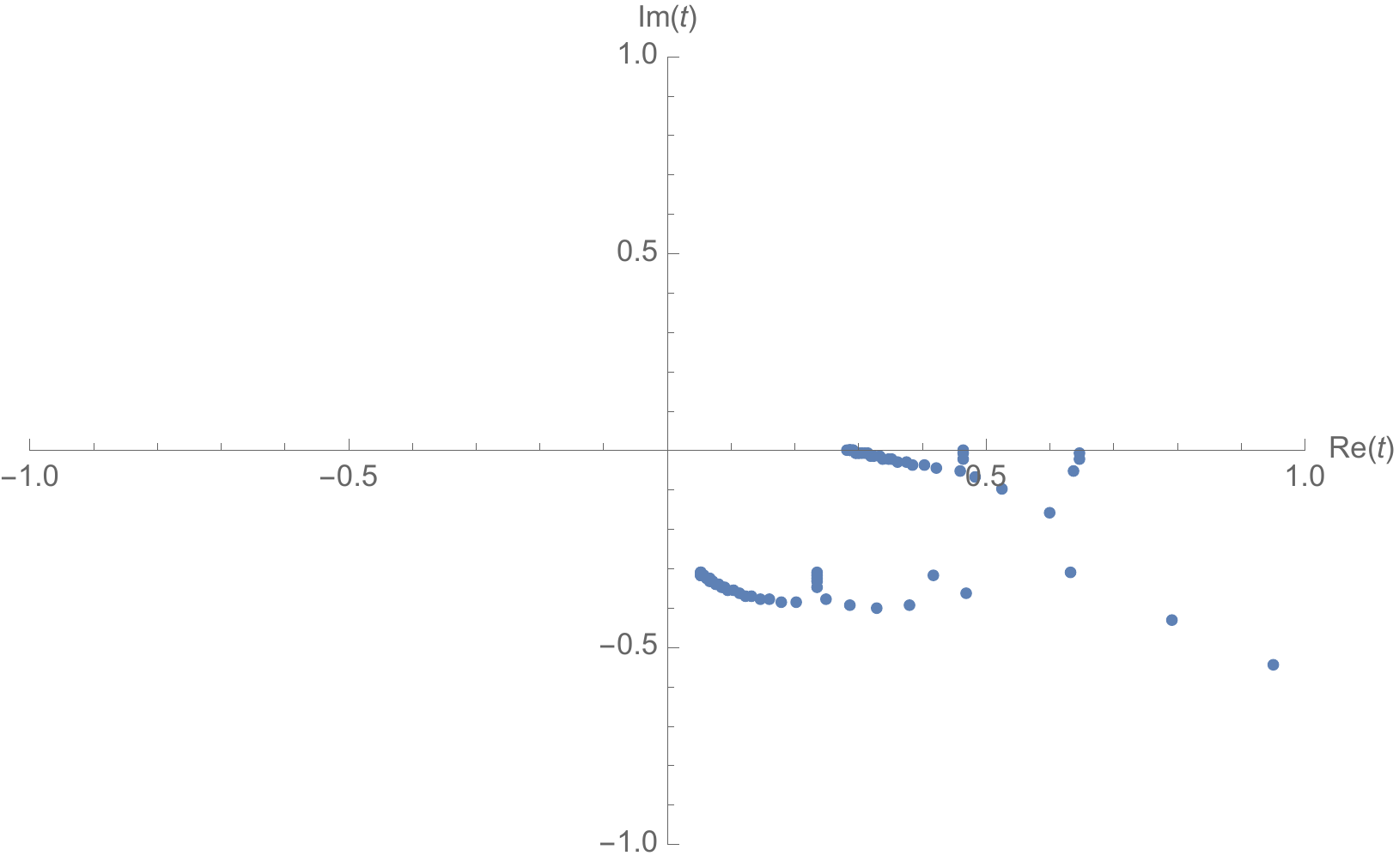}\label{fg:52k2n0s1}} \hspace{2ex}
  \subfloat[]{\includegraphics[width=0.5\textwidth]
    {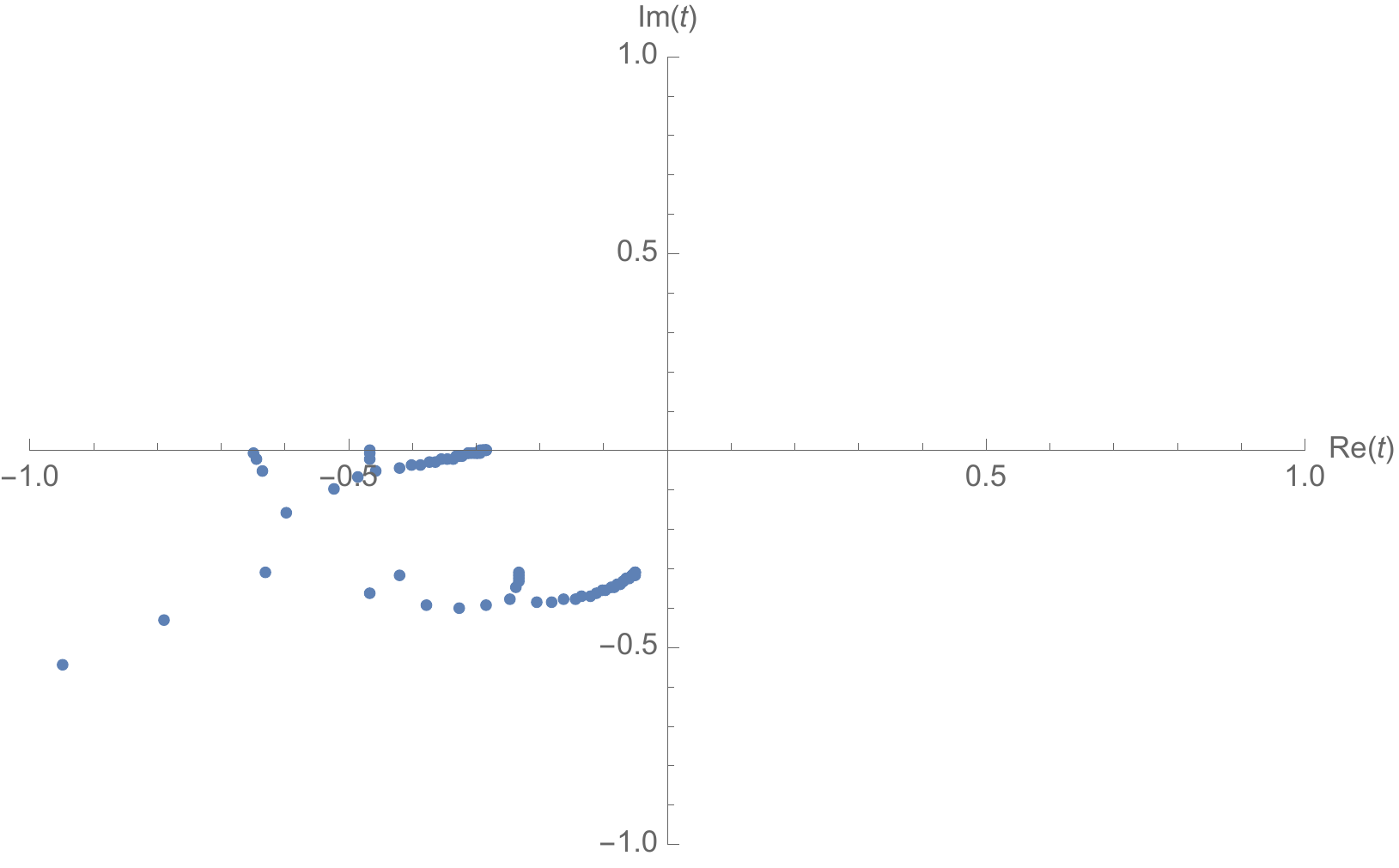}\label{fg:52k2n0s2}}\\
 \multicolumn{2}{c}{\subfloat[]{\includegraphics[width=0.5\textwidth]{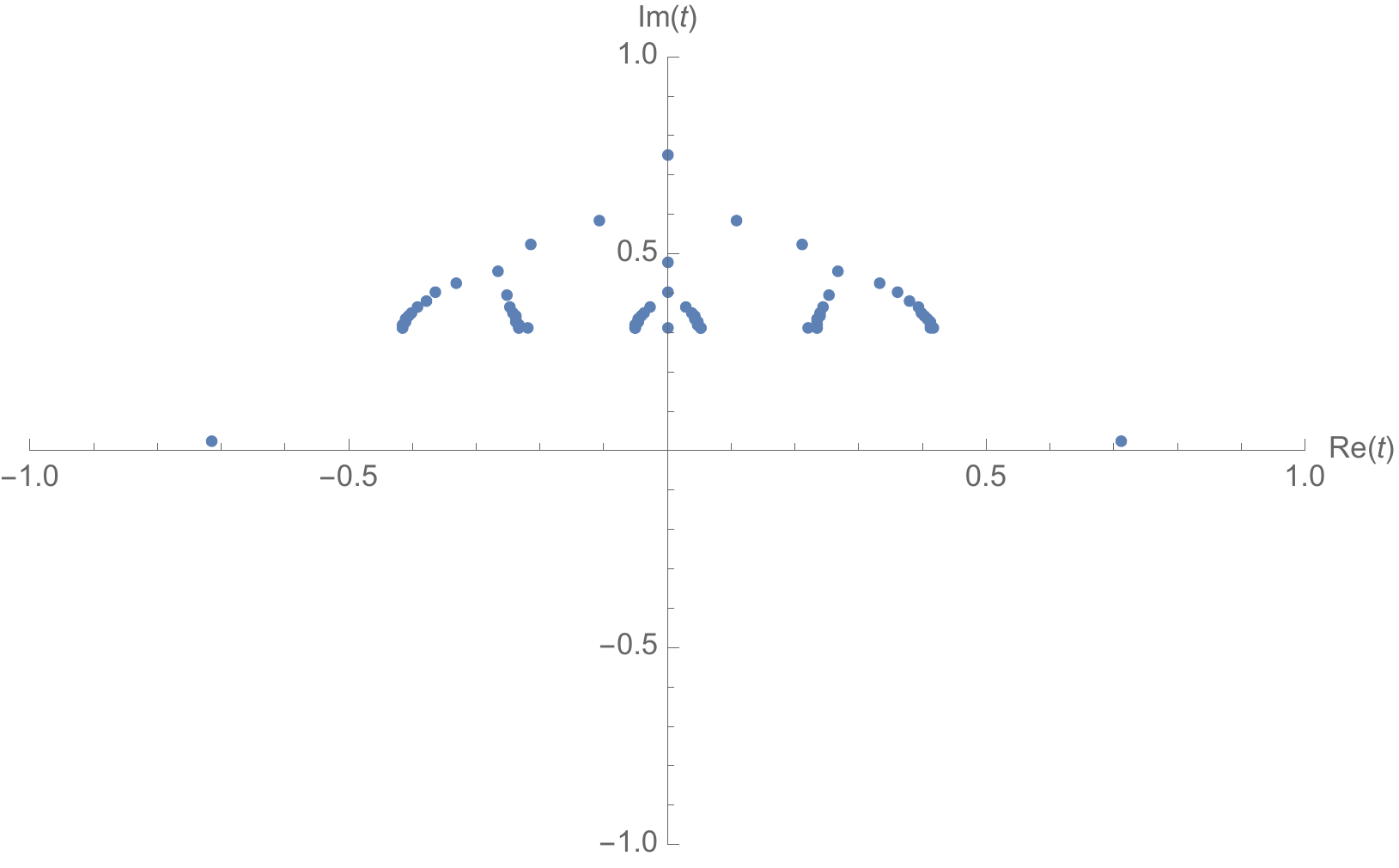}}}
 \end{array}$
  \caption{The distribution of poles of the Pad\'{e} approximant for
    $(a): \Phi_{\knot{5}_2}^{(k=2,\sigma_1)}(x,\tau)$,
    $(b): \Phi_{\knot{5}_2}^{(k=2,\sigma_2)}(x,\tau)$ and $(c): \Phi_{\knot{5}_2}^{(k=2,\sigma_3)}(x,\tau)$ with
    $x = 6/5$ and $N = 200$ terms.}
  \label{fig:52level2n0all}
\end{figure}

We verify these results numerically at level $k=2$.  We choose
$x =(-1)^n \frac{6}{5} $ ($n=0,1$) corresponding to
$u=2\log 6/5, n=0,1$ with power series truncated to $N=200$ terms.
The Stokes plane is divided into four regions similar to
Fig. \ref{fg:sing.k1.52}. For example, the positions of poles for the
Pad\'{e} approximant $\Phi_{\knot{5}_2}^{(k=2)}(x,\tau)$ at
$x = \frac{6}{5}$ are shown in Fig. \ref{fig:52level2n0all}.
Finally we present the numerical results in Tabs. \ref{tab:52k2n0} and
\ref{tab:52k2n1} respectively.
As we can see, the relative errors between two sides of
\eqref{eq:s1.52},\eqref{eq:s2.52},\eqref{eq:s3.52},\eqref{eq:s4.52}
(the first column) are always within the precision of the Borel
resummation (the second column), and in regions $I, II$ are always far
smaller than a potential $\tq$ (or $1/\tq$ in the lower half plane) or
a $\tx^{\pm 1}$ correction (the third and the fourth columns).  In
region $III$, the relative error between two sides of \eqref{eq:s2.52}
is relatively high, due to discrete singular points that break off
from the branch cut of the Borel transform of the truncated series
$\Phi_{\knot{5}_2}^{(k=2,\s_2)}(x;\tau)$ and stray into region $III$,
as seen in Fig.~\ref{fg:52k2n0s2}.  The relative error is still much
smaller than a $\tx^{\pm 1}$ correction, although it does not preclude
a potential $\tq^{- 1}$ correction. The latter, however, is not
possible.  Fig.~\ref{fg:52k2n0s2} indicates that
$\Phi_{\knot{5}_2}^{(k=2,\s_2)}(x;\tau)$ in regions $II$ and $III$
must be related by a Stokes automorphism that accounts for the three
Borel plane singularities in the negative real axis.  Indeed from
\eqref{eq:s2.52},\eqref{eq:s3.52} we find that
\begin{equation}
  s_{III}(\Phi)(x;\tau) = \mf{S}_{II\rightarrow III}(\tx)s_{II}(\Phi)(x;\tau),
  \quad \mf{S}_{II\rightarrow III}(\tx) =
  \begin{pmatrix}
    1&0&0\\
    -1-s(\tx)&1&0\\
    0&0&1
  \end{pmatrix}
\end{equation}
The $(2,1)$ entry of $\mf{S}_{II\rightarrow III}(\tx)$
\begin{equation}
  \mf{S}_{II\rightarrow III}(\tx)_{2,1} = -\tx-2-\tx^{-1}
\end{equation}
should encode the Stokes constants of these three Borel plane
singularities.  Any correction to the right hand side of \eqref{eq:s3.52} should
involve a correction of the coefficients of
$\mf{S}_{II\rightarrow III}(\tx)_{2,1}$ and thus must be of relative
order at least $\tx^{\pm 1}$.
Likewise, from \eqref{eq:s1.52},\eqref{eq:s4.52} we find that
\begin{equation}
  s_{I}(\Phi)(x;\tau) = \mf{S}_{IV\rightarrow I}(\tx)s_{IV}(\Phi)(x;\tau),
  \quad \mf{S}_{IV\rightarrow I}(\tx) =
  \begin{pmatrix}
    1&1+s(\tx)&0\\
    0&1&0\\
    0&0&1
  \end{pmatrix}
\end{equation}
The $(1,2)$ entry of $\mf{S}_{IV\rightarrow I}(\tx)$
\begin{equation}
  \mf{S}_{IV\rightarrow I}(\tx)_{1,2} = \tx+2+\tx^{-1}
\end{equation}
accounts for the three singularities of the Borel transform of
$\Phi_{\knot{5}_2}^{(k=2,\s_1)}(x;\tau)$ in the positive real axis
shown in Fig.~\ref{fg:52k2n0s1}.  Any correction to the right hand
side of \eqref{eq:s4.52} should involve a correction of the
coefficients of $\mf{S}_{IV\rightarrow I}(\tx)_{1,2}$ and thus must be
of relative order $\tx^{\pm 1}$, which is also excluded in
Tabs.~\ref{tab:52k2n0}, \ref{tab:52k2n1}.

\begin{table}
  \centering%
  \subfloat[Region $I$: $\tau = \frac{1}{7}\re^{\frac{\pi\ri}{6}}$]
  {\begin{tabular}{*{5}{>{$}c<{$}}}\toprule
     &|\frac{s_I(\Phi)(x;\tau)}{F_I(x;\tau)}- 1|
     &|\frac{s_I(\Phi)(x;\tau)}{s_I'(\Phi)(x;\tau)}- 1|
     &|\tq(\tau)|& \text{Min}(|\tx(x,\tau)|,|\tx(x,\tau)^{-1}|)\\\midrule
     \sigma_1& 1.4\times 10^{-17} & 1.7\times10^{-17}&&\\
     \sigma_2& 1.1\times 10^{-27} & 1.0\times 10^{-26}& \multirow{1}{*}{$1.7\times 10^{-5}$}
      & \multirow{1}{*}{$0.33$}\\
     \sigma_3& 1.9\times 10^{-21} & 1.9\times 10^{-21}&&\\\bottomrule
   \end{tabular}}\\
 \subfloat[Region $II$: $\tau = \frac{1}{7}\re^{\frac{5\pi\ri}{6}}$]
 {\begin{tabular}{*{5}{>{$}c<{$}}}\toprule
    &|\frac{s_{II}(\Phi)(x;\tau)}{F_{II}(x;\tau)}- 1|
    &|\frac{s_{II}(\Phi)(x;\tau)}{s_{II}'(\Phi)(x;\tau)}- 1|
    &|\tq(\tau)|& \text{Min}(|\tx(x,\tau)|,|\tx(x,\tau)^{-1}|)\\\midrule
      \sigma_1& 1.1\times 10^{-27} & 1.0\times10^{-26}&&\\
     \sigma_2& 1.4\times 10^{-17} & 1.7\times 10^{-17}& \multirow{1}{*}{$1.7\times 10^{-5}$}
      & \multirow{1}{*}{$0.33$}\\
     \sigma_3& 1.9\times 10^{-21} & 1.9\times 10^{-21}&&\\\bottomrule
   \end{tabular}}\\
\subfloat[Region $III$: $\tau = \frac{1}{7}\re^{-\frac{5\pi\ri}{6}}$]
 {\begin{tabular}{*{5}{>{$}c<{$}}}\toprule
    &|\frac{s_{III}(\Phi)(x;\tau)}{F_{III}(x;\tau)}- 1|
    &|\frac{s_{III}(\Phi)(x;\tau)}{s_{III}'(\Phi)(x;\tau)}- 1|
    & |\tq(\tau)^{-1}|& \text{Min}(|\tx(x,\tau)|,|\tx(x,\tau)^{-1}|)\\\midrule
      \sigma_1& 3.8\times 10^{-27} & 7.2\times10^{-26}&&\\
     \sigma_2& 1.3\times 10^{-2} & 1.5\times 10^{-3}& \multirow{1}{*}{$1.7\times 10^{-5}$}
      & \multirow{1}{*}{$0.36$}\\
     \sigma_3& 1.1\times 10^{-23} & 1.9\times 10^{-24}&&\\\bottomrule
  \end{tabular}}\\
 \subfloat[Region $IV$: $\tau = \frac{1}{7}\re^{-\frac{\pi\ri}{6}}$]
 {\begin{tabular}{*{5}{>{$}c<{$}}}\toprule
    &|\frac{s_{IV}(\Phi)(x;\tau)}{F_{IV}(x;\tau)}- 1|
    &|\frac{s_{IV}(\Phi)(x;\tau)}{s_{IV}'(\Phi)(x;\tau)}- 1|
    & |\tq(\tau)^{-1}|& \text{Min}(|\tx(x,\tau)|,|\tx(x,\tau)^{-1}|)\\\midrule
     \sigma_1& 1.3\times 10^{-2} & 1.5\times10^{-3}&&\\
     \sigma_2& 3.8\times 10^{-27} & 7.2\times 10^{-26}& \multirow{1}{*}{$1.7\times 10^{-5}$}
      & \multirow{1}{*}{$0.36$}\\
     \sigma_3& 1.1\times 10^{-23} & 1.9\times 10^{-24}&&\\\bottomrule
   \end{tabular}}
 \caption{We perform the numerical Borel resummation for
   $\Phi_{\knot{5}_2}^{(k=2)}(x,\tau)$ at $x = 6/5$ ($u=2\log 6/5$,
   $n=0$) with 200 terms, after choosing suitable $\tau$ in regions
   $I$ and $II$. We compare them with equations
   \eqref{eq:s1.52}\eqref{eq:s2.52}\eqref{eq:s3.52}\eqref{eq:s4.52}
   whose right hand side is denoted as $F_R(x,\tau)$. Meanwhile, we
   estimate the contribution of higher order terms by resumming
   $\Phi_{\knot{5}_2}^{(k=2)}(x,\tau)$ with 196 terms. The values of
   $|\tilde{q}(\tilde{q}^{-1})|$ and $|\tilde{x}^{\pm 1}|$ are also
   provided for comparison.
 }
  \label{tab:52k2n0}
\end{table}

\begin{table}
  \centering%
  \subfloat[Region $I$: $\tau = \frac{1}{7}\re^{\frac{\pi\ri}{6}}$]
  {\begin{tabular}{*{5}{>{$}c<{$}}}\toprule
     &|\frac{s_I(\Phi)(x;\tau)}{F_I(x;\tau)}- 1|
     &|\frac{s_I(\Phi)(x;\tau)}{s_I'(\Phi)(x;\tau)}- 1|
     &|\tq(\tau)|& \text{Min}(|\tx(x,\tau)|,|\tx(x,\tau)^{-1}|)\\\midrule
     \sigma_1& 3.9\times 10^{-18} & 4.0\times10^{-18}&&\\
     \sigma_2& 2.9\times 10^{-28} & 2.5\times 10^{-28}& \multirow{1}{*}{$1.7\times 10^{-5}$}
      & \multirow{1}{*}{$0.33$}\\
     \sigma_3& 2.2\times 10^{-21} & 7.3\times 10^{-21}&&\\\bottomrule
   \end{tabular}}\\
 \subfloat[Region $II$: $\tau = \frac{1}{7}\re^{\frac{5\pi\ri}{6}}$]
 {\begin{tabular}{*{5}{>{$}c<{$}}}\toprule
    &|\frac{s_{II}(\Phi)(x;\tau)}{F_{II}(x;\tau)}- 1|
    &|\frac{s_{II}(\Phi)(x;\tau)}{s_{II}'(\Phi)(x;\tau)}- 1|
    &|\tq(\tau)|& \text{Min}(|\tx(x,\tau)|,|\tx(x,\tau)^{-1}|)\\\midrule
      \sigma_1& 2.9\times 10^{-28} & 2.5\times10^{-28}&&\\
     \sigma_2& 3.9\times 10^{-18} & 4.0\times 10^{-18}& \multirow{1}{*}{$1.7\times 10^{-5}$}
      & \multirow{1}{*}{$0.33$}\\
     \sigma_3& 2.2\times 10^{-21} & 7.3\times 10^{-21}&&\\\bottomrule
   \end{tabular}}\\
 \subfloat[Region $III$: $\tau = \frac{1}{8}\re^{-\frac{5\pi\ri}{6}}$]
 {\begin{tabular}{*{5}{>{$}c<{$}}}\toprule
    &|\frac{s_{III}(\Phi)(x;\tau)}{F_{III}(x;\tau)}- 1|
    &|\frac{s_{III}(\Phi)(x;\tau)}{s_{III}'(\Phi)(x;\tau)}- 1|
    & |\tq(\tau)^{-1}|& \text{Min}(|\tx(x,\tau)|,|\tx(x,\tau)^{-1}|)\\\midrule
      \sigma_1& 2.4\times 10^{-29} & 1.7\times10^{-29}&&\\
     \sigma_2& 8.1\times 10^{-3} & 2.0\times 10^{-4}& \multirow{1}{*}{$3.5\times 10^{-6}$}
      & \multirow{1}{*}{$0.36$}\\
     \sigma_3& 2.7\times 10^{-25} & 1.1\times 10^{-24}&&\\\bottomrule
  \end{tabular}}\\
 \subfloat[Region $IV$: $\tau = \frac{1}{8}\re^{-\frac{\pi\ri}{6}}$]
 {\begin{tabular}{*{5}{>{$}c<{$}}}\toprule
    &|\frac{s_{IV}(\Phi)(x;\tau)}{F_{IV}(x;\tau)}- 1|
    &|\frac{s_{IV}(\Phi)(x;\tau)}{s_{IV}'(\Phi)(x;\tau)}- 1|
    & |\tq(\tau)^{-1}|& \text{Min}(|\tx(x,\tau)|,|\tx(x,\tau)^{-1}|)\\\midrule
     \sigma_1& 8.1\times 10^{-3} & 2.0\times10^{-4}&&\\
     \sigma_2& 2.4\times 10^{-29} & 1.7\times 10^{-29}& \multirow{1}{*}{$3.5\times 10^{-6}$}
      & \multirow{1}{*}{$0.36$}\\
     \sigma_3& 2.7\times 10^{-25} & 1.1\times 10^{-24}&&\\\bottomrule
   \end{tabular}}
 \caption{We perform the numerical Borel resummation for
   $\Phi_{\knot{5}_2}^{(k=2)}(x,\tau)$ at $x = -6/5$ ($u=2\log 6/5$,
   $n=1$) with 200 terms, after choosing suitable $\tau$ in regions
   $I$ and $II$. We compare them with equations
   \eqref{eq:s1.52},\eqref{eq:s2.52},\eqref{eq:s3.52},\eqref{eq:s4.52}
   whose right hand side is denoted as $F_R(x,\tau)$. Meanwhile, we
   estimate the contribution of higher order terms by resumming
   $\Phi_{\knot{5}_2}^{(k=2)}(x,\tau)$ with 196 terms. The values of
   $|\tilde{q}(\tilde{q}^{-1})|$ and $|\tilde{x}^{\pm 1}|$ are also
   provided for comparison.
 }
  \label{tab:52k2n1}
\end{table}

\section{Conclusion and discussion}
\label{sc:con}

In this short note we study the resurgent properties of the
$sl(2,\IC)$ Chern-Simons state integral model on knot complemnets
$S^3\backslash \knot{4}_1$, $S^3\backslash \knot{5}_2$ at generic
level $k\geq 1$.
When the level $k$ increases, the saddle point expansion of the $sl(2,\IC)$
Chern-Simons state integral becomes
increasingly more complicated.
But there are several interesting universal featrues.
First of all, the coefficients of the power series always live in the
trace field of the knot extended by the holonomy paramter $x$.
Second, these power series enjoy universal resurgent properies.  When
the holonomy deformed is weak so that $|\log x^k|\ll 1$, the
distribution of the singularities of the Borel transform is the same
and they are all related to the level $k=1$ singularites by a factor
of $1/k$.  In addition, the Stokes constants associated to these
singularities are independent of the level $k$.  We speculate that
this is related to the fact the Hilbert space of the quantum
Chern-Simons theory at level $k$ is simply a $k$ copy of the level one
theory, and that they all coincide with the same BPS invariants of the
dual 3d SCFT.

Some interesting questions remain unanswered following this work.
Both this work and its predecessor \cite{Garoufalidis:2020xec} focus
on the case when the holonomy is weak\footnote{There are some limited
  but unsystematic discussion for large holonomy in
  \cite{Garoufalidis:2020xec}.}.  It will be interesting to understand
the resurgent structure of the Chern-Simons theory in the entire
holonomy space.  Due to the identification of the Stokes constants
with the BPS invariants in the dual 3d SCFT
\cite{Garoufalidis:2020xec,Garoufalidis:2020nut}, the evolution of the
resurgent structure as the holonomy changes may enjoy features similar
to the Wall Crossing Formulas of Kontsevich and Soibelman
\cite{Kontsevich:2008fj}.

Second, it was shown in \cite{Garoufalidis:2021osl} that the resurgent
structures presented in
\cite{Garoufalidis:2020xec,Garoufalidis:2020nut} are not complete:
They involve only asymptotic series $\varphi^{(k=1,\s_j)}(x;\tau)$
($j\geq 1$) associated to non-Abelian flat connections, while there
still remains the asymptotic series $\varphi^{(k=1,\s_0)}(x;\tau)$
associated to the Abelian flat connection.  It was revealed in
\cite{Garoufalidis:2021osl} with examples of $\knot{4}_1$ and
$\knot{5}_2$ that there are non-trivial Stokes automorphisms that
relate $\varphi^{(k=1,\s_0)}(x;\tau)$ to
$\varphi^{(k=1,\s_j)}(x;\tau)$ ($j\geq 1$) but not the other way
around.  This is similar to the discovery in the case of compact three
manifolds in \cite{Gukov2016}.  It would be interesting to also
generalise this connection to generic levels.

In fact, the power series $\varphi^{(k=1,\s_0)}(x;\tau)$ were computed
by expanding colored Jones polynomial near $q\rightarrow 1$.  Both
$\varphi^{(k=1,\s_0)}(1;\tau)$ and $\varphi^{(k=1,\s_j)}(1;\tau)$ with
holonomy turned off appear in the Refined Quantum Modularity
Conjecture \cite{Zagier:2010qmf,Garoufalidis:2021lcp} in the case of
S-transformation (see also \cite{Dimofte:2015kkp,Dimofte:2012qj}).
For a generic $sl(2,\IZ)$ transformation
$(\begin{smallmatrix} a&b\\c&d\end{smallmatrix})$, the power series
should be replaced by $\varphi^{(k=c,\s_0)}(1;\tau)$ and
$\varphi^{(k=c,\s_j)}(1;\tau)$ or their Galois transformation in the
trace field.  The power series $\varphi^{(k=c,\s_0)}(x;\tau)$ can be
computed by expanding colored Jones polynomial near
$q\rightarrow \exp2\pi\ri/c$.  We conjecture then that
$\varphi^{(k,\s_0)}(x;\tau)$ constructed this way is also related to
$\varphi^{(k,\s_j)}(x;\tau)$ ($j\geq 1$) computed in this work by
non-trivial Stokes automorphisms, where the Stokes constants coincide
with those in \cite{Garoufalidis:2021osl}.

We have been focusing on the generalisation of
\cite{Garoufalidis:2020xec,Garoufalidis:2020nut} to $sl(2,\IC)$
Chern-Simons theory with higher levels $k\geq 1$.  Other directions of
generalisation are also possible.  For instance one can consider
$sl(N,\IC)$ Chern-Simons theory with $N\geq 2$.  State integral models
have also been written down in \cite{Dimofte:2013iv}.
Alternatively, one can also study the resurgent structures of the
Chern-Simons theory at level $k=0$.  This is a special case, and by
the 3d-3d correspondence, it is dual to the 3d SCFT on $S^2\times S^1$
\cite{Dimofte:2011py}, whose partition function is nothing else but
the supersymmetric indices of the 3d SCFT.  In other words, one could
study the saddle point expansion of the indices and their resurgent
structures. See related work in
\cite{Garoufalids:2022pmi}. Furthermore, three manifolds obtained from
Dehn filling of hyperbolic knot complements are also interesting
objects to study \cite{Gang:2017cwq,Choi:2022dju}.

Finally, following the arguments in \cite{Gaiotto2008cd} (see a
similar application of the same idea \cite{Ito:2018eon}), the Stokes
automorphism together with the asymptotic behavior define for us a
Riemann-Hilbert problem for the Borel resummed asymptotic series.  The
latter may be solved by a TBA-like equation.  It is interesting to
explore if such a TBA-like equation can be written down for Borel
resummed asymptotic series in complex Chern-Simons theory.  As
showcased in \cite{Ito:2018eon}, such an equation could help answer
the question of the evolution of the resurgent structure as we change
the moduli of the system, which are the holonomy parameter $x$ here.

\section*{Acknowledgements}

We thank Sunjin Choi, Dongmin Gang, Stavros Garoufalidis, Kimyeong Lee, Sungjay Lee, Marcos Mari\~no for stimulating
discussions. Main results of this work have been presented by ZD in Incheon workshop "New Frontiers in Quantum Field Theory and String Theory" and string theory seminar at LPENS. ZD would like to thank the audience for suggestions and feedback. ZD is supported by KIAS Grant PG076902. JG is supported by the Startup Funding no.~3207022203A1 and no.~4060692201/011 of the Southeast University.

\appendix

\section{Tetrahedron partition function at level $k$}
\label{sc:dilog}

The partition function of $sl(2,\IC)$ Chern-Simons theory at level $k$
on an ideal tetrahedron is given by
\begin{equation}
  \mc{Z}_\bb^{(k)}[\Delta](\mu,n) = \prod_{(r,s)\in \Gamma(k;n)}\Phi_\bb(c_\bb
  -\frac{1}{k}(\mu+\ri \bb r + \ri\bb^{-1}s))
\end{equation}
where
\begin{equation}
  \Gamma(k;n) = \{(r,s)\in\IZ^2\;|\; 0\leq r,s<k, r-s \equiv n
  (\text{mod}\;k)\}
\end{equation}
and $c_\bb = \frac{\ri}{2}(\bb+\bb^{-1})$.
$\Phi_\bb(x)$ is Faddeev's quantum dilogarithm.
This defines a meromorphic
function of $\mu\in\IC$ for each $n\in\IZ_k$, and it is defined for
all values of $\bb$ with $\bb^2$ in the cut plane
$\IC' = \IC\backslash \IR_{\leq 0}$.

When $\imag \bb >0$ or $\imag\bb <0$, it has the factorisation form
\begin{equation}
  \mc{Z}_\bb^{(k)}[\Delta](\mu,n) = (qx^{-1};q)_\infty (\tq^{-1}\tx^{-1};\tq^{-1})_\infty
\end{equation}
where we use the notation
\begin{equation}
  \begin{gathered}
    q = \exp \frac{2\pi\ri}{k}(\bb^2+1),\quad \tq = \exp
    -\frac{2\pi\ri}{k}(\bb^{-2}+1),\\
    x = \exp \left(\frac{2\pi\bb\mu}{k} - \frac{2\pi\ri n}{k}\right),\quad
    \tx = \exp \left(\frac{2\pi\bb^{-1}\mu}{k} + \frac{2\pi\ri n}{k}\right),
  \end{gathered}
\end{equation}
and $(a,q)_\infty$ is the $q$-Pochhammer symbol defined to be
$\prod_{j=1}^\infty(1-a q^j)$ if $|q|<1$ or
$1/(q^{-1}a;q^{-1})_\infty$ if $|q|>1$.

\subsection{Fundamental properties}

\begin{itemize}
\item Inversion relation
  \begin{equation}
    \mc{Z}^{(k)}_\bb[\Delta](c_\bb+x,n)
    \mc{Z}^{(k)}_\bb[\Delta](c_\bb-x,-n) = (-1)^n
    \re^{\frac{\pi\ri}{k}(x^2-n^2)}\eta_k^{-1},
  \end{equation}
  where
  \begin{equation}
    \eta_k = \re^{\pi\ri(k+2c_{\bb}^2 k^{-1})/6}.
  \end{equation}
  Note that this implies the special value
  \begin{equation}
    \mc{Z}^{(k)}_\bb[\Delta](c_\bb,0)^2 = \eta_k^{-1}.
  \end{equation}
\item Quasi-periodicity (Faddeev's difference equations).  For
  $x,\bb\in \IC$, $\imag(\bb) \neq 0$, $n\in\IZ$
  \begin{equation}
    \begin{aligned}
      &\mc{Z}^{(k)}_\bb[\Delta]\left(x-\ri\bb^{\pm 1},n\pm 1\right)
      =\mc{Z}^{(k)}_\bb[\Delta](x,n)
      \left(1-q_{\pm}x_{\pm}^{-1}\right)^{-1}\\
      &\mc{Z}^{(k)}_\bb[\Delta]\left(x+\ri\bb^{\pm 1},n\mp 1\right)
      =\mc{Z}^{(k)}_\bb[\Delta](x,n) \left(1-x_{\pm}^{-1}\right).
    \end{aligned}
    \label{eq:Db-per}
  \end{equation}
  where $q_+ = q, q_- = \tq$, $x_+ = x$, $x_- = \tx$.
\item Asymptotic behavior
  \begin{align}
    \mc{Z}_\bb^{(k)}[\Delta](\mu,n) \sim
    \begin{cases}
      \re^{\frac{\pi\ri}{k}(c_\bb-\mu)^2}\quad &\real(\mu) \ll 0\\
      1 \quad &\real(\mu) \gg 0.
    \end{cases}
                \label{eq:Zasymp}
  \end{align}
\end{itemize}

\subsection{Zeros and poles}

For $\imag(\bb) > 0$, the tetrahedron partition function
$\mc{Z}_\bb^{(k)}[\Delta](x,n)$ has zeros
\begin{equation}
  \begin{cases}
    x = 2c_\bb + \ri \bb^{-1} l +\ri \bb\, m\\
    n = l-m \; (\text{mod}\; k)
  \end{cases}
\end{equation}
and poles
\begin{equation}
  \begin{cases}
    x = -\ri \bb^{-1}l - \ri \bb\,m\\
    n = m-l \; (\text{mod}\; k)
  \end{cases}
\end{equation}
for $l,m\in\IZ_{\geq 0}$, and the residue is
\begin{equation}
  \frac{k}{2\pi\bb^{-1}}\frac{(q;q)_\infty}{(\tq;\tq)_\infty}
  \frac{1}{(q;q)_m}\frac{1}{(\tq^{-1};\tq^{-1})_l}.
\end{equation}

\subsection{Integral identities}

The tetrahedron partition function has the following property of
Fourier transformation,
\begin{equation}
  \frac{1}{k}\sum_{n\in\IZ_k}\int_{\IR+c_\bb} \rd u
  \mc{Z}_\bb^{(k)}[\Delta](u,n)\re^{\frac{2\pi\ri}{k}(uw-nm)}
  =\mc{Z}_\bb^{(k)}[\Delta](w,m)(-1)^m
  \re^{-\frac{\pi\ri}{k}((w-c_{\bb})^2-m^2)+\frac{\pi\ri}{12}
    (k+8c_{\bb}^2 k^{-1})}.
  \label{eq:FourierFQD}
\end{equation}

\subsection{Semi-classical limit}

In the double scaling limit
\begin{equation}
  \bb\rightarrow 0,\quad \mu\rightarrow \infty,\quad \bb\mu\;\text{fixed}
\end{equation}
the tetrahedron partition function has the asymptotic expansion
\begin{equation}
  \log \mc{Z}_\bb^{(k)}[\Delta](\mu,m) =
  \sum_{\ell=0}^\infty \frac{(2\pi\ri\bb^2)^{\ell-1}}{\ell!}
  \sum_{j\in\IZ_k}B_\ell(j/k)\Li_{2-\ell}\left(
    \re^{\frac{2\pi\ri}{k}(j+m)}\re^{-\frac{2\pi\bb\mu}{k}}
  \right).
  \label{eq:ZDasymp}
\end{equation}
Alternatively, define
\begin{equation}
  \DD_\bb^{(k)}(u,m) = \mc{Z}_\bb^{(k)}[\Delta](c_\bb-\sqrt{k}u,m).
\end{equation}
In the double scaling limit
\begin{equation}
  \bb\rightarrow 0,\quad u\rightarrow \infty,\quad \bb u\;\text{fixed},
\end{equation}
it has the asymptotic expansion
\begin{equation}
  \log \DD_\bb^{(k)}(u,m) = \sum_{\ell=0}^\infty
  \frac{(2\pi\ri\bb^2)^{\ell-1}}{\ell!}\sum_{j\in\IZ_k}
  B_\ell(1-\frac{1}{2k}-\frac{j}{k})
  \Li_{2-\ell}(\re^{-\frac{\pi\ri}{k}(1+2j-2m)}\re^{\frac{2\pi\bb
      u}{\sqrt{k}}}).
  \label{eq:Dbasymp}
\end{equation}

\section{Power series of state integrals at different levels}

\subsection{Knot $\knot{4}_1$}
\label{sc:phi.41}

At level $k=2$, the power series are
\begin{align}
  (\omega_{\knot{4}_1}^{(k=2,\s_j)}(x))^{-1}
  \varphi_{\knot{4}_1}^{(k=2,\s_j)}(x;\frac{\tau}{2\pi\ri}) = 1 +
  a^{(\s_j)}_1(x) \tau + a^{(\s_j)}_2(x)\tau^2+ a^{(\s_j)}_3(x)\tau+\cO(\tau^4),
\end{align}
where $a^{(\s_j)}_1(x)$ is given in \eqref{eq:phi41k2} and some
additional coefficients are
\begin{subequations}
  \begin{align}
    a^{(\s_j)}_2(x) =
    &(1 + 12 x + 46 x^2 - 576 x^3 + 1413 x^4 + 1608 x^5 + 202 x^6 + 
      7788 x^7 + 3258 x^8 - 34572 x^9\nn
    &+ 1094 x^{10} + 27816 x^{11} - 
      3555 x^{12} + 27816 x^{13} + 1094 x^{14} - 34572 x^{15}
      + 3258 x^{16} \nn
    &+ 7788 x^{17} + 202 x^{18} + 1608 x^{19} + 1413 x^{20} - 576 x^{21} + 
      46 x^{22} + 12 x^{23} + 
      x^{24})/\nn
    &(4608 (-1 - x + x^2)^3 (1 - x + x^2)^3 (-1 + x + x^2)^3 (1 + 
      x + x^2)^3),\\
    a^{(\s_j)}_3(x) =
    &-((5 - 8550 x - 10329 x^2 - 84870 x^3 + 534081 x^4 + 156600 x^5 + 
      907732 x^6 + 7978680 x^7\nn
    &+ 4743126 x^8 - 47249100 x^9 + 
      10187454 x^{10} + 68274900 x^{11} - 8491270 x^{12} + 
      156024000 x^{13}\nn
    &+ 16794912 x^{14} - 565084800 x^{15} + 
      12734439 x^{16} + 387780030 x^{17} - 42351651 x^{18}\nn
    &+ 387780030 x^{19} + 12734439 x^{20} - 565084800 x^{21} + 
      16794912 x^{22} + 156024000 x^{23}\nn
    &- 8491270 x^{24} + 
      68274900 x^{25} + 10187454 x^{26} - 47249100 x^{27} + 4743126 x^{28} + 
      7978680 x^{29}\nn
    &+ 907732 x^{30} + 156600 x^{31} + 534081 x^{32} - 
      84870 x^{33} - 10329 x^{34} - 8550 x^{35} + 5 x^{36})\nn
    &(-1 - x^2 + 
      x^4 + 2 x^2 Y_j))/(3317760 (-1 - x + x^2)^5 (1 - x + x^2)^5 (-1 +
      x + x^2)^5 (1 + x + x^2)^5).
  \end{align}
\end{subequations}

At level $k=3$, the power series are
\begin{align}
  (\omega_{\knot{4}_1}^{(k=3,\s_j)}(x))^{-1}
  \varphi_{\knot{4}_1}^{(k=3,\s_j)}(x;\frac{\tau}{2\pi\ri}) = 1 +
  c^{(\s_j)}_1(x) \tau + c^{(\s_j)}_2(x)\tau^2+ \cO(\tau^3),
\end{align}
where
\begin{align}
  c_1^{(\s_j)}(x) =
  & (1 +8 x +17 x^2 +8(2+3\zeta) x^3-24(1+\zeta) x^4
    -24(1+\zeta) x^5
    -4(1+6\zeta) x^6\nn
  &+ 8(-1+3\zeta) x^7
    + 36(-1+2\zeta) x^8
    + 2(47-48\zeta) x^9
    + 24(1+2\zeta) x^{10}
    +142 x^{11} \nn
  &+ (95-24\zeta) x^{12}
    + 8(4+15\zeta) x^{13}
    + +3(53-8\zeta) x^{14}
    - 194 x^{15}
    + 8 (5+\zeta) x^{16}\nn
  &- 2(89+48\zeta) x^{17}
    + 2(25+36\zeta) x^{18}
    -8 (1-3\zeta) x^{19}
    -2(7+12\zeta) x^{20}
    + 2(17-12\zeta) x^{21}\nn
  &+ 8(2-3\zeta) x^{22}
    + 2(5+12\zeta) x^{23}
    -17 x^{24} -8 x^{25} -
    x^{26} \nn
  &+ 2 x^3 (1 + 8 x + 17 x^2 + 3 x^3 - 20 x^4 - 29 x^5 + 6 x^6 + 8 x^7 - 
    26 x^8 + 139 x^9 - 28 x^{10} \nn
  &+ 139 x^{11} - 26 x^{12} + 8 x^{13} + 
    6 x^{14} - 29 x^{15} - 20 x^{16} + 3 x^{17} + 17 x^{18} + 8 x^{19}
    + x^{20})Y_j)/\nn
  &(72(1 + x^2) (1 - 3 x^3 + x^6)^2 (1 + x^3 + x^6)^2),\\
  c_2^{(\s_j)}(x) =
  &(1 + 16 (5 + 8 \zeta) x + (33 - 128 \zeta) x^2 + 2 (207 + 440 \zeta) x^3 - 16 (129 + 103 \zeta) x^4\nn
  &- 
    2 (-167 + 1144 \zeta) x^5 + (3101 + 6256 \zeta) x^6 - 
    112 (-64 + 49 \zeta) x^7 + (-5923 + 2928 \zeta) x^8 \nn
  &- 2 (-14241 + 3544 \zeta) x^9 - 16 (484 + 459 \zeta) x^{10} + 
    2 (40457 + 6680 \zeta) x^{11}\nn
  &+ 2 (-6751 + 13744 \zeta) x^{12} - 
    32 (1406 + 615 \zeta) x^{13} - 2 (63311 + 27248 \zeta) x^{14}\nn
  &+ 2 (6179 + 48568 \zeta) x^{15} - 16 (635 + 1871 \zeta) x^{16} - 
    2 (54965 + 10456 \zeta) x^{17}\nn
  &+ (228909 - 59120 \zeta) x^{18} - 
    16 (1402 + 1109 \zeta) x^{19} + (329133 + 141328 \zeta) x^{20}\nn
  &- 
    2 (72397 + 45320 \zeta) x^{21} - 16 (-128 + 345 \zeta) x^{22} - 
    10 (4033 + 824 \zeta) x^{23}\nn
  &+ 2 (-32943 + 33488 \zeta) x^{24} + 
    32 (-964 + 269 \zeta) x^{25} - 2 (26287 + 25328 \zeta) x^{26}\nn
  &- 
    2 (-36257 + 1720 \zeta) x^{27} + 16 (-249 + 11 \zeta) x^{28} + 
    2 (20473 + 8920 \zeta) x^{29}\nn
  &+ (-11011 - 7248 \zeta) x^{30} + 
    16 (680 + 121 \zeta) x^{31} + (-2275 - 4496 \zeta) x^{32}\nn
  &+ 
    2 (1247 + 1016 \zeta) x^{33} + 16 (-18 + 119 \zeta) x^{34} - 
    2 (233 + 440 \zeta) x^{35} + (161 + 128 \zeta) x^{36}\nn
  & - 16 (3 + 8 \zeta) x^{37} + 
    x^{38} - 
    16 (1 + 2 \zeta) x^4 (8 - 8 x + 55 x^2 - 111 x^3 - 135 x^4 + 
    336 x^5 \nn
  &- 224 x^6 + 310 x^7 - 724 x^8 - 346 x^9 + 390 x^{10} + 
    2778 x^{11} - 1108 x^{12} - 3486 x^{13} + 2569 x^{14} \nn
  &- 1109 x^{15} + 
    2569 x^{16} - 3486 x^{17} - 1108 x^{18} + 2778 x^{19} + 390 x^{20} - 
    346 x^{21} - 724 x^{22}\nn
  &+ 310 x^{23} - 224 x^{24} + 336 x^{25} - 
    135 x^{26} - 111 x^{27} + 55 x^{28} - 8 x^{29} + 8 x^{30}) Y)/\nn
  &(10368 (1 +
    x^2) (1 - 3 x^3 + x^6)^3 (1 + x^3 + x^6)^3);
\end{align}

\subsection{Knot $\knot{5}_2$}
\label{sc:phi.52}

At level $k=2$, the power series are
\begin{align}
  (\omega_{\knot{5}_2}^{(k=2,\s_j)}(x))^{-1}
  \varphi_{\knot{5}_2}^{(k=2,\s_j)}(x;\frac{\tau}{2\pi\ri}) = 1
  + a_1^{(\s_j)}\tau + a_2^{(\s_j)}\tau^2 + \cO(\tau^3),
\end{align}
where
\begin{subequations}
  \begin{align}
    a_1^{(\s_j)} =
    &(1 + 6 x + 14 x^2 - 30 x^3 - 245 x^4 - 216 x^5 + 944 x^6 + 1428 x^7 - 
      1231 x^8 - 2346 x^9 + 1326 x^{10}\nn
    &+ 3042 x^{11} - 1873 x^{12} - 
      6168 x^{13} - 3400 x^{14} - 2556 x^{15} - 3935 x^{16} - 4950 x^{17} - 
      5474 x^{18}\nn
    &- 4950 x^{19} - 3935 x^{20} - 2556 x^{21} - 3400 x^{22} - 
      6168 x^{23} - 1873 x^{24} + 3042 x^{25} + 1326 x^{26}\nn
    &- 2346 x^{27} - 
      1231 x^{28} + 1428 x^{29} + 944 x^{30} - 216 x^{31} - 245 x^{32} - 30 x^{33} + 
      14 x^{34} + 6 x^{35} \nn
    &+ x^{36} + (-1 - 3 x  - 10 x^2 + 15 x^3 +
      262 x^4 + 381 x^5  - 964 x^6  - 2118 x^7  + 697 x^8  + 
      2886 x^9 \nn
    &+ 469 x^{10} - 873 x^{11} + 1656 x^{12} + 4797 x^{13} + 
      2911 x^{14} + 903 x^{15} + 899 x^{16} + 984 x^{17} + 1185
      x^{18}\nn
    &+ 984 x^{19} + 899 x^{20} + 903 x^{21} + 2911 x^{22} + 4797 x^{23} + 
      1656 x^{24} - 873 x^{25} + 469 x^{26} + 2886 x^{27} \nn
    &+ 697 x^{28} - 2118 x^{29} - 964 x^{30} + 381 x^{31} + 262 x^{32} + 15 x^{33} - 
      10 x^{34} - 3 x^{35} - x^{36}) Y_j\nn
    &+ (x^2 + 3 x^3 + 9 x^4 - 
      27 x^5 - 273 x^6 - 318 x^7 + 973 x^8 + 
      1737 x^9 - 916 x^{10} - 2070 x^{11}\nn
    &+ 193 x^{12} + 
      288 x^{13} - 2922 x^{14} - 6345 x^{15} - 4154 x^{16} - 
      1221 x^{17} - 825 x^{18} - 1221 x^{19}\nn
    &- 4154 x^{20} - 
      6345 x^{21} - 2922 x^{22} + 288 x^{23} + 193 x^{24} - 
      2070 x^{25} - 916 x^{26} + 1737 x^{27}\nn
    &+ 973 x^{28} - 
      318 x^{29} - 273 x^{30} - 27 x^{31} + 9 x^{32} + 
      3 x^{33} + 
      x^{34}) Y_j^2)/\nn
    &(24 (1 + 3 x + 3 x^2 + 3 x^3 + x^4) (1 - 2 x - x^2 + 
      8 x^3 - 11 x^4 + 8 x^5 - x^6 - 2 x^7 + x^8)^2 \nn
    &(1 + 2 x - x^2 - 
      8 x^3 - 11 x^4 - 8 x^5 - x^6 + 2 x^7 + x^8)^2),\\
    a_2^{(\s_j)} =
    &(1 + 9 x + 183 x^2 + 27 x^3 - 4394 x^4 - 6915 x^5 + 24656 x^6 + 
      55122 x^7 - 7429 x^8 - 73344 x^9 \nn
    &- 227211 x^{10} - 497652 x^{11} + 
      35060 x^{12} + 948795 x^{13} + 659308 x^{14} + 607485 x^{15} + 
      3506489 x^{16} \nn
    &+ 4523433 x^{17} - 7706054 x^{18} - 16876941 x^{19} + 
      1410725 x^{20} + 16099737 x^{21} - 1232206 x^{22} \nn
    &- 18943725 x^{23} - 
      8872703 x^{24} + 6776679 x^{25} + 10490726 x^{26} + 6776679 x^{27} - 
      8872703 x^{28} \nn
    &- 18943725 x^{29} - 1232206 x^{30} + 16099737 x^{31} + 
      1410725 x^{32} - 16876941 x^{33} - 7706054 x^{34} \nn
    &+ 4523433 x^{35} + 
      3506489 x^{36} + 607485 x^{37} + 659308 x^{38} + 948795 x^39 + 
      35060 x^{40} - 497652 x^{41}\nn
    &- 227211 x^{42} - 73344 x^{43} - 7429 x^{44} + 
      55122 x^{45} + 24656 x^{46} - 6915 x^{47} - 4394 x^{48} + 27
      x^{49}\nn
    &+183 x^{50} + 
      9 x^{51} + x^{52} + (-1 - 9 x - 318 x^2 + 546 x^3 + 7812 x^4 + 
      5967 x^5 - 42735 x^6\nn
    &- 59742 x^7 + 74478 x^8 + 179832 x^9 + 
      140847 x^{10} + 36363 x^{11} - 270391 x^{12} - 424173 x^{13}\nn
    &+ 219178 x^{14} + 505557 x^{15} - 3042062 x^{16} - 4952124 x^{17} + 
      5787235 x^{18} + 12365862 x^{19}\nn
    &- 1529844 x^{20} - 11109657 x^{21} + 
      1982588 x^{22} + 14755839 x^{23} + 7961410 x^{24} - 3333816
      x^{25}\nn
    &-7012037 x^{26} - 3333816 x^{27} + 7961410 x^{28} + 14755839 x^{29} + 
      1982588 x^{30} - 11109657 x^{31}\nn
    &- 1529844 x^{32} + 12365862 x^{33} + 
      5787235 x^{34} - 4952124 x^{35} - 3042062 x^{36} + 505557
      x^{37}\nn
    &+219178 x^{38} - 424173 x^{39} - 270391 x^{40} + 36363 x^{41} + 
      140847 x^{42} + 179832 x^{43} + 74478 x^{44} \nn
    &- 59742 x^{45} - 
      42735 x^{46} + 5967 x^{47} + 7812 x^{48} + 546 x^{49} - 318 x^{50} - 
      9 x^{51} - x^{52}) Y \nn
    &+x^2 (1 + 9 x + 319 x^2 - 531 x^3 - 7770 x^4 - 6363 x^5 + 43157 x^6 + 
      66417 x^7 - 72273 x^8 \nn
    &- 213864 x^9 - 169562 x^{10} + 14532 x^{11} + 
      375824 x^{12} + 532452 x^{13} - 180770 x^{14} - 559596 x^{15}\nn
    &+2833078 x^{16} + 4801236 x^{17} - 5971218 x^{18} - 14640504 x^{19} - 
      223658 x^{20} + 13591086 x^{21} \nn
    &+ 3954688 x^{22} - 7274442 x^{23} - 
      8512366 x^{24} - 7274442 x^{25} + 3954688 x^{26} + 13591086
      x^{27}\nn
    &-223658 x^{28} - 14640504 x^{29} - 5971218 x^{30} + 4801236 x^{31} + 
      2833078 x^{32} - 559596 x^{33} \nn
    &- 180770 x^{34} + 532452 x^{35} + 
      375824 x^{36} + 14532 x^{37} - 169562 x^{38} - 213864 x^{39} - 
      72273 x^{40} \nn
    &+ 66417 x^{41} + 43157 x^{42} - 6363 x^{43} - 7770 x^{44} - 
      531 x^{45} + 319 x^{46} + 9 x^{47} + x^{48}) Y^2)/\nn
    &(1152 (1 + 3 x + 3 x^2 + 3 x^3 + x^4) (1 - 2 x - x^2 + 8 x^3 - 
      11 x^4 + 8 x^5 - x^6 - 2 x^7 + x^8)^3 \nn
    &(1 + 2 x - x^2 - 8 x^3 - 
      11 x^4 - 8 x^5 - x^6 + 2 x^7 + x^8)^3)
  \end{align}
\end{subequations}

\printindex

\bibliographystyle{amsmod} 
\bibliography{Papers}

\end{document}